\documentclass[dvips]{article}
\usepackage{supertabular,lscape,epsfig}
\usepackage{amssymb}
\usepackage{amsmath}

\DeclareSymbolFont{ppa}{OT1}{ppl}{m}{it}
\DeclareMathSymbol{\vv}{\mathalpha}{ppa}{'166}

\begin{document}

\newcommand{\ie}{{\it i.e.},\,}
\newcommand{\etal}{{\it et al.\ }}
\newcommand{\eg}{{\it e.g.},\,}
\newcommand{\cf}{{\it cf.\ }}
\newcommand{\vs}{{\it vs.\ }}
\newcommand{\MS}{{\rm M}\ifmmode_{\odot}\else$_{\odot}$\fi}

\newcommand{\Abstract}[2]{{\footnotesize\begin{center}ABSTRACT\end{center}
\vspace{1mm}\par#1\par
\noindent
{~}{\it #2}}}

\newcommand{\TabCap}[2]{\begin{center}\parbox[t]{#1}{\begin{center}
  \small {\spaceskip 2pt plus 1pt minus 1pt T a b l e}
  \refstepcounter{table}\thetable \\[2mm]
  \footnotesize #2 \end{center}}\end{center}}

\newcommand{\TableSep}[2]{\begin{table}[p]\vspace{#1}
\TabCap{#2}\end{table}}

\newcommand{\FigCap}[1]{\footnotesize\par\noindent Fig.\  %
  \refstepcounter{figure}\thefigure. #1\par}

\newcommand{\TableFont}{\footnotesize}
\newcommand{\TableFontIt}{\ttit}
\newcommand{\SetTableFont}[1]{\renewcommand{\TableFont}{#1}}

\newcommand{\MakeTable}[4]{\begin{table}[htb]\TabCap{#2}{#3}
  \begin{center} \TableFont \begin{tabular}{#1} #4 
  \end{tabular}\end{center}\end{table}}

\newcommand{\MakeTableSep}[4]{\begin{table}[p]\TabCap{#2}{#3}
  \begin{center} \TableFont \begin{tabular}{#1} #4 
  \end{tabular}\end{center}\end{table}}

\newenvironment{references}%
{
\footnotesize \frenchspacing
\renewcommand{\thesection}{}
\renewcommand{\in}{{\rm in }}
\renewcommand{\AA}{Astron.\ Astrophys.}
\newcommand{\AAS}{Astron.~Astrophys.~Suppl.~Ser.}
\newcommand{\ApJ}{Astrophys.\ J.}
\newcommand{\ApJS}{Astrophys.\ J.~Suppl.~Ser.}
\newcommand{\ApJL}{Astrophys.\ J.~Letters}
\newcommand{\AJ}{Astron.\ J.}
\newcommand{\IBVS}{IBVS}
\newcommand{\PASP}{P.A.S.P.}
\newcommand{\Acta}{Acta Astron.}
\newcommand{\MNRAS}{MNRAS}
\renewcommand{\and}{{\rm and }}
\section{{\rm REFERENCES}}
\sloppy \hyphenpenalty10000
\begin{list}{}{\leftmargin1cm\listparindent-1cm
\itemindent\listparindent\parsep0pt\itemsep0pt}}%
{\end{list}\vspace{2mm}}

\def\TYLDA{~}
\newlength{\DW}
\settowidth{\DW}{0}
\newcommand{\dw}{\hspace{\DW}}

\newcommand{\refitem}[5]{\item[]{#1} #2%
\def\REFARG{#3}\ifx\REFARG\TYLDA\else, {\it#3}\fi
\def\REFARG{#4}\ifx\REFARG\TYLDA\else, {\bf#4}\fi
\def\REFARG{#5}\ifx\REFARG\TYLDA\else, {#5}\fi.}

\newcommand{\Section}[1]{\section{#1}}
\newcommand{\Subsection}[1]{\subsection{#1}}
\newcommand{\Acknow}[1]{\par\vspace{5mm}{\bf Acknowledgements.} #1}
\pagestyle{myheadings}

\newfont{\bb}{ptmbi8t at 12pt}
\newcommand{\xrule}{\rule{0pt}{2.5ex}}
\newcommand{\xxrule}{\rule[-1.8ex]{0pt}{4.5ex}}
\def\thefootnote{\fnsymbol{footnote}}
\begin{center}
{\Large\bf Resonant Excitation of Nonradial Modes in RR~Lyr Stars}
\vskip1cm
{Rafa\l~~M.~~N~o~w~a~k~o~w~s~k~i$^1$~~ and~~ 
Wojciech~~A.~~D~z~i~e~m~b~o~w~s~k~i$^{1,2}$}
\vskip3mm
{$^1$ Nicolaus Copernicus Astronomical Center, ul.~Bartycka~18, 
00-716~Warsaw, Poland\\
e-mail: rafaln@camk.edu.pl\\
$^2$ Warsaw University Observatory, Al.~Ujazdowskie~4, 00-478~Warsaw, Poland\\
e-mail: wd@astrouw.edu.pl}
\end{center}

\Abstract{We study a nonlinear development of radial pulsation instability to 
a resonant excitation of nonradial modes. Our theory covers the cases of 
axisymmetric ${(m=0)}$ modes as well as ${(m,-m)}$ pairs. Adopting a simplified 
treatment of the radial and nonradial mode coupling we find that the 
asymptotic state is a pulsation with constant amplitudes and we evaluate the 
relative amplitude of the nonradial component. 

Observable consequence of the $m=0$ mode excitation is a small period change 
and a more significant amplitude change, especially in the case of a dipole 
mode (${l=1}$). Such a mode has a fairly large excitation probability. 

Significant amplitude and phase modulation is predicted in the case of 
excitation of a ${m=\pm1}$ pair. We suggest that this may explain Blazhko-type 
modulation in RR~Lyr stars. If this model is correct, the modulation period 
is determined by the rotation rate and the Brunt-V\"ais\"al\"a frequency in 
the deepest part of the radiative interior.}{Stars: oscillations -- Stars: 
variables: RR~Lyr}

\Section{Introduction}
Among a variety of types of pulsating stars, RR~Lyr variables seem to be the 
most important ones. They are significant in various fields of astronomy: 
distance measurements, evolution of Population~II and low mass stars, 
structure of Galaxy and, obviously, stellar pulsation theory. Thus, it is 
important to understand the nature of all phenomena connected with their 
pulsations. One can distinguish a few subclasses of RR~Lyr stars: RRab stars 
pulsating in the fundamental radial mode with large amplitude; RRc stars -- 
the first overtone pulsators with small amplitude and sinusoidal light curves; 
RRd stars that pulsate in both fundamental and first overtone modes. 

The most intriguing phenomenon that has been waiting for theoretical 
explanation for almost a century is the Blazhko effect. It is characterized by 
periodic modulation of oscillation amplitude and phase on a longer time scale. 
The effect was discovered by Blazhko (1907). Basic properties of observed 
Blazhko-type modulations are summarized by Szeidl (1988) and Kov\'acs (1993, 
2000). According to these authors, about 20--30\% of all RRab stars exhibit 
this phenomenon. The modulation is often quite strong, pulsation amplitude may 
change by a factor of up to 3 during the Blazhko cycle. Fourier analysis of 
light curves with the Blazhko effect shows two peaks, symmetrically spaced on 
both sides of the main peak corresponding to the basic oscillation frequency, 
as well as similar peaks around harmonics of the main frequency. This feature 
of Fourier spectrum is by some treated as a definition of the phenomenon. 
However, we should stress that the Fourier analyses are not available for most 
of RRab stars reported to show the Blazhko effect. 

Whether RRc stars exhibit this type of modulations was a matter of 
speculations for years. Recently such variables were discovered in the LMC by 
the MACHO project (see Kurtz \etal 2000, Alcock \etal 2000). However, 
incidence of the Blazhko phenomenon among RRc stars is much lower than in the 
RRab case and is estimated at about 2\%. 

Early explanations of the Blazhko effect postulated the existence of strong 
magnetic field with the axis inclined to the axis of rotation. This field was 
thought to cause a nonspherical distortion of the pulsation. Stellar rotation 
led to the amplitude modulation seen by observers. A mathematical model of 
magnetically modified pulsation was developed by Shibahashi (1995) (see also 
Shibahashi 2000). There are many basic problems connected with this model (see 
Kov\'acs 1993, 2000) and we will not deal with it in this paper. 

Alternative interpretations invoke nonlinear interactions between oscillation 
modes. Two approaches have been adopted: numerical simulations of 
fully-nonli\-near pulsation and an approximate formalism of amplitude equations 
(AEs). The latter is the only available formalism for nonradial oscillations. 
It proved particularly useful to study effects of resonances (see \eg 
Dziembowski 1982, Buchler and Goupil 1984, Van~Hoolst 1992, 1994). The 
AE-formalism finds its applications in various fields of physics. Our 
application bears most resemblance with work of Wersinger \etal (1980) who 
studied nonlinear interaction of plasma waves in the case of parametric 
resonance. These authors showed that solutions of AEs may be stationary 
(fixed-point), periodically modulated (periodic limit cycle) as a result of 
Hopf-bifurcation, or even chaotic. In the case of pulsating stars parametric 
resonance is likely responsible for limiting of oscillation amplitudes in 
$\delta$-Sct stars (Dziembowski and Kr\'olikowska 1985). 

Moskalik (1986) proposed 2:1 resonance between the fundamental and the third 
overtone modes as a possible explanation of the Blazhko effect which would be 
a manifestation of the periodic limit cycle solution of the corresponding AEs. 
However, the values of parameters necessary to obtain modulated solutions seem 
to be outside the range of RR~Lyr models (Kov\'acs 1993 and references 
therein). Moreover, there is no sign of modulation in hydrodynamic simulations 
(Kov\'acs and Buchler 1988). 

Nowadays, the most promising hypotheses involve excitation of nonradial 
mo\-des. Properties of such modes in giant pulsators (RR~Lyr, Cepheids) were 
studied already over twenty years ago (Dziembowski 1977a, Osaki 1977) but 
until the early 90s these papers were largely out of interest of 
astronomers. 

Van~Hoolst (1992) derived AEs for resonance between two modes with nearly 
equal frequencies (the 1:1 resonance) in the adiabatic approximation. He 
calculated polytropic stellar model and found a nonradial mode with frequency 
close to that of the second radial overtone. For these two modes he calculated 
some parameters in AEs. For other parameters he considered wide ranges of 
values. Depending on the values of parameters he found different types of 
solutions which include single mode pulsation, double mode pulsation with 
constant amplitudes, pulsation with periodically or chaotically modulated 
amplitudes. 

Kov\'acs (1993) and Van~Hoolst and Waelkens (1995) proposed the 1:1 resonance 
between the radial fundamental mode and a nonradial mode as a possible 
explanation of the Blazhko effect. This type of resonance is not possible 
between low order radial modes. Van~Hoolst \etal (1998, hereafter VDK) showed 
that in realistic RR~Lyr models, nonradial mode spectra in the range of the 
fundamental and the first overtone radial modes are very dense. Hence the 
frequency distance between the radial and the nearest nonradial mode is very 
small. VDK studied stability of monomode radial pulsation in a representative 
RR~Lyr star model. They found high probability of resonant excitation of the 
nonradial mode. The highest probability occures for modes with ${l=1,5,6}$ and 
${l=1,4}$ at fundamental and first overtone modes, respectively. Dziembowski 
and Cassisi (1999, hereafter DC) repeated similar calculations for a set of 
realistic RR~Lyr models covering the whole range of effective temperatures 
corresponding to instability strip in RR~Lyr stars luminosity region and found 
that conclusions of VDK are generally valid. They found, in particular, that 
the probability of resonant excitation of nonradial modes is about 0.5, which 
is not far from the observed incidence rate of the Blazhko effect among RRab 
stars. According to their calculations, the effect should occur among first 
overtone pulsators nearly as often as among RRab stars, which is in a strong 
contradiction with observations. However, they stressed that excitation of a 
nonradial mode does not necessarily imply amplitude modulations. 

Nonlinear consequences of nonradial mode excitation due to the 1:1 resonance 
have never been studied in application to RR~Lyr stars. The Blazhko effect may 
be a manifestation of periodic limit cycle solution of AEs, but it may also be 
caused by nonradial mode excitation with constant amplitude. The latter 
possibility requires $m\ne0$. In this case, the Blazhko period is related to 
but not identical with the rotation period. 

In this paper we extend the formalism of AEs used by VDK to study solutions 
with finite amplitudes of nonradial modes. We keep simple scaling of resonant 
coefficients adopted by DC. 

The plan of the paper is as follows. In Section~2 we outline the basic linear 
properties of nonradial modes in evolutionary models of RR~Lyr stars. The 
properties are described in details by VDK and DC. In Section~3 we transform 
AEs of VDK into the form used in further part of this paper, by adopting some 
simplifications, \eg above mentioned scaling of coefficients. We study 
stationary solutions and their stability in Section~4, as well as time 
dependent solutions and attractors in Section~5. Section~6 is devoted to 
estimating the mode amplitudes and period changes caused by the resonant 
interaction in all DC's models, as well as the observational effect of ${m=0}$ 
mode presence. In Section~7 we explain how the ${(m,-m)}$ pair excitation may 
cause the observed amplitude and phase modulation. Section~8 includes 
conclusions and discussion of the results. 

\Section{Nonradial Modes in RR~Lyr Models}
RR~Lyr stars are Horizontal Branch helium burning giants, characterized by 
strong mass concentration in the center. VDK showed that in such stars the 
oscillation equations can be solved analytically in the deep interior, using 
an asymptotic approximation. The solution may be matched to the numerical 
solution in the envelope. Applying this procedure one can determine basic 
properties of nonradial modes. The most important is that there is always 
dense spectrum of nonradial modes near the two lowest degree radial mode 
frequencies. For a given $l$, the mode frequency decreases with increasing 
$n$. This is so because, owing to very high values of the Brunt-V\"ais\"al\"a 
frequency in the interior, all nonradial modes are in fact $g$-modes of high 
degree. When rotation is ignored all modes with the same $l,n$ have the same 
frequency. Non-zero rotation means still denser spectrum for each $l$. In the 
envelope nonradial modes closest in frequency to a given radial mode have 
their radial eigenfunctions almost the same as the radial mode eigenfunctions. 
This remains true as long as the mode frequency is much higher than the Lamb 
frequency (${L_l=c(l+0.5)/r}$ where $c$ is the local sound velocity). 

Radial modes are of course purely acoustic and they propagate only in the 
envelope, where the Brunt-V\"ais\"al\"a frequency is lower than their 
eigenfrequencies. The outer layers determine properties of radial modes and 
the interior plays no role. 

From linear nonadiabatic calculations we find time dependence of perturbations 
in the form ${\exp({\rm i}\omega_j+\kappa_j)}$ where $\omega_j$ is the 
eigenfrequency and $\kappa_j$ is the linear growth rate of mode numbered by 
$j$. The growth rate may be expressed as (see \eg Unno \etal 1989) 
$$\kappa_j=\frac{W_j}{2I_j\omega_j}\eqno(1)$$
where $I_j$ is the inertia of the mode and $W_j$ is the work integral, \ie
the integral over star's volume of nonadiabatic operator acting on the 
eigenfunction. In the case of the radial mode it is enough to integrate over 
the envelope, because the contribution from the deep interior is negligible. 
In the case of the nonradial modes one may calculate separately the integrals 
over the deep interior (gravity waves propagation zone) and over the envelope 
(acoustic waves propagation zone). Moreover, the eigenfrequencies of both 
modes are nearly the same, and one can put $\omega_0$ instead of $\omega_l$ in 
the formula for the nonradial mode growth rate. With above assumptions one 
obtains 
\setcounter{equation}{1}
\begin{eqnarray}
\kappa_0&=&\frac{W_0}{2I_0\omega_0},\\
\kappa_l&=&\frac{W_0+W_g}{2I_l\omega_0}\equiv\kappa_{l,0}+\kappa_g.
\end{eqnarray}
The quantity ${\kappa_g=W_g/(2I_l\omega_0)}$, which is always negative, 
measures the damping rate in the deep interior. 

DC calculated three sets of evolutionary models of RR~Lyr stars (for different 
masses and metallicities). They used the procedure described by VDK for these 
models and calculated properties of the two lowest radial modes and the 
corresponding resonant nonradial modes for a few lowest $l$'s. In the vicinity 
of the fundamental mode frequency, nonradial modes with ${l\ge7}$ are damped 
very strongly and they are never excited due to 1:1 resonance with the radial 
mode. The same is true for the modes with ${l\ge6}$ in the vicinity of the 
first overtone frequency. 

In all numerical calculations in this paper we use data for models from DC. 

\Section{Amplitude Equations}
The amplitude equations used in this paper are based on those derived by 
Van~Hoolst (1994) and used by VDK in their study of stability of radial 
pulsation in RR~Lyr stars. Here we first derive our simplified form for the 
case of resonant coupling between a radial (${l=0}$) and an axisymmetric 
nonradial mode (${l>0}$, ${m=0}$). Then we will derive corresponding equations 
for the case of  coupling to a nonradial mode pair of the same degrees $l$ and 
opposite azimuthal degrees, $m$ and ${-m}$. We will see that there are no 
essential differences between these two cases. Some details of the derivation 
are deferred to Appendix~A. 

We introduce complex pulsation amplitudes $\tilde{a}$ associated with an 
individual oscillation mode ($j$), which are defined by the following 
expression for the displacement 
$$\frac{\delta\boldsymbol{r}}{r_0}=
\pmb{\xi}_j(\boldsymbol{r},t)=\frac{1}{2}\tilde{a}_j\,{\rm e}^{{\rm i}\omega_j t}
\boldsymbol{w}_j(\boldsymbol{r})+cc,\eqno(4)$$
where $r_0$ is the Lagrangian radial coordinate, $\boldsymbol{w}_j$ is vector 
eigenfunction, and $\omega_j$ is eigenfrequency of the mode. We adopt 
normalization 
$$w_{j,r}=Y_{l_j,m_j}\sqrt{4\pi}\eqno(5)$$
at the surface, which implies that $|\tilde{a}|$ is the {\it rms} amplitude of 
the surface radius. 

\subsection{The $\bb{m=0}$ Case}
The time dependence of amplitudes ${a_j=\tilde{a}_j\exp({\rm i}\omega_jt)}$ is 
described by AEs. For the 1:1 resonance between the radial mode and 
axisymmetric nonradial mode the relevant AEs are (VDK) 
\setcounter{equation}{5}
\begin{eqnarray}
\frac{{\rm d} a_0}{{\rm d} t}&=&
(\kappa_0+{\rm i}\omega_0)a_0+S_0^0a_0|a_0|^2+S_0^la_0|a_l|^2+
\nonumber\\
&&+R_0^1\,a_0^*a_l^2+R_0^2a_l|a_l|^2,\\
\frac{{\rm d} a_l}{{\rm d} t}&=&
(\kappa_l+{\rm i}\omega_l)a_l+S_l^0a_l|a_0|^2+S_l^la_l|a_l|^2+
\nonumber\\
&&+R_l^1\,a_l^*\,a_0^2+R_l^2|a_l|^2a_0+
R_l^3a_l^2\,a_0^*,
\end{eqnarray}
where $S_j^k,\ R_j^k$ denote saturation and resonant coupling coefficients, 
respectively. For odd $l$ coefficients $R_0^2,\ R_l^2$ and $R_l^3$ vanish. In 
our work we devote special attention to the case of ${l=1}$ and thus we neglect 
all the terms involving these coefficients. We believe that application of so 
simplified system to even $l$'s will not result in large error because these 
terms are of higher order in the amplitude of the nonradial modes. We will see 
that the amplitudes of ${l>1}$ modes are always much smaller than that of the 
radial mode. We can now introduce real amplitudes $A_j$ and phases $\phi_j$ in 
the form ${a_j=A_j\exp({\rm i}\omega_jt+\phi_j)}$ and separate Eqs.~(6) and (7) 
into the real and the imaginary parts. This yields 
\begin{eqnarray}
\frac{{\rm d} A_0}{{\rm d} t}&=&\Big[\kappa_0+\Re(S_0^0)A_0^2+\Re(S_0^l)A_l^2+
\nonumber\\
&&+\big(\Re(R_0^1)\cos\Gamma-\Im(R_0^1)\sin\Gamma\big)A_l^2\Big]A_0,\\
\frac{{\rm d} A_l}{{\rm d} t}&=&\Big[\kappa_l+\Re(S_l^0)A_0^2+\Re(S_l^l)A_l^2+
\nonumber\\
&&+\big(\Re(R_l^1)\cos\Gamma+\Im(R_l^1)\sin\Gamma\big)A_0^2\Big]A_l,\\
A_0\frac{{\rm d}\phi_0}{{\rm d} t}&=&\Big[\Im(S_0^0)A_0^2+\Im(S_0^l)A_l^2+
\nonumber\\
&&+\big(\Re(R_0^1)\sin\Gamma+\Im(R_0^1)\cos\Gamma\big)A_l^2\Big]A_0,\\
A_l\frac{{\rm d}\phi_l}{{\rm d} t}&=&\Big[\Im(S_l^0)A_0^2+\Im(S_l^l)A_l^2+
\nonumber\\
&&+\big(-\Re(R_l^1)\sin\Gamma+\Im(R_l^1)\cos\Gamma\big)A_0^2\Big]A_l,
\end{eqnarray}
where we introduced the relative phase,
$$\Gamma\equiv2(\Delta\omega t+\phi_l-\phi_0),\eqno(12)$$
and the detuning parameter, ${\Delta\omega\equiv\omega_l-\omega_0}$. The time 
derivatives of the phases $\phi_j$ give nonlinear corrections to the linear 
mode frequencies. Terms with saturation coefficients in Eqs.~(10) and (11) 
give nonresonant corrections and may be neglected because they move the whole 
spectrum of nonradial modes without significant change of frequency distance 
between consecutive nonradial modes. Moreover, we assume that the resonant 
coupling coefficients may be regarded purely imaginary as in the adiabatic 
case (Van~Hoolst 1992). Although pulsation is never strictly adiabatic, we 
follow here a suggestion of VDK who argued that real parts of these 
coefficients are much smaller than imaginary parts. 

All coupling coefficients are proportional to the complicated integrals 
containing products of four eigenfunctions for the modes involved. These 
integrals can be separated into integrals over radial and angular coordinates. 
The latter are easy to calculate analytically, because they contain well known 
spherical harmonics. The former are not possible to be determined analytically. 
Nevertheless, some useful relations between coefficients can be derived. This 
is due to the fact that the radial eigenfunctions of both radial and nonradial 
modes are almost the same in the outer layers and integration of products of 
different eigenfunctions gives the same result. In Appendix~A.1 it is found, in 
particular, that 
$$S_0^l=2S_0^0,\qquad S_l^l=H_lS_l^0,\eqno(13)$$
where ${H_l=\int\limits_{-1}^1 \tilde{P}_l^4(x){\rm d} x}$ and $\tilde{P}_l^4$ 
is normalized Legendre polynomial. The values of $H_l$ for a few low $l$'s are 
given in Table~1. 
\MakeTable{|c|c|}{\textwidth}{Numerical values of $H_l$}
{\hline
$l$ & $H_l$\\
\hline
1 & 0.9~~~~\\
2 & 1.071\\
3 & 1.180\\
4 & 1.260\\
5 & 1.322\\
6 & 1.373\\ 
\hline}

Like the growth rates (see Eqs.~2 and 3), all the coupling coefficients are 
proportional to the inverse of mode inertia. Thus, they satisfy the relations 
$$\frac{\Re(S_0^0)}{\kappa_0}=\frac{\Re(S_l^0)}{2\kappa_{l,0}}\equiv-
\alpha,\eqno(14)$$
$$\frac{\Im(R_0^1)}{\kappa_0}=\frac{\Im(R_l^1)}{\kappa_{l,0}}\equiv 
R.\eqno(15)$$
Let us note that $\alpha$ determines the amplitude of single-mode radial 
pulsation. 

When we introduce the dimensionless time ${\tau\equiv\kappa_0t}$, the 
dimensionless frequency distance ${\Delta\sigma\equiv\Delta\omega/\kappa_0}$, 
and the following quantities 
$$\gamma\equiv-\frac{\kappa_g}{\kappa_{l,0}},\qquad 
I\equiv\frac{I_0}{I_l}=\frac{\kappa_{l,0}}{\kappa_0},\eqno(16)$$
we can, with the help of Eqs.~(2), (3), (14), and (15), rewrite Eqs.~(8)--(11) 
in the form 
\setcounter{equation}{16}  
\begin{eqnarray}
\frac{{\rm d} A_0}{{\rm d} \tau}&=&
\left[1-\alpha\left(A_0^2+2A_l^2\right)-R\sin\Gamma\,A_l^2\right]A_0,\\
\frac{{\rm d} A_l}{{\rm d} \tau}&=&
I\left[1-\gamma-2\alpha\left(A_0^2+H_lA_l^2\right)+
R\sin\Gamma\,A_0^2\right]A_l,\\
A_0\frac{{\rm d} \phi_0}{{\rm d} \tau}&=&R\cos\Gamma\,A_l^2A_0,\\
A_l\frac{{\rm d} \phi_l}{{\rm d} \tau}&=&IR\cos\Gamma\,A_0^2A_l.
\end{eqnarray}
We neglected the real parts of the resonant coefficients and the imaginary 
parts of the saturation coefficients, as was explained above. From the last 
two equations and Eq.~(12), with an additional assumption of non-zero values 
of both amplitudes, we obtain 
$$\frac{{\rm d} \Gamma}{{\rm d} \tau}=
2\left(\Delta\sigma+\frac{{\rm d}\phi_l}{{\rm d}\tau}-
\frac{{\rm d}\phi_0}{{\rm d} \tau}\right)=
2\left(\Delta\sigma+R\cos\Gamma\,\left(IA_0^2-A_l^2\right)\right).\eqno(21)$$
Eqs.~(17), (18), and (21) form a complete set of equations and we will use 
them to study double-mode solutions. The equations are invariant under the 
transformation 
$$(\Delta\sigma,\Gamma)\rightarrow(-\Delta\sigma,\pi-\Gamma).\eqno(22)$$
Thus, we will restrict ourselves to ${\Delta\sigma\ge0}$.

\subsection{The $\bb{m\ne0}$ Case}
Resonant interaction between radial and nonradial modes with nearly 
equal frequencies is only possible when the nonradial mode has azimuthal 
number $m$ equal zero. In other cases, all resonant coupling coefficients 
vanish, as was shown by Van~Hoolst (1994). Nevertheless, modes with ${m\ne0}$ 
can interact with a radial mode due to the resonance between the radial mode 
and a pair of nonradial modes of the same value of $l$ and opposite values of 
$m$. Analogical situation of resonance between modes of rotationally split 
triplet ${l=1}$ is described by Buchler \etal (1995). The difference is that 
in our case the ${m=0}$ mode is replaced by the radial mode. These authors 
call this kind of resonance the 1:1:1 resonance but we think a more 
appropriate name is the 2:1+1 resonance. The reason is that the three 
frequencies satisfy the resonant condition 
${2\omega_0\approx\omega_m+\omega_{-m}}$. 

The AEs for the above-mentioned type of resonance have the form given \eg by 
Buchler \etal (1995). The relations between coupling coefficients similar to 
those in the ${m=0}$ case are derived in Appendix~A.2. The relative phase 
$\Gamma$ in this case is given by 
$$\Gamma=2\Delta\omega t+\phi_-+\phi_+-2\phi_0\eqno(23)$$
where ${\Delta\omega=(\omega_-+\omega_+)/2-\omega_0}$. Such a definition of 
$\Gamma$ and $\Delta\omega$ is consistent with the ${m=0}$ case. In further 
studies we will assume that rotation is slow and except for the linear 
eigenfrequencies it does not change the dynamical properties of the nonradial 
modes. The radial eigenfunctions of these modes are $m$-independent. Thus, we 
have a full symmetry between coefficients in equations for amplitudes and 
phases of both nonradial modes. This means, in particular, that 
$$\kappa_-=\kappa_+\equiv\kappa_l,\qquad S_-=S_+\equiv S_l,
\qquad R_-=R_+\equiv R_l\eqno(24)$$
where quantities with subscript $l$ are the same as in the ${m=0}$ case. We 
stress that in the present case there is only one non-vanishing resonant 
coefficient also for even $l$'s. Similarly as in the ${m=0}$ case we 
introduce dimensionless quantities and we obtain 
\setcounter{equation}{24}
\begin{eqnarray}
\frac{{\rm d} A_0}{{\rm d} \tau}&=&
\left[1-\alpha\left(A_0^2+2A_-^2+2A_+^2\right)
-2R\sin\Gamma\,A_-A_+\right]A_0,\\
\frac{{\rm d} A_-}{{\rm d} \tau}&=&
I\left[1-\gamma-2\alpha\left(A_0^2+H_l^m(A_-^2+2A_+^2)\right)\right]\!A_-
+IR\sin\Gamma\,A_0^2A_+,\quad\\
\frac{{\rm d} A_+}{{\rm d} \tau}&=&
I\left[1-\gamma-2\alpha\left(A_0^2+H_l^m(2A_-^2+A_+^2)\right)\right]\!A_+
+IR\sin\Gamma\,A_0^2A_-,\quad\\
A_0\frac{{\rm d} \phi_0}{{\rm d} \tau}&=&2R\cos\Gamma\,A_-A_+A_0,\\
A_-\frac{{\rm d} \phi_-}{{\rm d} \tau}&=&IR\cos\Gamma\,A_0^2A_+,\\
A_+\frac{{\rm d} \phi_+}{{\rm d} \tau}&=&IR\cos\Gamma\,A_0^2A_-,
\end{eqnarray}
where ${H_l^m=\int\limits_{-1}^1\!\tilde{P}_{l,m}^4(x)\,{\rm d} x}$, 
$\tilde{P}_{l,m}$ is normalized associated Legendre function and 
Eqs.~(23) and (28)--(30) allow us to write equation for the phase $\Gamma$ 
$$\frac{{\rm d}\Gamma}{{\rm d}\tau}=2\Delta\sigma+R\cos\Gamma
\left[IA_0^2\left(\frac{A_+}{A_-}+\frac{A_-}{A_+}\right)-4A_-
A_+\right].\eqno(31)$$
Eqs.~(25)--(27) together with Eq.~(31) form a complete system of equations 
which allow us to determine the amplitudes of the modes participating in 
oscillations. 

When we multiply Eq.~(26) by $A_-$, Eq.~(27) by $A_+$, and subtract them,
we obtain 
$$\frac{{\rm d}}{{\rm d}\tau}\left(A_-^2-A_+^2\right)=2I\left[1-\gamma-
2\alpha A_0^2-2\alpha H_l^m(A_-^2+A_+^2)\right]\left(A_-^2-
A_+^2\right).\eqno(32)$$
It will be seen further that a typical value of $A_0^2$ for stationary 
solutions is about $1/\alpha$ (exact equality is for single-mode solution). 
This means negative value of the square bracket in Eq.~(32) and exponential 
decrease of the difference of the two amplitudes. Thus, we limit ourselves to 
the ${A_-=A_+}$ solution. 

Let us now focus on the case of modes with ${l=1}$, ${m=\pm1}$. Then we have 
${H_1^1=3/5}$. When we introduce a pseudo amplitude $A_l$ 
$$A_-=A_+\equiv\frac{A_l}{\sqrt{2}}.\eqno(33)$$
then Eqs.~(25)--(27) and (31) are identical with Eqs.~(17), (18), and (21), 
where ${H_1=9/10}$. Thus, we may consider the pair of nonradial modes ${l=1}$, 
${m=\pm1}$ as an equivalent of the mode ${l=1}$, ${m=0}$. This fact has a 
simple physical and geometrical explanation: spherical harmonic $Y_{1,0}$ 
is equivalent to the sum of spherical harmonics ${Y_{1,-1}+Y_{1,1}}$ divided 
by $\sqrt2$ in the system of coordinates with the $z$-axis inclined by 
$\pi/2$. The pair of modes ${l=1}$, ${m=\pm1}$ excited with the same 
amplitudes form a distortion of the star that has the same shape as the 
distortion associated with the mode ${l=1}$, ${m=0}$ but with the axis of 
symmetry lying in the plane of equator. Since all physical mode 
characteristics are $m$-independent such a pair interacts with the radial mode 
in the same way as the ${l=1}$, ${m=0}$ mode. Therefore, the amplitude $A_l$ 
determined as a solution of Eqs.~(17), (18), and (21) gives also the 
amplitudes of the nonradial modes ${A_\pm=A_l/\sqrt2}$. 

We treat the interaction of the radial mode with the ${l=1}$, ${m=0}$ mode and 
with the ${l=1}$, ${m=\pm1}$ pair independently. Strictly speaking this is 
justified only for the instability analysis. The small quadratic effects in 
the rotation rate cause that ${(\omega_{1,1}+\omega_{1,-1})/2\ne\omega_{1,0}}$ 
and these two cases are indeed independent. However, at finite amplitudes of 
the nonradial modes the interaction between the modes within the triplet is 
likely to be important. It is possible that the whole triplet will be exited 
but this should not have large consequences for amplitude prediction. 

Similarity of the 1:1 and the 2:1+1 resonances, involving the ${(m,-m)}$ 
doublet, extends to ${l>1}$. The only difference concerns the $H_l$ 
coefficient (see Eqs.~26 and 27). 

It is also possible for a pair of nonradial modes belonging to different 
multiplets (\ie of different radial orders) to be excited. Then the the 
amplitudes of the two modes cannot be assumed equal. The difference is mainly 
due to the difference in mode inertia, which is strongly frequency sensitive. 
We define ${I_\pm}$ similarly as in Eq.~(16), that is we set 
$$I_\pm=\frac{I_0}{I_{l,\pm m}},\eqno(34)$$
and, ignoring the difference in the coefficients at $H_l^m$, we get
$$\frac{A_+}{A_-}=\sqrt{\frac{I_+}{I_-}}.\eqno(35)$$
If this ratio is not too different from unity then, upon introducing
$$I\equiv\sqrt{I_-I_+},\qquad A_l^2\equiv2A_+A_-,\eqno(36)$$
we reduce equations describing such interaction to Eqs.~(17), (18), and (21) 
which describe the two-mode resonance. 

\Section{Stationary Solutions}
Stationary, or fixed-point solutions of the amplitude equations are obtained 
from Eqs.~(17), (18), and (21) where we set ${{\rm d} A_j/{\rm 
d}\tau=0}$. Then the nonlinear frequencies are given by ${\omega_j+{\rm 
d}\phi_j/{\rm d}\tau}$, and, like the amplitudes, are constant. 

\subsection{Single-Mode Solutions}
Eqs.~(17), (18) admit single mode solutions
$$(A_0^0)^2=\frac{1}{\alpha},\quad (A_l^0)^2=0,\eqno(37)$$
and
$$(A_0)^2=0,\quad(A_l)^2=\frac{1-\gamma}{2\alpha H_l}.\eqno(38)$$
It is of some interest to investigate stability of these solutions in the 
absence of resonances. Perturbation of Eqs.~(17) and (18) with ${R\equiv0}$ 
leads to a simple eigenvalue problem for the growth rate $\beta$. We find for 
the radial mode solution ${\beta=-2}$ and ${-(1+\gamma)}$. Thus, this is a 
stable solution. For the nonradial mode solution we find ${\beta=-1-3\gamma}$ 
and ${1-(1-\gamma)/H_l}$. The latter eigenvalue is less than zero, if 
${\gamma<1-H_l}$. Since ${\gamma>0}$, this may happen only for ${l=1}$ because 
as we may see in Table~1 ${H_l>1}$ for ${l>1}$. This means that in principle 
we could expect that some of RR~Lyr stars are ${l=1}$ pulsators. However, we 
find it unlikely because even if $\gamma$ is very small the growth of the 
${l=1}$ mode is much slower than that of ${l=0}$. 

A survey of stability of radial pulsation to excitation of resonant nonradial 
modes in RR~Lyr star models was given by DC. However, there was a need to 
repeat the calculations because of an error in the adopted form of the 
stability criterion. The correct criterion derived by VDK is 
$$\Delta\omega^2>|R_l^1|^2(A_0^0)^2-g^2\eqno(39)$$
where 
$$g=\kappa_l+\Re(S_l^0)(A_0^0)^2\eqno(40)$$
is the nonlinear growth rate of the nonradial mode, and other symbols are the 
same as in Eq.~(9). In our dimensionless notation Eq.~(39) has the form 
$$\Delta\sigma^2>I^2\left[\frac{R^2}{\alpha^2}-(1+\gamma)^2\right],\eqno(41)$$
where we used Eq.~(37) and the nonlinear growth rate of the nonradial mode is 
easily obtained from Eq.~(18). In statistical approach, the above criterion 
implies that the probability of instability of radial-mode fixed-point 
solution is given by 
$$P_l=\frac{I\sqrt{R^2/\alpha^2-(1+\gamma)^2}}{\Delta\sigma_{\rm max}}\eqno(42)$$
where $\Delta\sigma_{\rm max}$ is the maximum absolute value of the detuning 
parameter, which in the case of dense spectra of nonradial modes in RR~Lyr 
stars is given by a half of the frequency distance between consecutive 
nonradial modes, ${(\sigma_{l,n}-\sigma_{l,n+1})/2}$. 

DC calculated probabilities of excitation of nonradial modes in the three sets 
of RR~Lyr models. However, they used erroneously ${-I\gamma}$ (in our 
notation) as the nonlinear damping rate of the nonradial mode. From our AEs we 
get the damping rate ${-I(1+\gamma)}$, which implies lower probabilities of 
excitation of nonradial modes, in particular in those cases where $\gamma$ is 
very small and $\alpha$ is comparable to $R$. This is what happens in the 
first overtone vicinity. 

\begin{figure}[htb]
\centerline{\includegraphics[width=10cm]{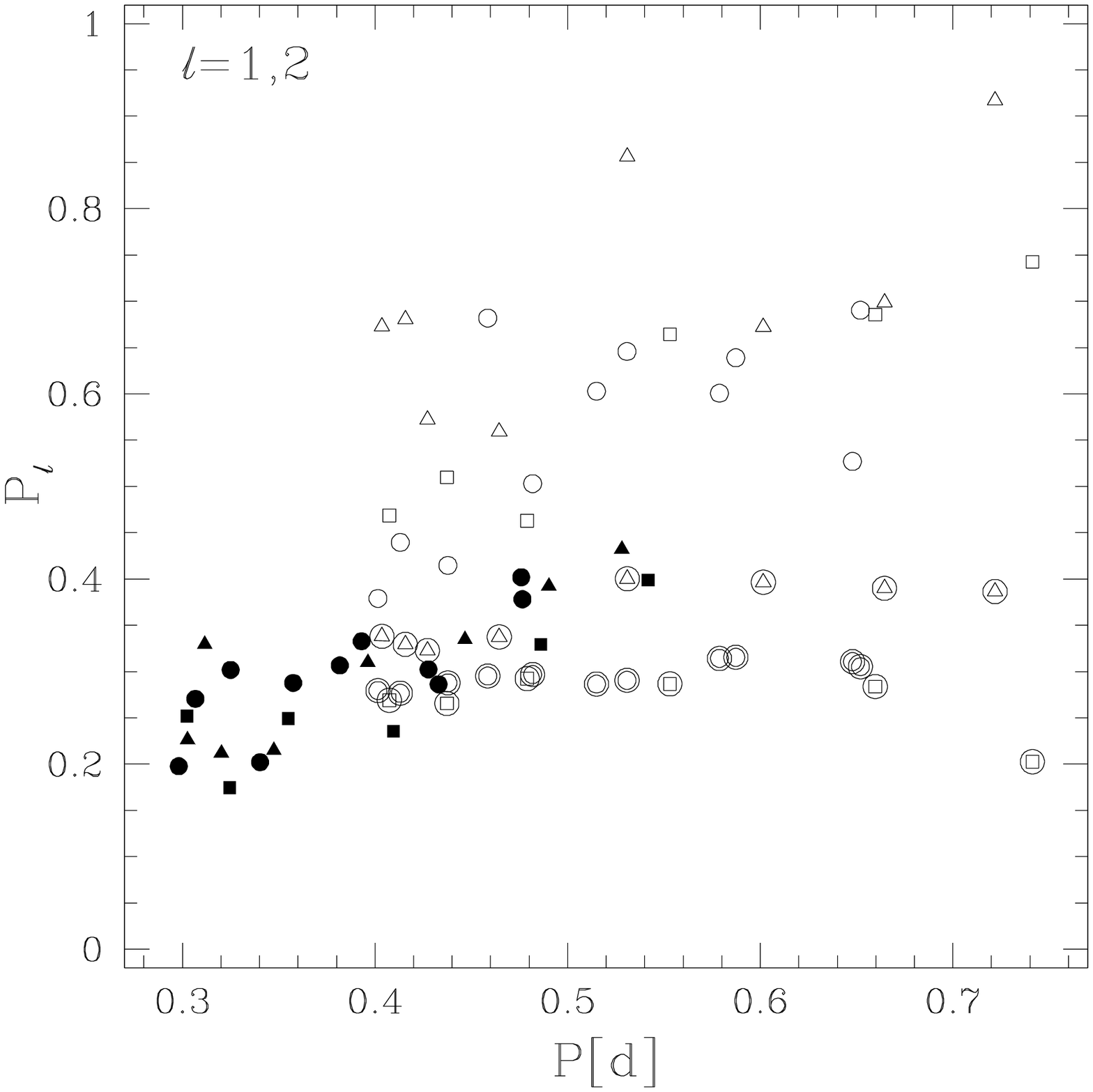}}
\FigCap{Probabilities of excitation of nonradial modes with ${l=1,2}$ in the 
models of DC. Solid and empty symbols denote modes in first overtone and in 
vicinity of fundamental mode, respectively. Encircled symbols indicate ${l=2}$ 
(only for fundamental mode vicinity). Symbols are the same as in DC and refer 
to evolutionary tracks of different initial parameters. Squares and circles 
denote models characterized by ZAHB composition ${Z=0.001}$, ${Y_{\rm 
HB}=0.243}$ and masses 0.65 and 0.67~\MS, respectively. These quantities for 
models denoted by triangles are ${Z=0.0002}$, ${Y_{\rm HB}=0.24}$ and 
0.74~\MS.} 
\end{figure}
\begin{figure}[htb]
\centerline{\includegraphics[width=10cm]{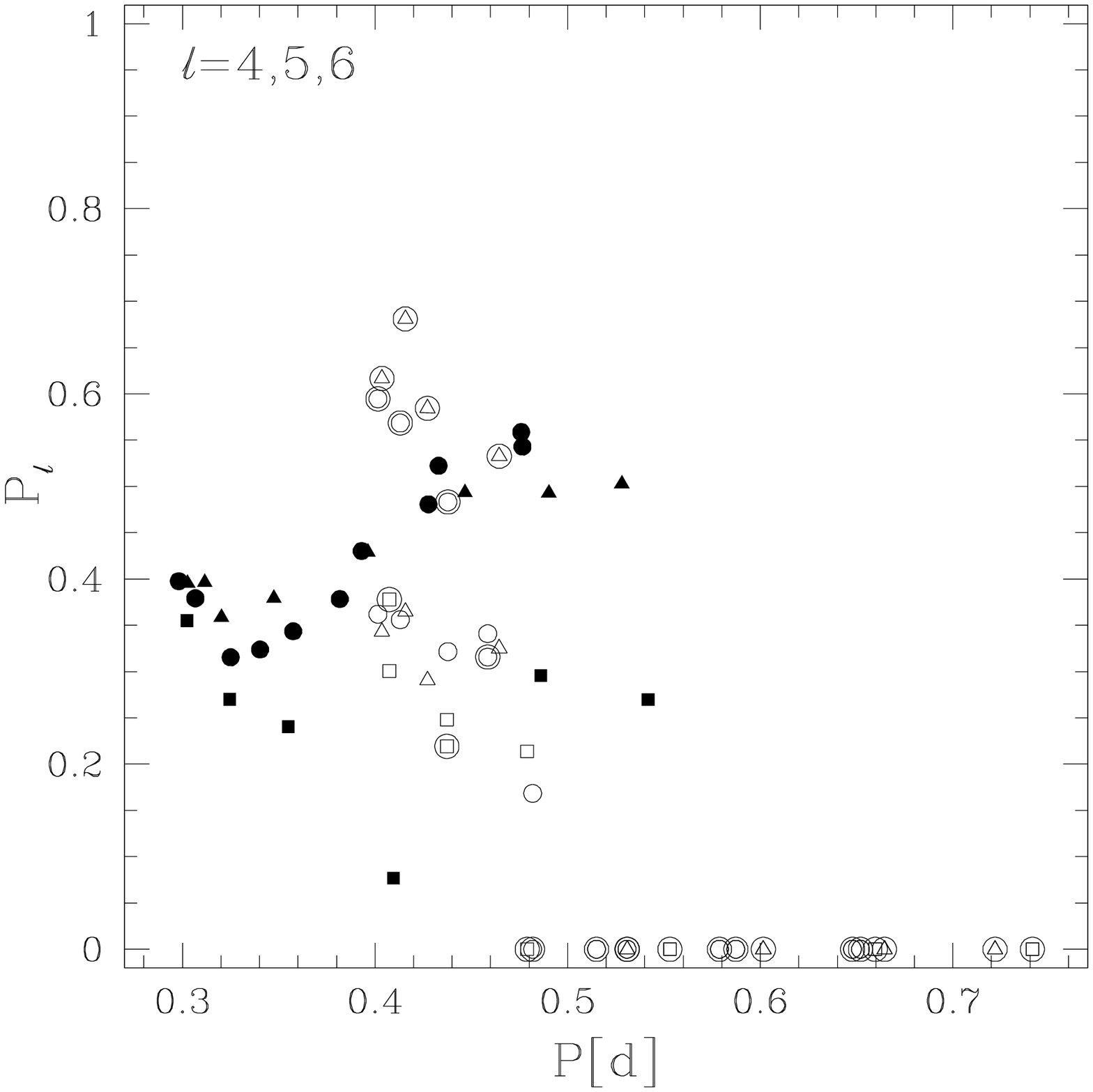}}
\FigCap{The same as in Fig.~1, but for the modes with ${l=4}$ 
(overtone), ${l=5}$ (fundamental) and ${l=6}$ (fundamental, encircled).}
\end{figure}
We calculated these probabilities once again for the DC RR~Lyr models using 
the $R,\gamma,I,\Delta\sigma_{\rm max}$ parameters inferred from the data 
given in DC. We adopted ${\alpha=200}$ for the fundamental mode pulsation and 
1200 for the first overtone, which, according to Eq.~(37), correspond to 
${A_0^0=0.0707}$ and ${A_0^0=0.0289}$, respectively. These values may be 
regarded typical for RRab and RRc stars, respectively, and we will use them 
for all numerical calculations in this paper. Our calculations confirm DC's 
result that the most probably unstable are the ${l=1,5,6}$ modes in the 
fundamental mode vicinity and the ${l=1,4}$ modes in the first overtone 
vicinity. However, in the former vicinity we now find relatively higher 
probability of  excitation for the ${l=2}$ modes. The probabilities are 
presented in Figs.~1 and 2. 

\subsection{Double-Mode Solutions}
When all amplitudes are constant and non-zero, we also have constant $\Gamma$, 
as can be seen in Eqs.~(17), (18), and (21) or in Eqs.~(25)--(27) and (31). 
Constant $\Gamma$ implies the nonlinear frequencies to fulfill exactly the 
resonance condition, \ie phase-lock phenomenon. For the 1:1 resonance it means 
the equality of the two frequencies, while for the 2:1+1 resonance it means 
that the three frequencies are equidistant with the radial mode frequency 
being in the center of the triplet (see. Eqs.~12 and 23, respectively). 
Buchler \etal (1997) give general discussion of the phase-lock for other 
resonances. 

Using Eqs.~(17), (18), and (21) we can express the amplitudes and the detuning 
parameter in terms of the relative phase as follows 
\setcounter{equation}{42}
\begin{eqnarray}
(A_0^{\rm dm})^2&=&
\frac{R\sin\Gamma_{\rm dm}(1-\gamma)-2\alpha(H_l-1+\gamma)}{2\alpha^2(2-H_l)-
(R\sin\Gamma_{\rm dm})^2},\\
(A_l^{\rm dm})^2&=&\frac{\alpha(1+\gamma)-R\sin\Gamma_{\rm dm}}{2\alpha^2(2-H_l)-
(R\sin\Gamma_{\rm dm})^2},\\
\Delta\sigma&=&R\cos\Gamma_{\rm dm}\left[(A_l^{\rm dm})^2-I(A_0^{\rm dm})^2\right].
\end{eqnarray}
In general it is impossible to obtain an explicit expression for $\Gamma_{\rm dm}$ 
as a function of $\Delta\sigma$. Thus, fixed-point solutions have to be 
studied numerically. However, one important property of the solutions can be 
obtained analytically. Let us consider transition between single- 
and double-mode fixed-point, \ie the vicinity of ${A_l^{\rm dm}=0}$. Then 
${R\sin\Gamma=\alpha(1+\gamma)}$ and ${A_0^{\rm dm}=A_0^0}$ (see Eqs.~43, 44, and 37). 
With the help of these expressions we get from Eq.~(45) 
$$|\Delta\sigma|=\Delta\sigma_d\equiv\frac{IR}{\alpha}\sqrt{1-\frac{\alpha^2}{R^2}(1+\gamma)^2}=
I\sqrt{\frac{R^2}{\alpha^2}-(1+\gamma)^2}.\eqno(46)$$
Note that it is the same critical value of $\Delta\sigma$ as given by 
Eq.~(41). It means that the nonradial mode appears with zero amplitude 
at the onset of instability. Further, it may be shown analytically that for 
${|\Delta\sigma|>\Delta\sigma_d}$ there are no double-mode solutions for which 
${A_l/A_0\ll1}$. The analysis is not reproduced here but the property will be 
seen in our numerical examples. 

The value of $\Delta\sigma_d$ is critical for the consequences of the 
resonances considered here. For ${|\Delta\sigma|>\Delta\sigma_d}$ the 
single-mode fixed-point solution is stable. For 
${|\Delta\sigma|<\Delta\sigma_d}$ it is unstable, but a double-mode fixed-point 
occurs. It will be shown that this double-mode fixed-point solution is almost 
always stable. 

\subsection{Stability of Double-Mode Fixed-Point Solution}
Let us note that Eqs.~(17), (18), (21) can be presented in the form 
\setcounter{equation}{46}
\begin{eqnarray}
\frac{{\rm d} A_0^2}{{\rm d}\tau}&=&
2\left(1-\alpha A_0^2-2\alpha A_l^2-R\sin\Gamma A_l^2\right)A_0^2,\\
\frac{{\rm d} A_l^2}{{\rm d}\tau}&=&
2I\left(1-\gamma-2\alpha A_0^2-2H_l\alpha A_l^2+R\sin\Gamma A_0^2\right)A_l^2,\\
\frac{{\rm d}\sin\Gamma}{{\rm d}\tau}&=&2\cos\Gamma\left[\Delta\sigma+R\cos\Gamma(IA_0^2-A_l^2)\right].
\end{eqnarray}
Let us now consider a small perturbation of the double-mode fixed-point
solution,
$$A_0^2=(A_0^{\rm dm})^2+\delta A_0^2,\quad A_l^2=(A_l^{\rm dm})^2+\delta 
A_l^2,\eqno(50)$$
$$\sin\Gamma=\sin\Gamma_{\rm dm}+\delta\sin\Gamma,\quad\cos\Gamma=
\cos\Gamma_{\rm dm}+\delta\cos\Gamma.\eqno(51)$$
Now we insert the above quantities to Eqs.~(47)--(49) and neglect terms of 
second and higher orders with respect to small perturbations. In this way we 
obtain a system of linear differential equations which can be written in a 
matrix form 
$$\frac{{\rm d}{\bf X}}{{\rm d}\tau}={\bf A}{\bf X}\eqno(52)$$
where
$${\bf X}=
\left[\begin{array}{c}\delta A_0^2\\ \delta A_l^2\\ \delta\sin\Gamma
\end{array}\right]
\eqno(53)$$
and
$${\bf A}=2\left[\begin{array}{ccc}
-\alpha A_0^2 & -(2\alpha+R\sin\Gamma)A_0^2 & -RA_0^2A_l^2\\
I(-2\alpha+R\sin\Gamma)A_l^2 & -2IH_l\alpha A_l^2 & IRA_0^2A_l^2\\
IR(1-\sin^2\Gamma) & -R(1-\sin^2\Gamma) & -R\sin\Gamma(IA_0^2-A_l^2)
\end{array}\right].\eqno(54)$$
We made use of ${\cos^2\Gamma=1-\sin^2\Gamma}$, 
${\cos\Gamma\,\delta\cos\Gamma=-\sin\Gamma\,\delta\sin\Gamma}$ relations and 
omitted sub- and superscripts $dm$ denoting double-mode fixed-point solution. 

Eq.~(52) implies time dependence of the perturbations in the 
exponential form, $\exp(\lambda\tau)$. There are three eigenvalues 
determined by the following cubic equation 
$$\lambda^3+a_1\lambda^2+a_2\lambda+a_3=0\eqno(55)$$
where
\setcounter{equation}{55}
\begin{eqnarray}
a_1&=&\alpha(A_0^2+2IH_l A_l^2)+R\sin\Gamma(IA_0^2-A_l^2),\\
a_2&=&I\alpha^2A_0^2A_l^2(2H_l-4)+\alpha R\sin\Gamma(IA_0^2-A_l^2)(A_0^2+2IH_l A_l^2)+
\nonumber\\&&+IR^2A_0^2A_l^2(2-\sin^2\Gamma),\\
a_3&=&IA_0^2A_l^2\big[R\sin\Gamma(IA_0^2-A_l^2)\left(R^2-\alpha^2(4-2H_l)\right)+
\nonumber\\
&&+R^2(1-\sin^2\Gamma)\alpha\left(A_0^2(1+2I)+2A_l^2(1+IH_l)\right)\big].
\end{eqnarray}
The stationary solution is stable when all three eigenvalues have negative 
real parts. The condition for the stability is then given by Hurwitz criteria 
\begin{eqnarray}
a_1&>&0,\\
a_1a_2-a_3&>&0,\\
a_3&>&0.
\end{eqnarray}
We will check these inequalities numerically to determine stability of each of 
our stationary solutions. 

\subsection{Numerical Examples}
In this Section we will show examples of double-mode fixed-point solutions for 
the models from DC and will discuss their stability. The solution is 
characterized by the parameters $\gamma$, $R$, $\alpha$, $I$, $\Delta\sigma$. 
For the detuning parameter $\Delta\sigma$ we will consider the range of values 
from $0$ to $\Delta\sigma_{\rm max}$. 

We choose representative examples of nonradial modes which have the highest 
excitation probability in the vicinity of the fundamental mode and the first 
overtone. 

\begin{figure}[htb!]
\centerline{\includegraphics[width=0.48\textwidth]{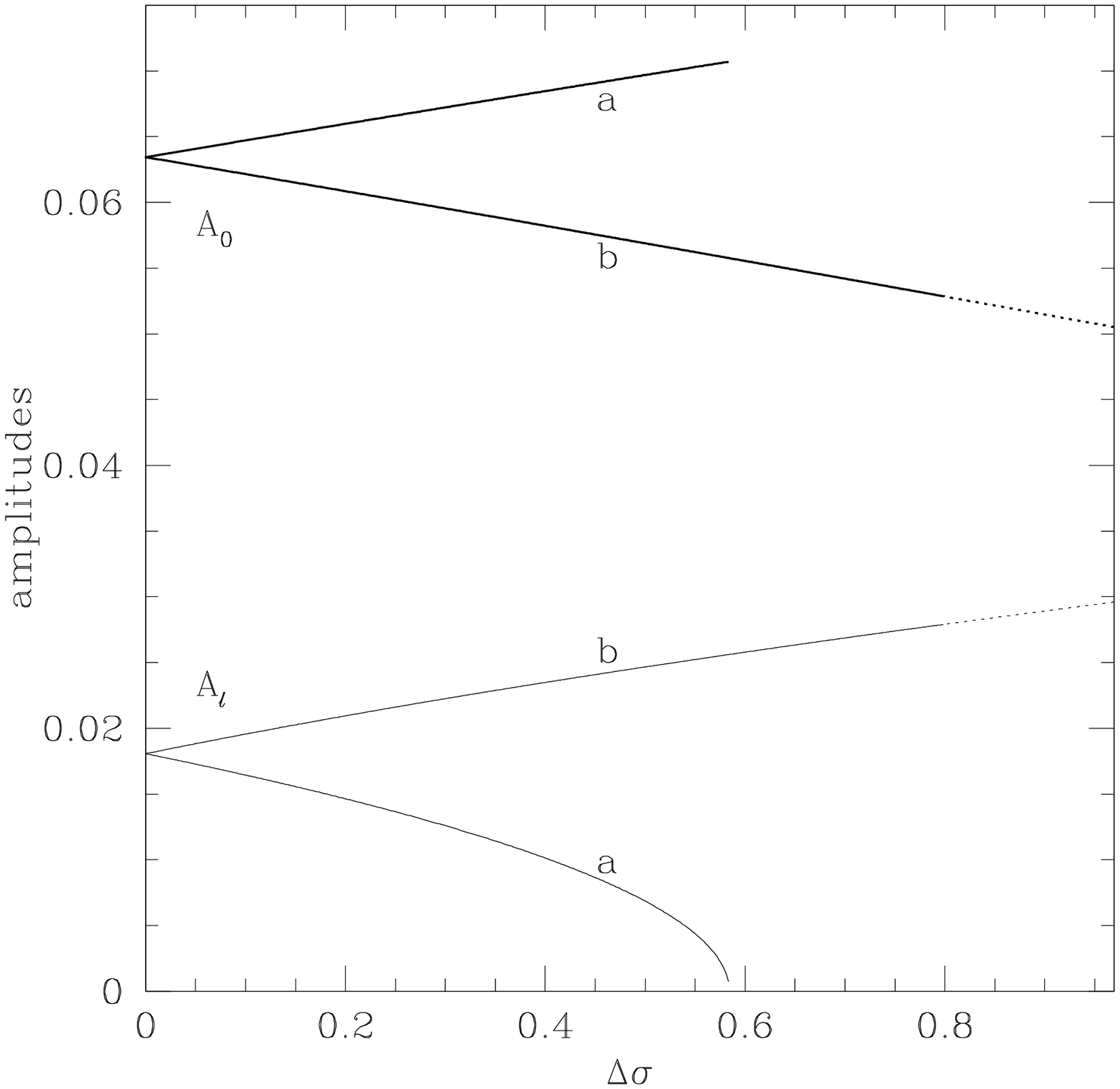}\hskip6mm\includegraphics[width=0.48\textwidth]{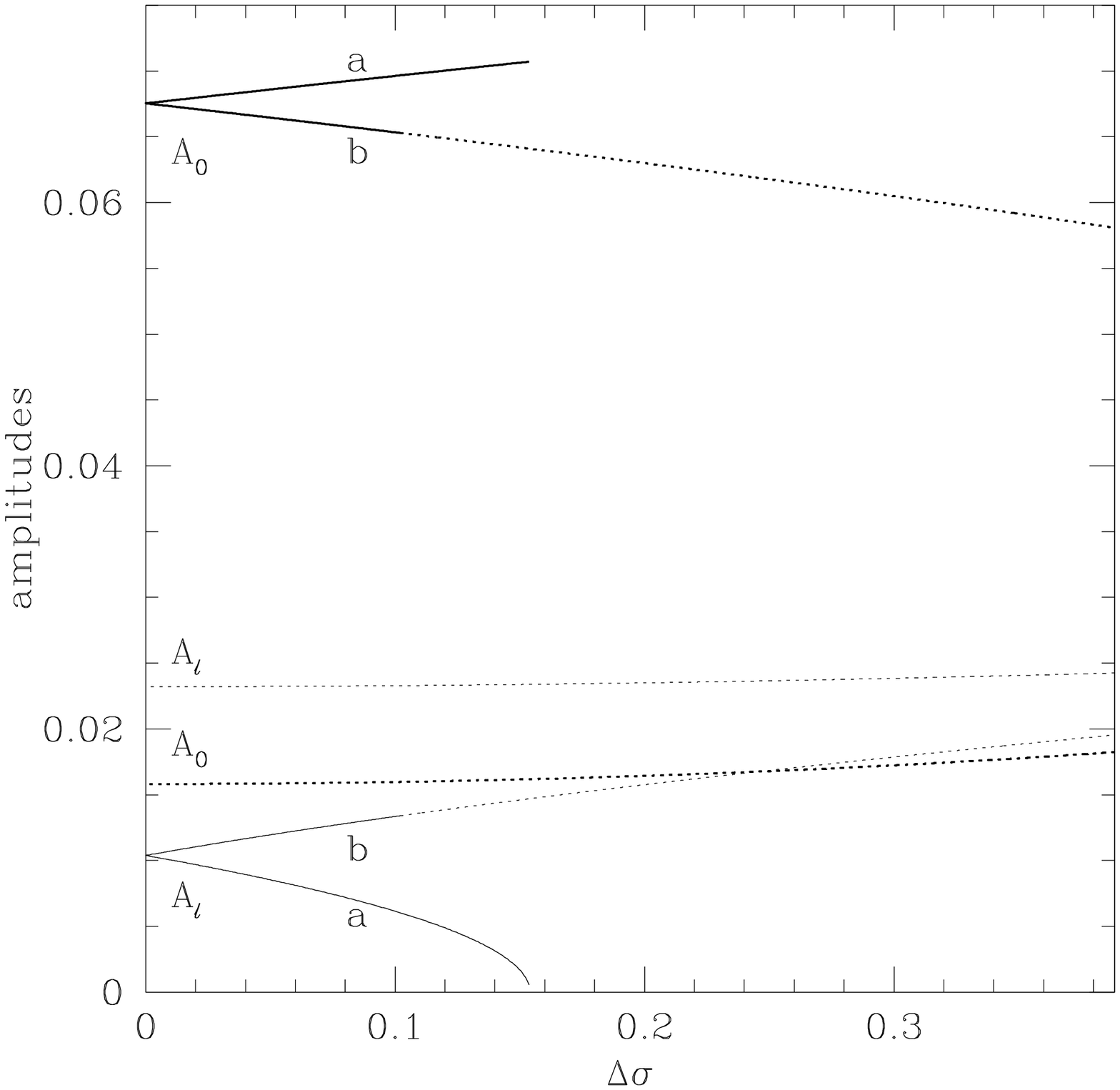}}
\centerline{\includegraphics[width=0.48\textwidth]{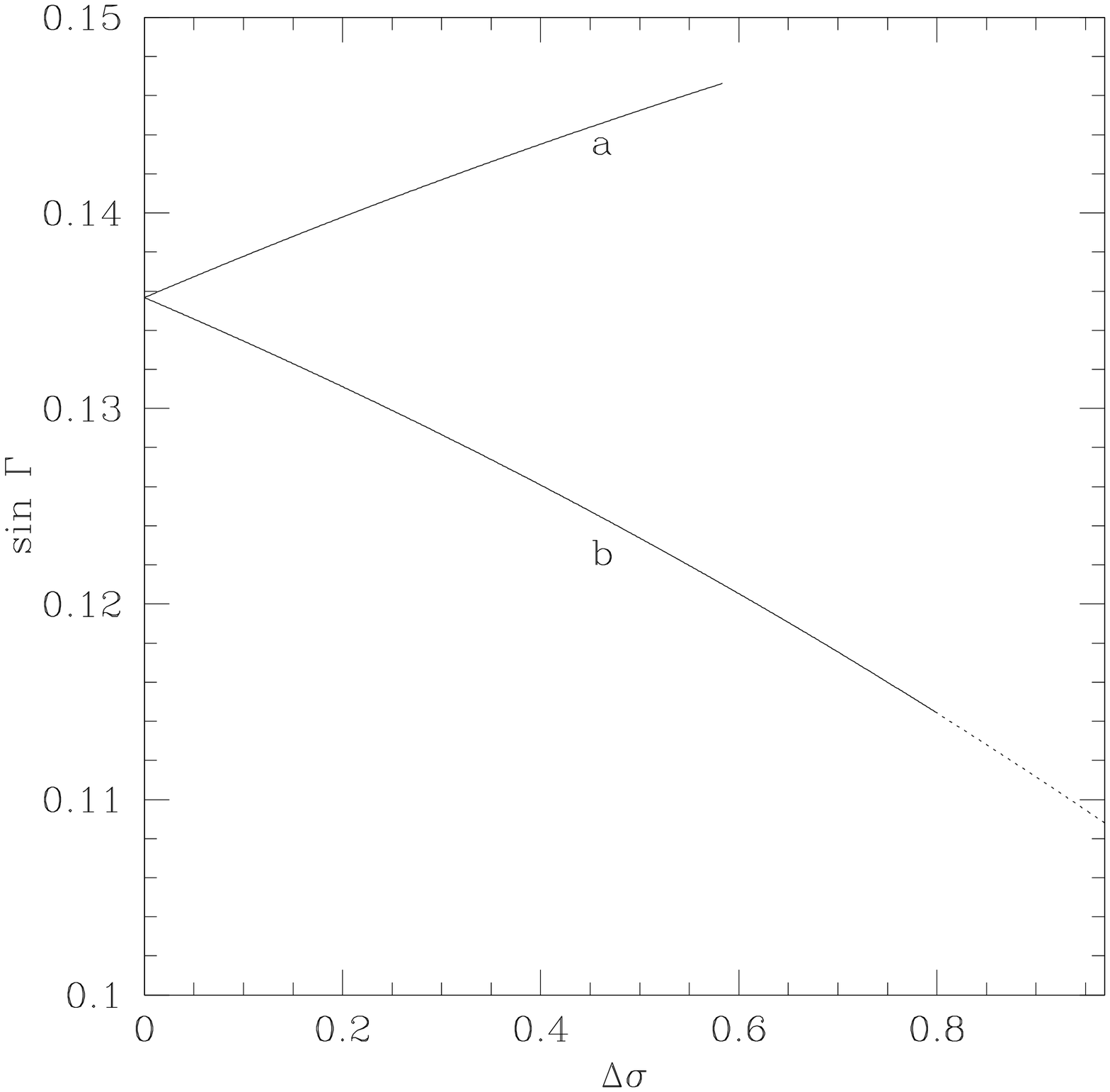}\hskip6mm\includegraphics[width=0.48\textwidth]{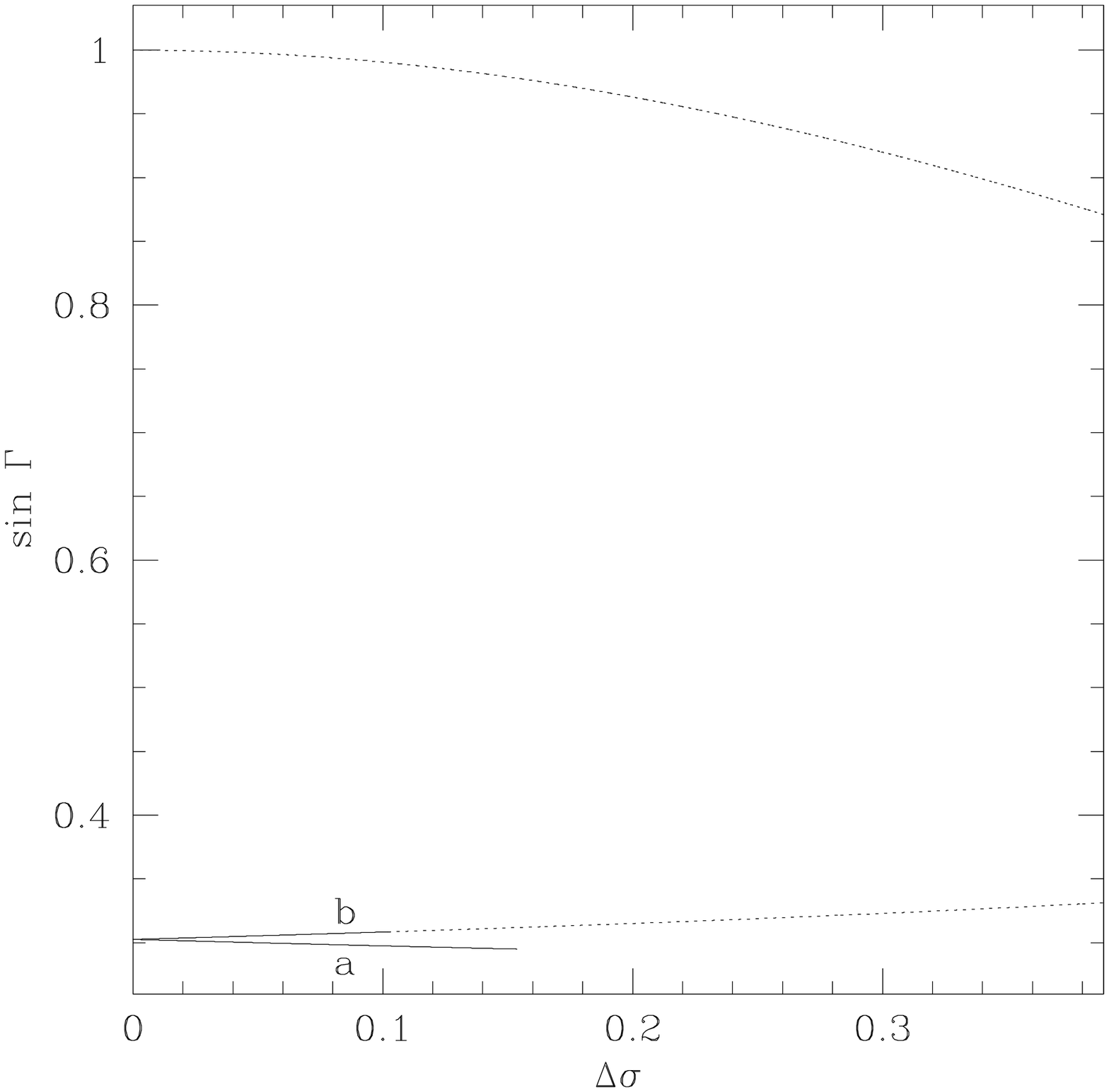}}
\FigCap{Fixed-point solutions for the resonance between the fundamental mode 
and the nonradial modes with ${l=1}$ (left panels) and ${l=2}$ (right panels). 
The $\Delta\sigma$-dependence of amplitudes (upper panels) and $\sin\Gamma$ 
(lower panels) are shown. Solid and dotted lines denote stable and unstable 
solutions, respectively. The range of $\Delta\sigma$ corresponds to 
$\Delta\sigma_{\rm max}$, \ie to the density of the spectra of nonradial 
modes. The adopted coefficients are ${R=1456}$, ${\gamma=0.068}$, ${I=0.0811}$ 
for ${l=1}$ and ${R=1362}$, ${\gamma=1.0099}$, ${I=0.0237}$ for ${l=2}$.}
\end{figure}
\begin{figure}[htb!]
\centerline{\includegraphics[width=0.48\textwidth]{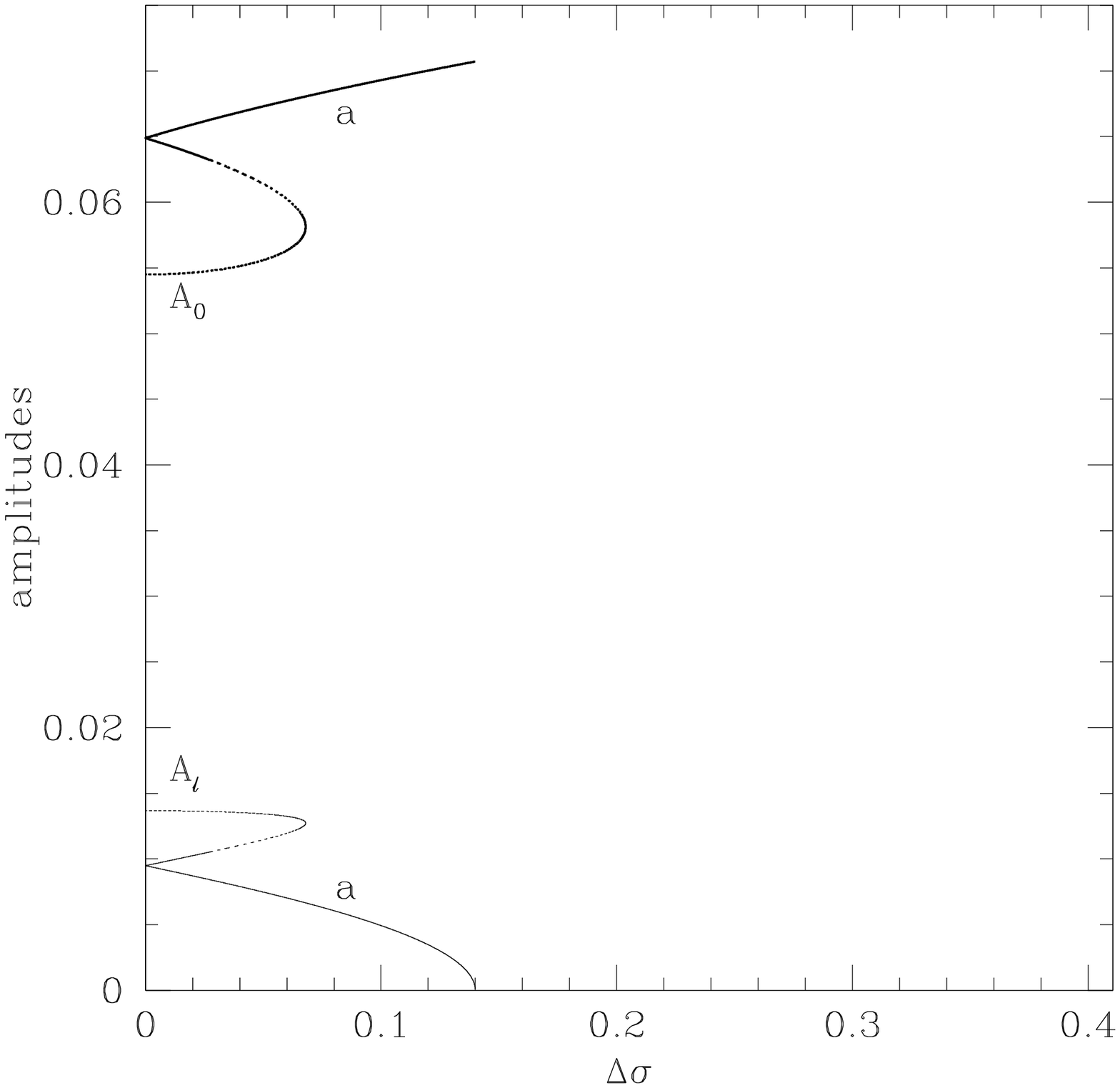}\hskip6mm\includegraphics[width=0.48\textwidth]{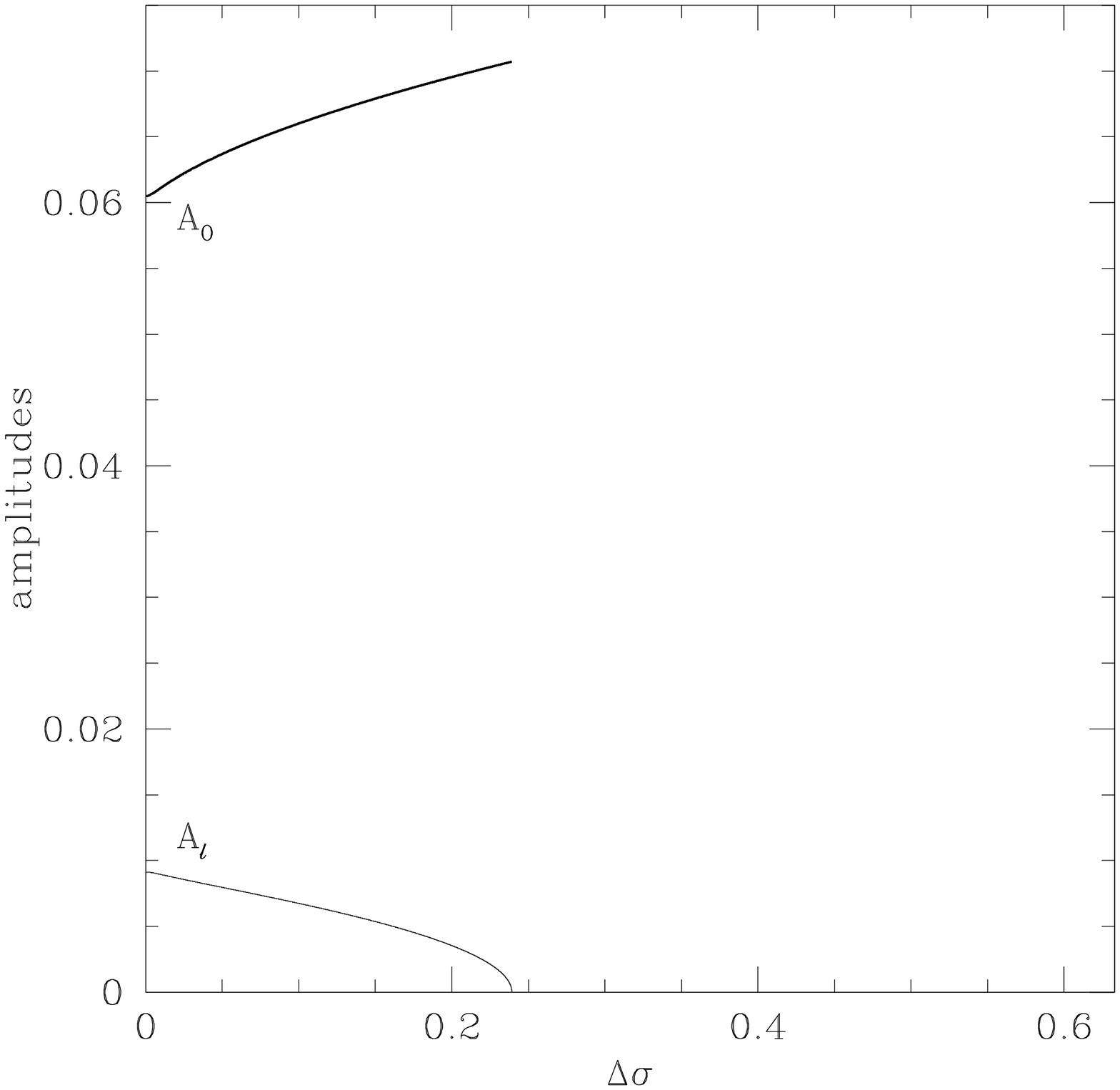}}
\centerline{\includegraphics[width=0.48\textwidth]{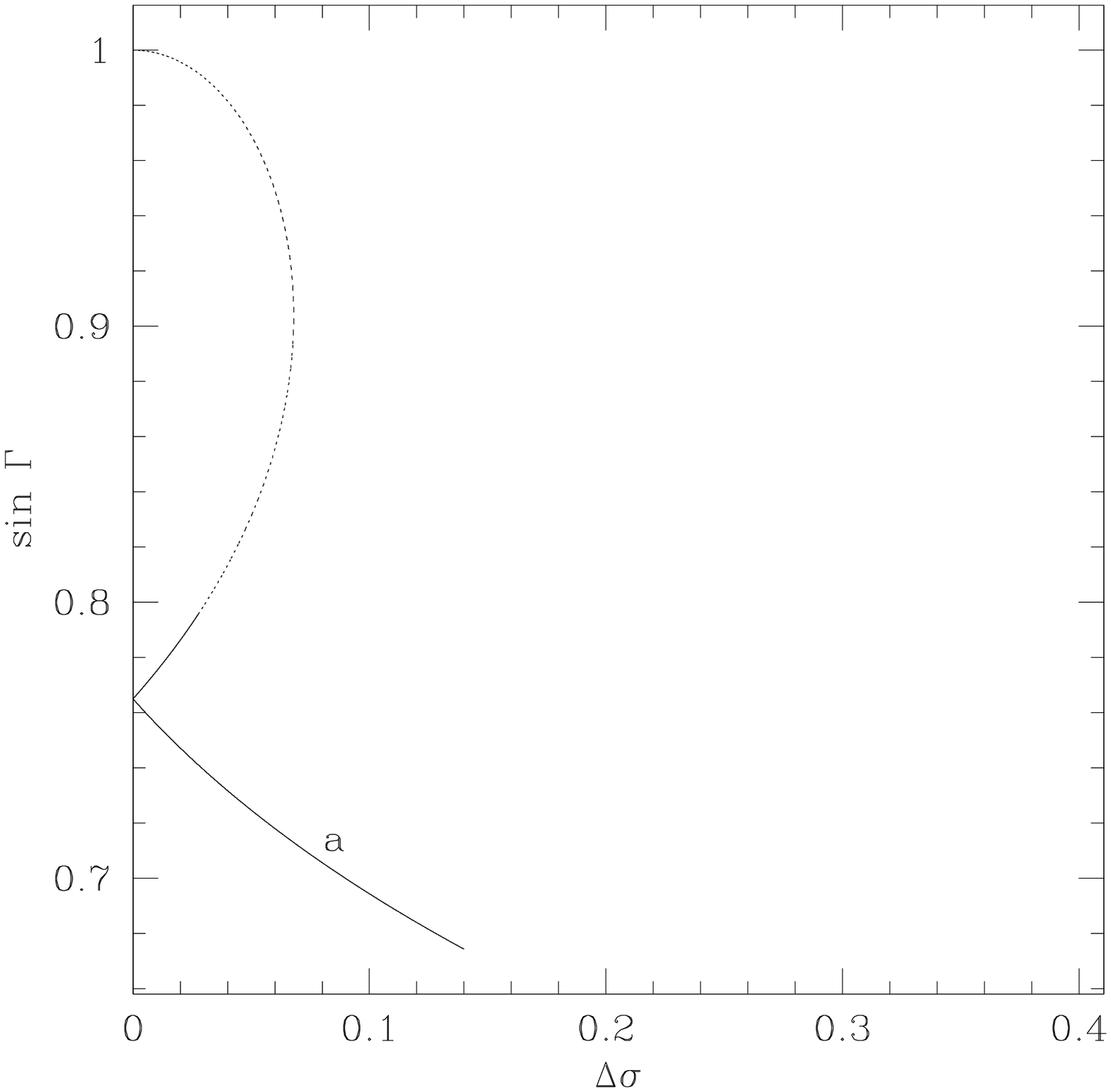}\hskip6mm\includegraphics[width=0.48\textwidth]{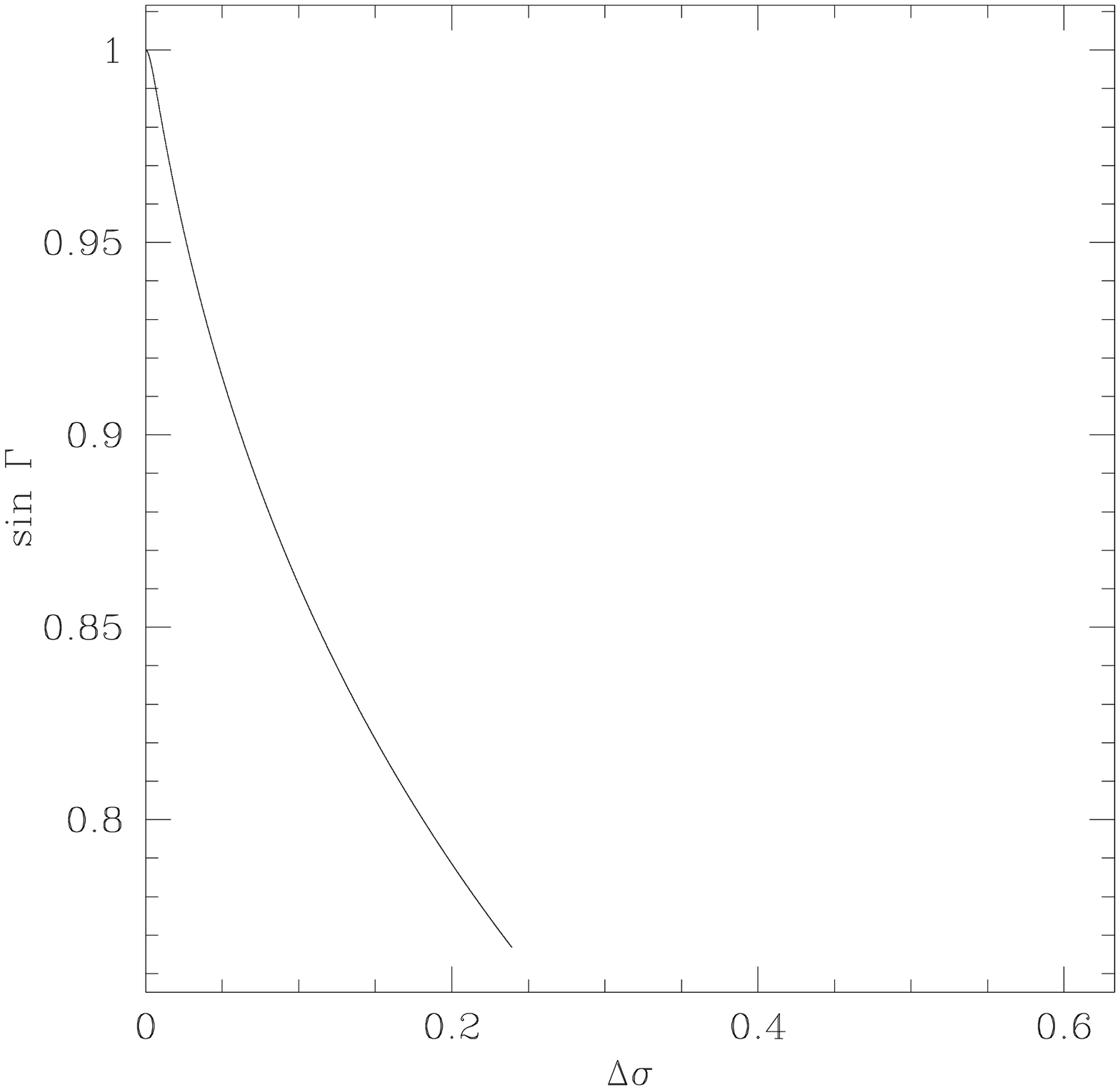}}
\FigCap{The same as in Fig.~3 but for the nonradial modes with ${l=5}$ (left 
panels) and ${l=6}$ (right panels). The adopted coefficients are ${R=1771}$, 
${\gamma=4.9738}$, ${I=0.0214}$ for ${l=5}$ and ${R=2814}$, ${\gamma=9.7903}$, 
${I=0.0265}$ for ${l=6}$.}
\end{figure}

\begin{figure}[htb!]
\centerline{\includegraphics[width=0.48\textwidth]{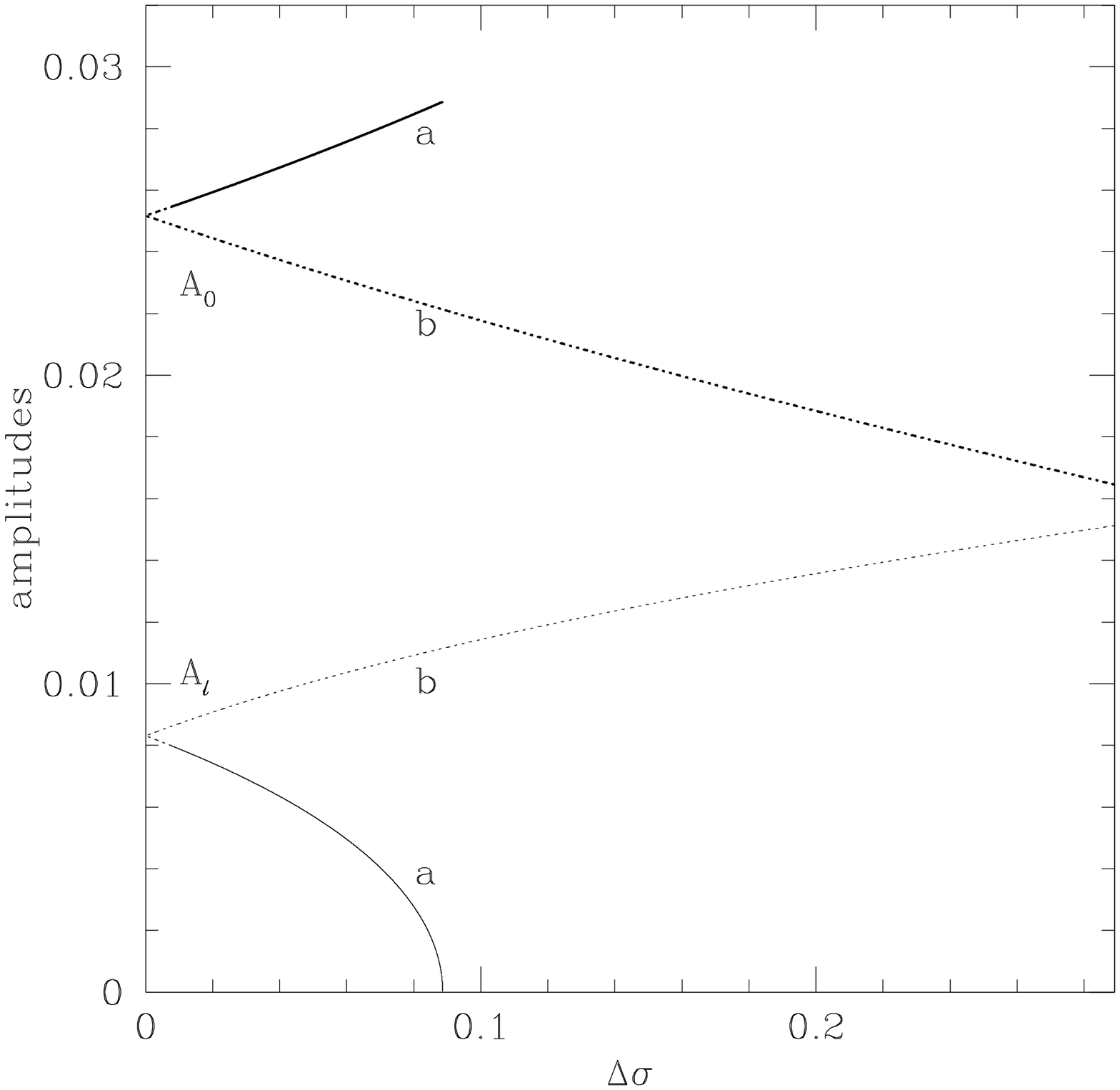}\hskip6mm\includegraphics[width=0.48\textwidth]{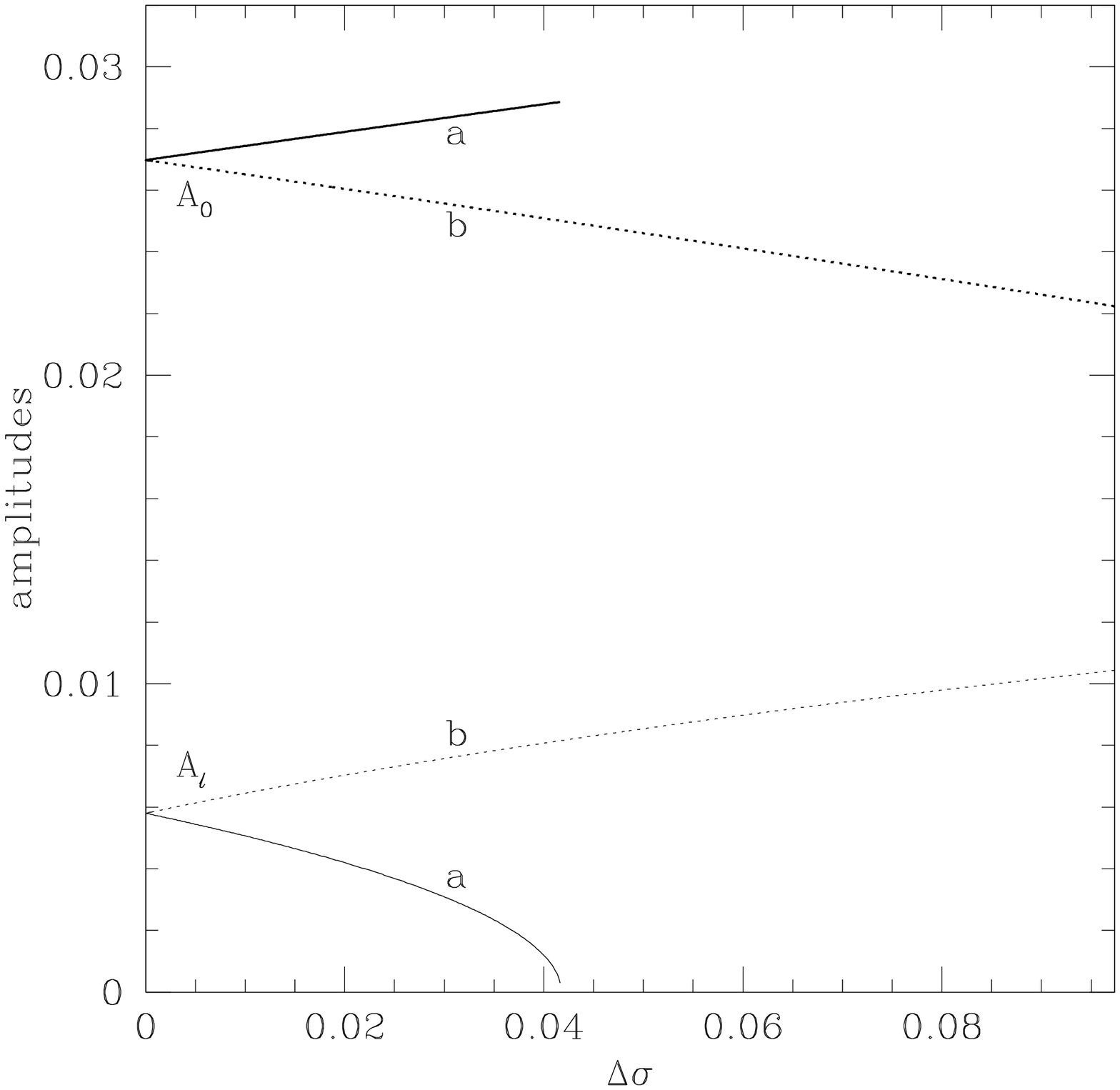}}
\centerline{\includegraphics[width=0.48\textwidth]{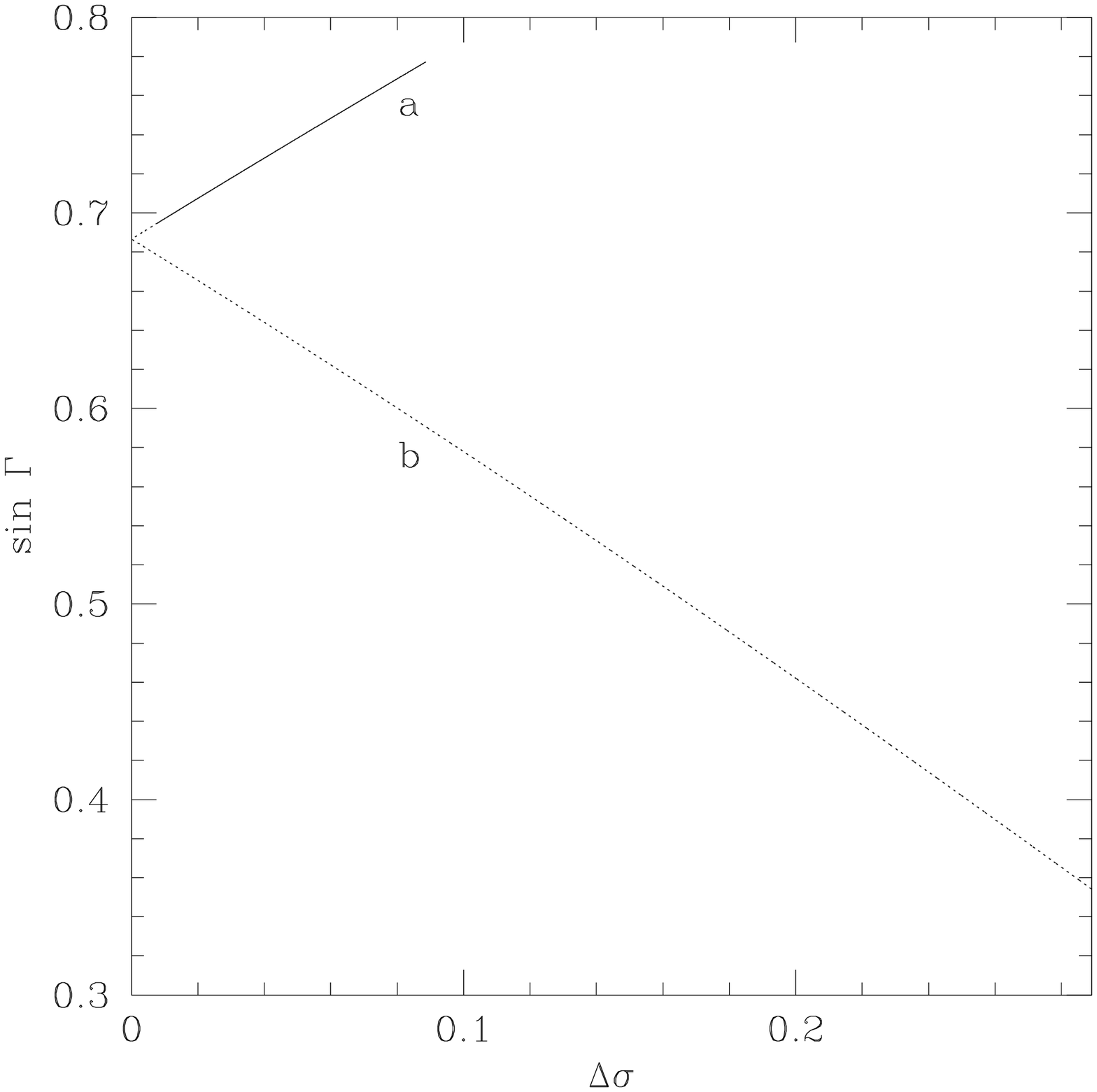}\hskip6mm\includegraphics[width=0.48\textwidth]{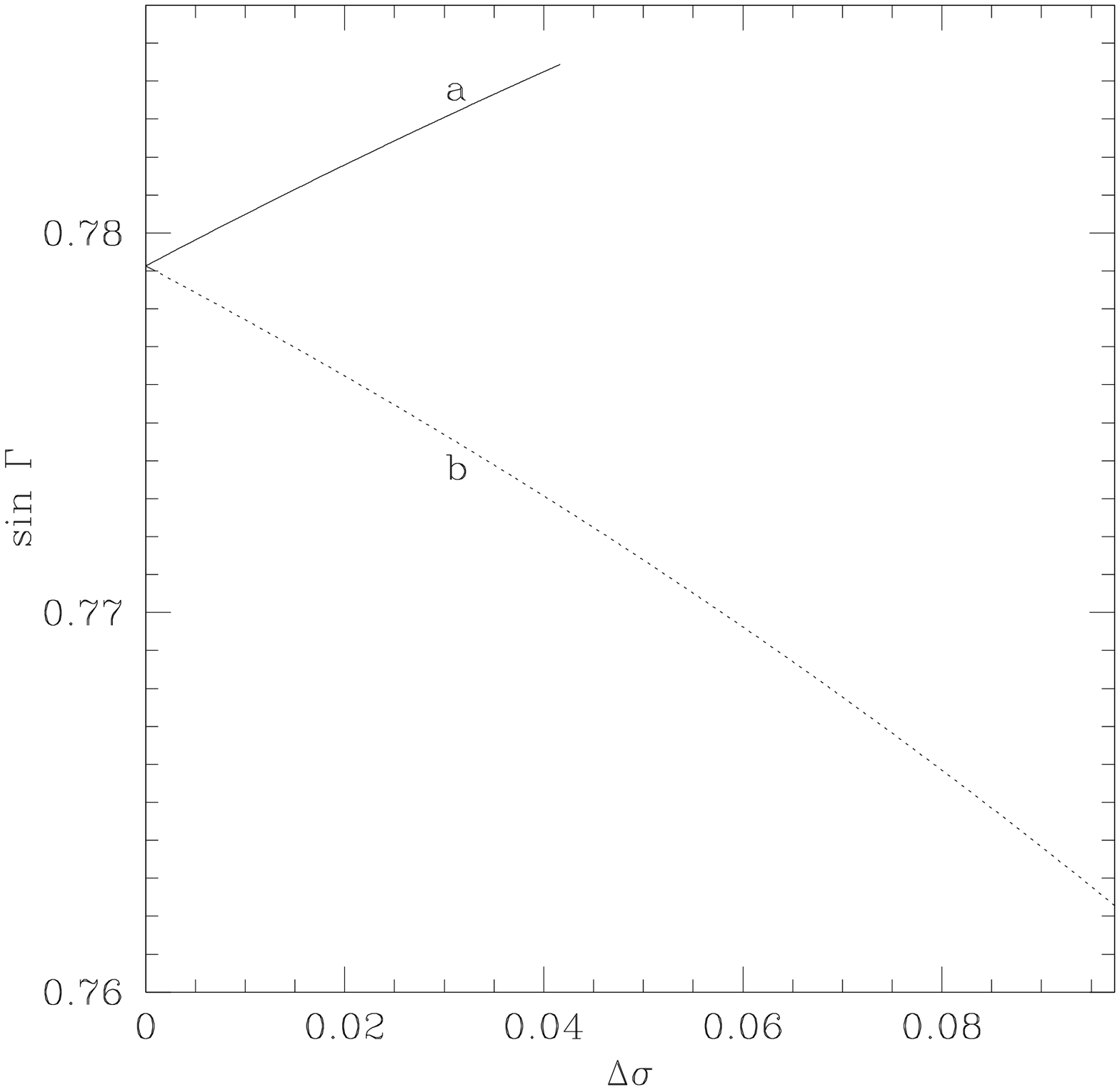}}
\FigCap{The same as in Figs.~3 and 4 but for the first overtone and the 
nonradial modes with ${l=1}$ (left panels) and ${l=4}$ (right panels). The 
following values of the coefficients were adopted: ${R=1550}$, 
${\gamma=0.044}$, ${I=0.1090}$ for ${l=1}$ and ${R=1748}$, ${\gamma=0.1427}$, 
${I=0.0462}$ for ${l=4}$.}
\end{figure}

In Figs.~3--5 we plot $A_0,A_l$, and $\sin\Gamma$ as functions of 
$\Delta\sigma$. As one can see there are at least two branches of stable 
solutions in most cases. In the figures the branches are marked with $a$ and 
$b$. The third branch that is present in some cases is always unstable and we 
will not be interested in it. The $a$-branch exists for 
${\Delta\sigma<\Delta\sigma_d}$, \ie in the region where the single-mode 
fixed-point is unstable, as discussed in Sections~4.1 and 4.2. This branch is 
stable except for a very small region near ${\Delta\sigma=0}$ in the case of 
the ${l=1}$ mode at the first overtone (Fig.~5, left panels). We are not sure 
whether this instability is real or rather an artifact of our simplifications. 

The conclusion is that there exists one or two stable stationary solutions for 
all possible values of $\Delta\sigma$ because in the region where $a$-branch 
does not exist, the single-mode fixed-point is stable. Which one of these 
solutions, if any, is chosen by the system will be studied by numerical 
integration of the amplitude equations. 

\Section{Time Dependent Solutions}
In this Section we present time dependent solutions obtained by means of 
numerical integration of AEs for the same values of coefficients that were 
used in the stationary solution studies. We focus on the $l=1$ cases but the 
results are similar in other cases. 

\subsection{Examples}
Figs.~6 and 7 show the solutions for the resonance between the fundamental and 
${l=1}$ modes for the selected values of the detuning parameter and different 
initial parameters. In Fig.~6 the detuning parameter of 0.3 is in the range 
where the $a$-branch exists. The two stationary solutions are 
${A_0^a\approx0.067}$, ${A_l^a\approx0.013}$, ${\sin\Gamma_a\approx0.142}$ and 
${A_0^b\approx0.059}$, ${A_l^b\approx0.022}$, ${\sin\Gamma_b\approx0.129}$ 
(see Fig.~3, left panels). Both solutions are stable. In the left panels we 
show the development of the instability beginning with very small pulsation 
amplitudes. The final (asymptotic) state is an $a$-branch solution. In the 
right panels the initial conditions are close to a $b$-branch solution which 
is another asymptotic state of the system. 

\begin{figure}[htb!]
\centerline{\includegraphics[width=0.48\textwidth]{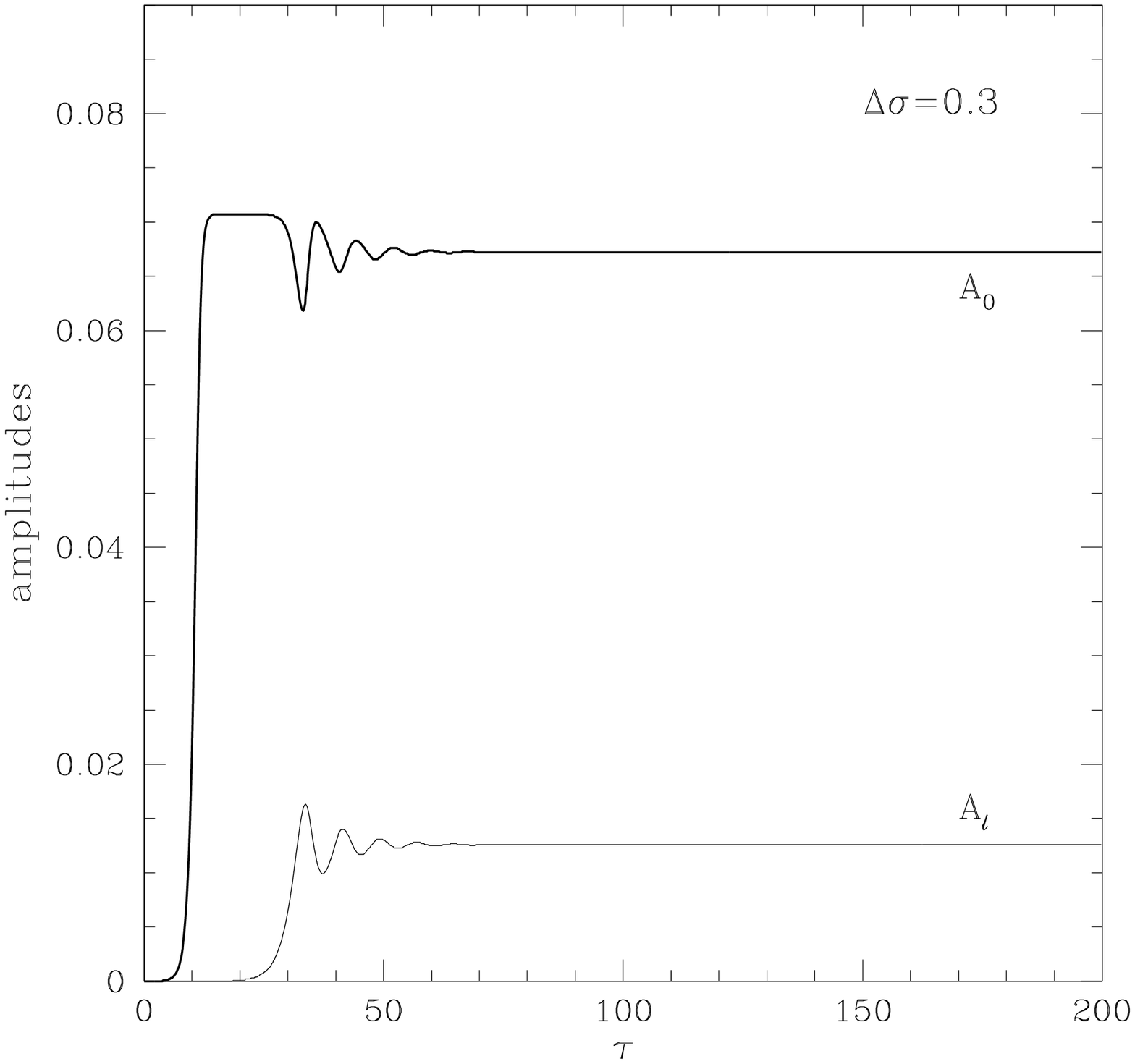}\hskip6mm\includegraphics[width=0.48\textwidth]{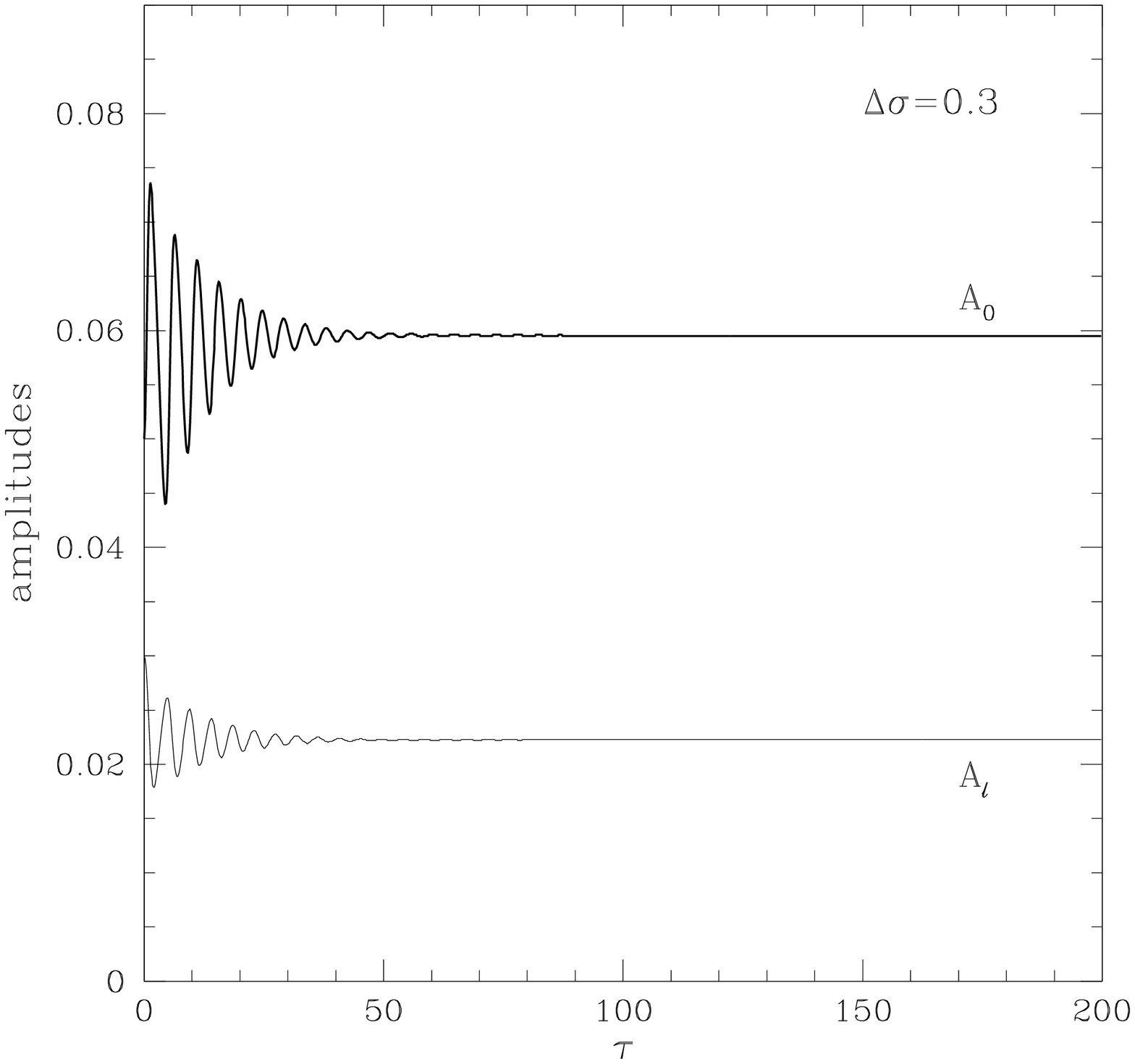}}
\centerline{\includegraphics[width=0.48\textwidth]{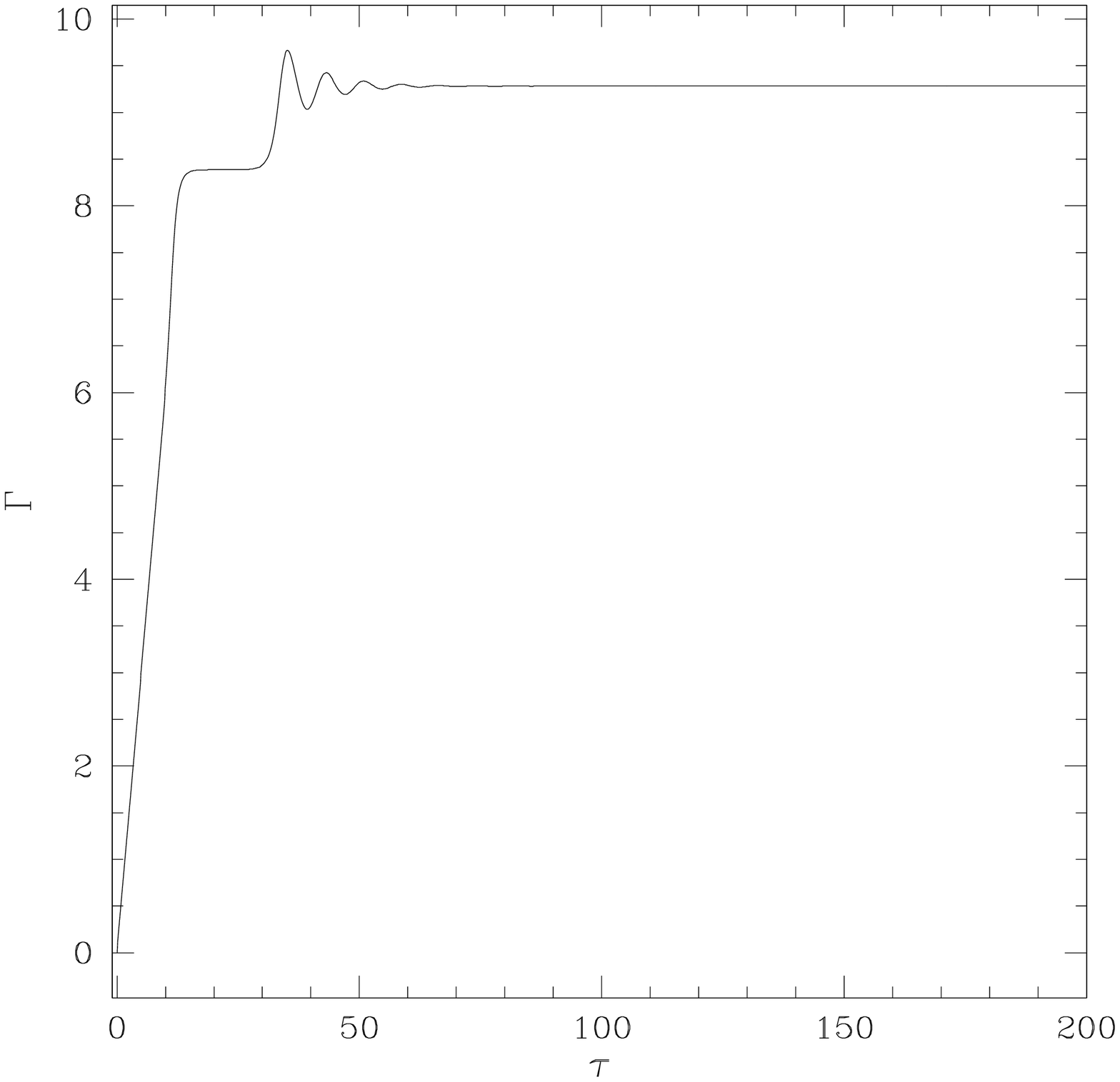}\hskip6mm\includegraphics[width=0.48\textwidth]{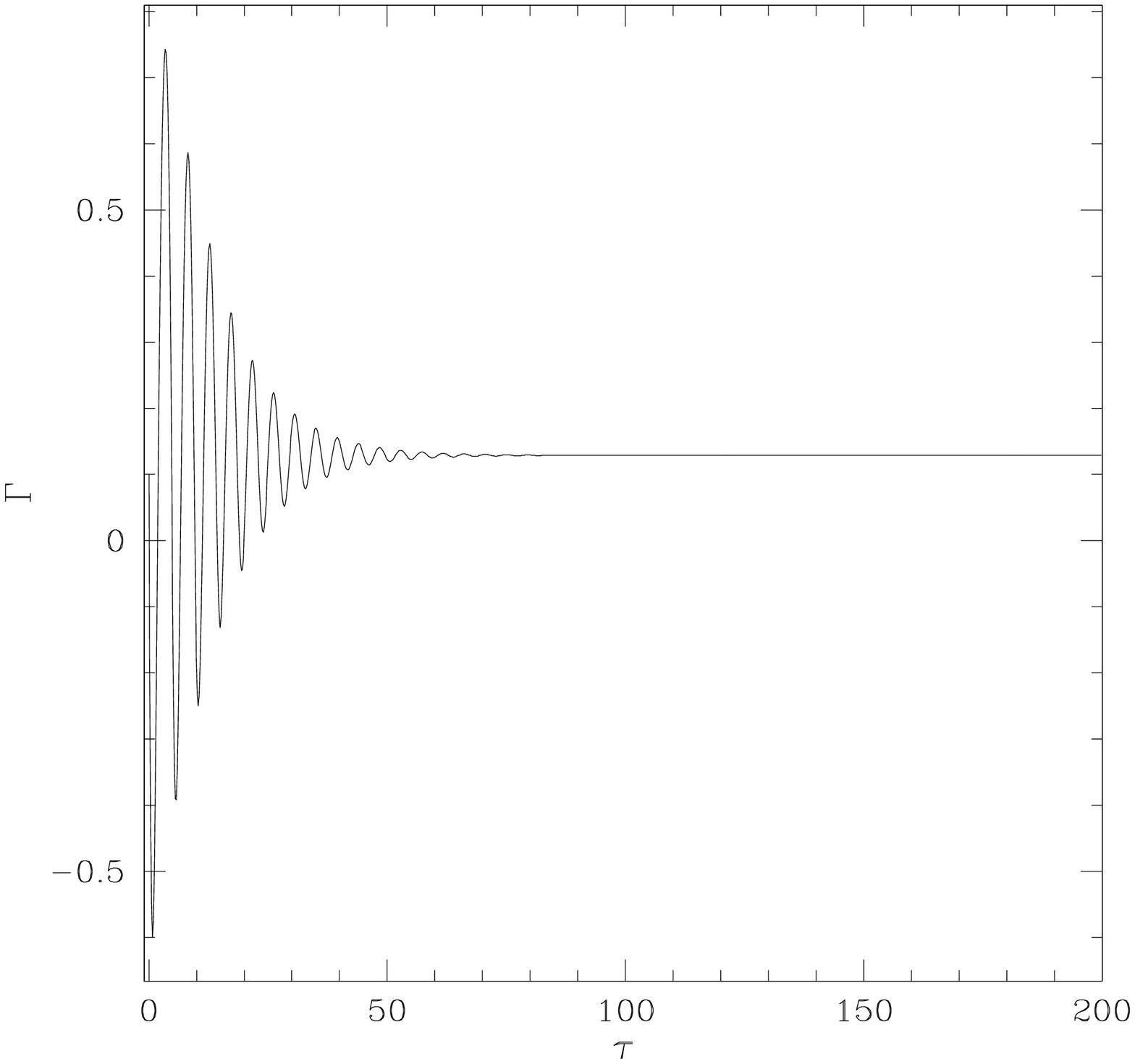}}
\FigCap{Time dependent solutions for the same coefficients as used in the left 
panels of Fig.~3, and the detuning parameter 0.3. The initial parameters are 
${A_0=A_l=0.00001}$, ${\Gamma=0}$ (left panels) and ${A_0=0.05}$, 
${A_l=0.03}$, ${\Gamma=0.1}$ (right panels).} 
\end{figure}

In Fig.~7 the detuning parameter is 0.65 and, as can be seen from Fig.~3, 
left panels, it is in the range where the $a$-branch does not exist. Instead 
we have stable single-mode solution ${A_0^0=0.0707}$, ${A_l^0=0}$. The second 
stable solution is the $b$-branch solution ${A_0^b\approx0.055}$, 
${A_l^b\approx0.026}$, ${\sin\Gamma_b=0.119}$. The initial conditions in the 
left panels are very close to the $b$-branch solution and this solution is the 
asymptotic state. In the right panels we start only slightly farther from the 
$b$-branch than in the left panels, but the asymptotic state is completely 
different, \ie it is the single-mode solution. 

\begin{figure}[htb!]
\vglue-6mm
\centerline{\includegraphics[width=0.48\textwidth]{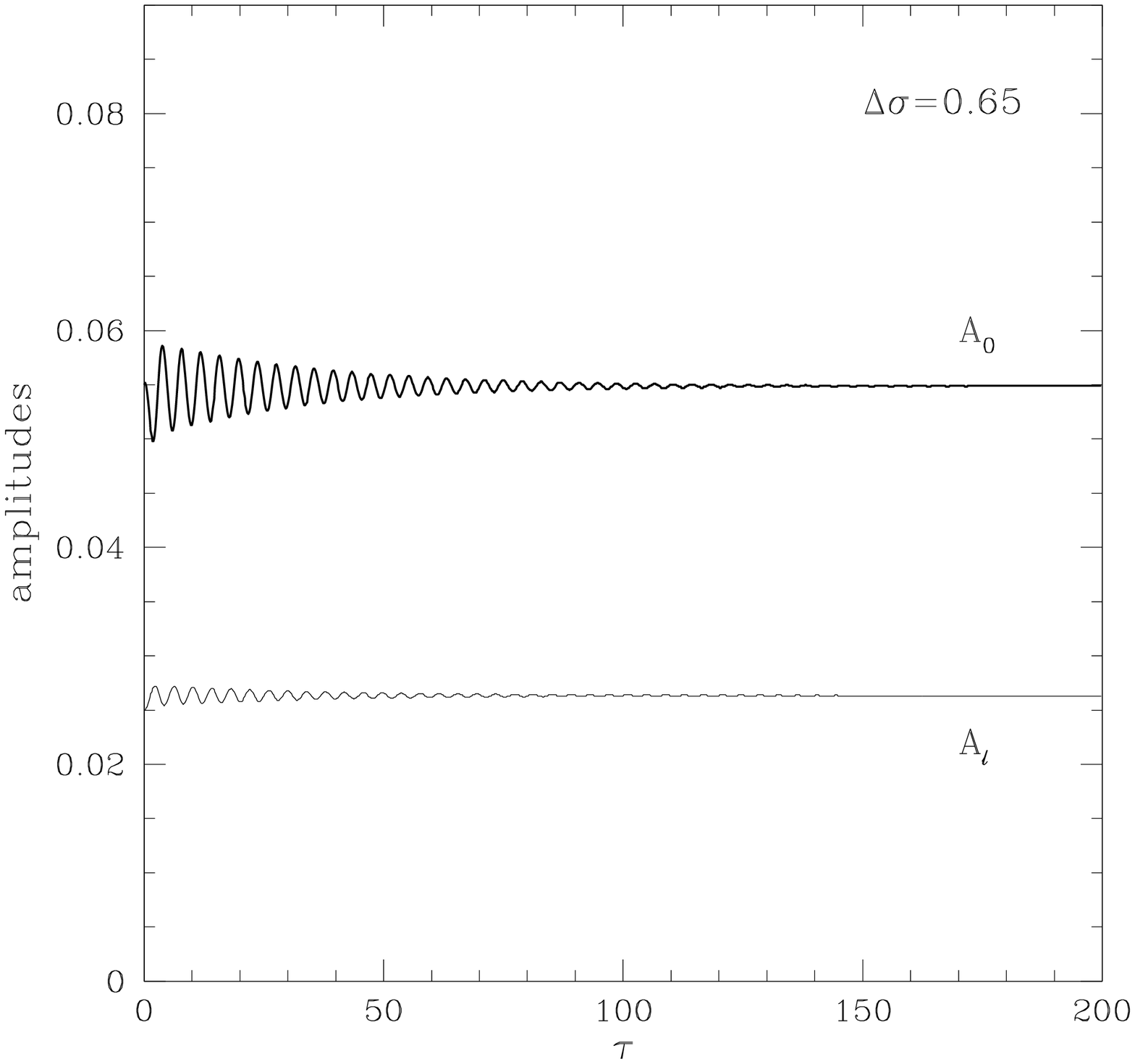}\hskip6mm\includegraphics[width=0.48\textwidth]{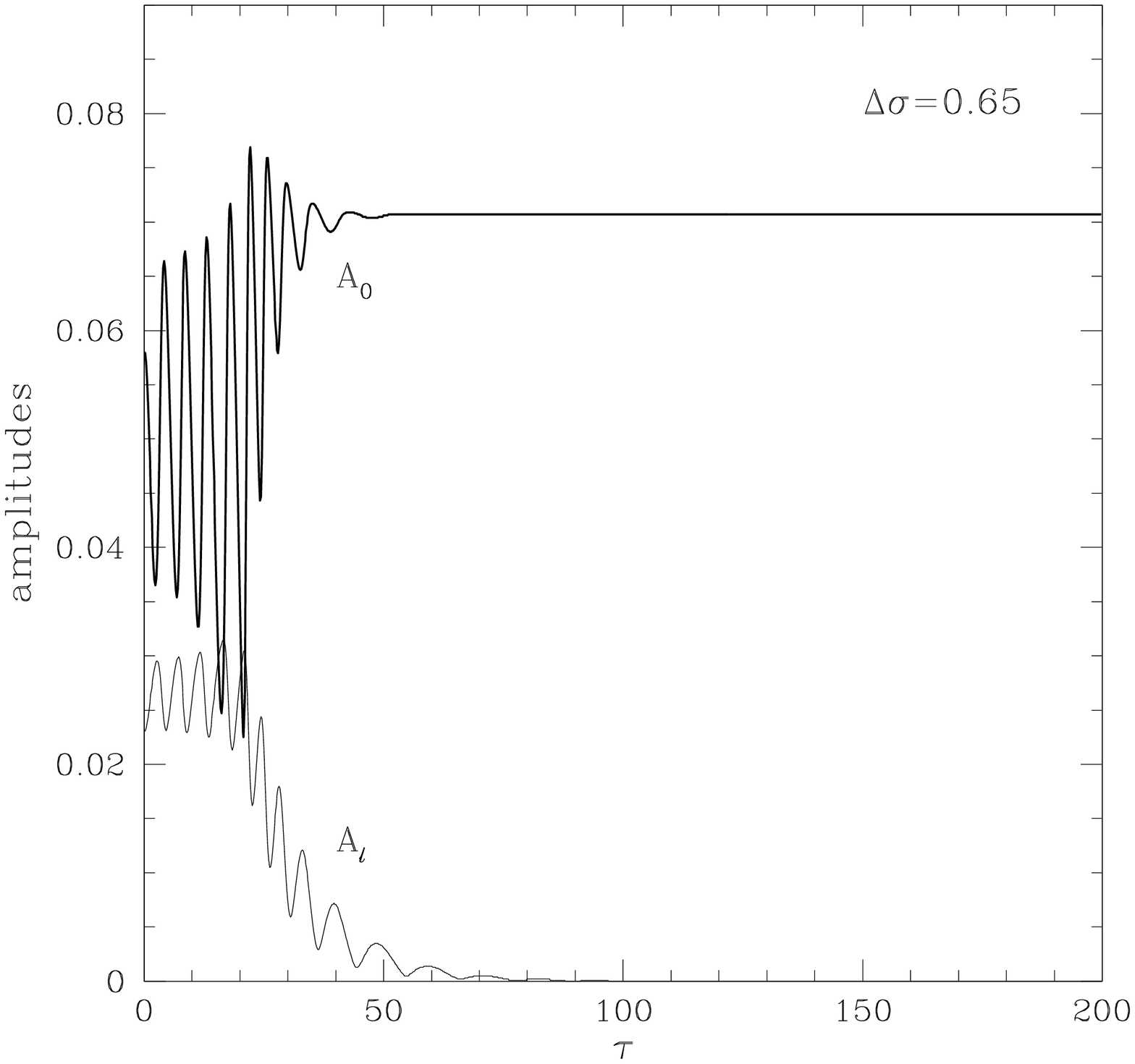}}
\centerline{\includegraphics[width=0.48\textwidth]{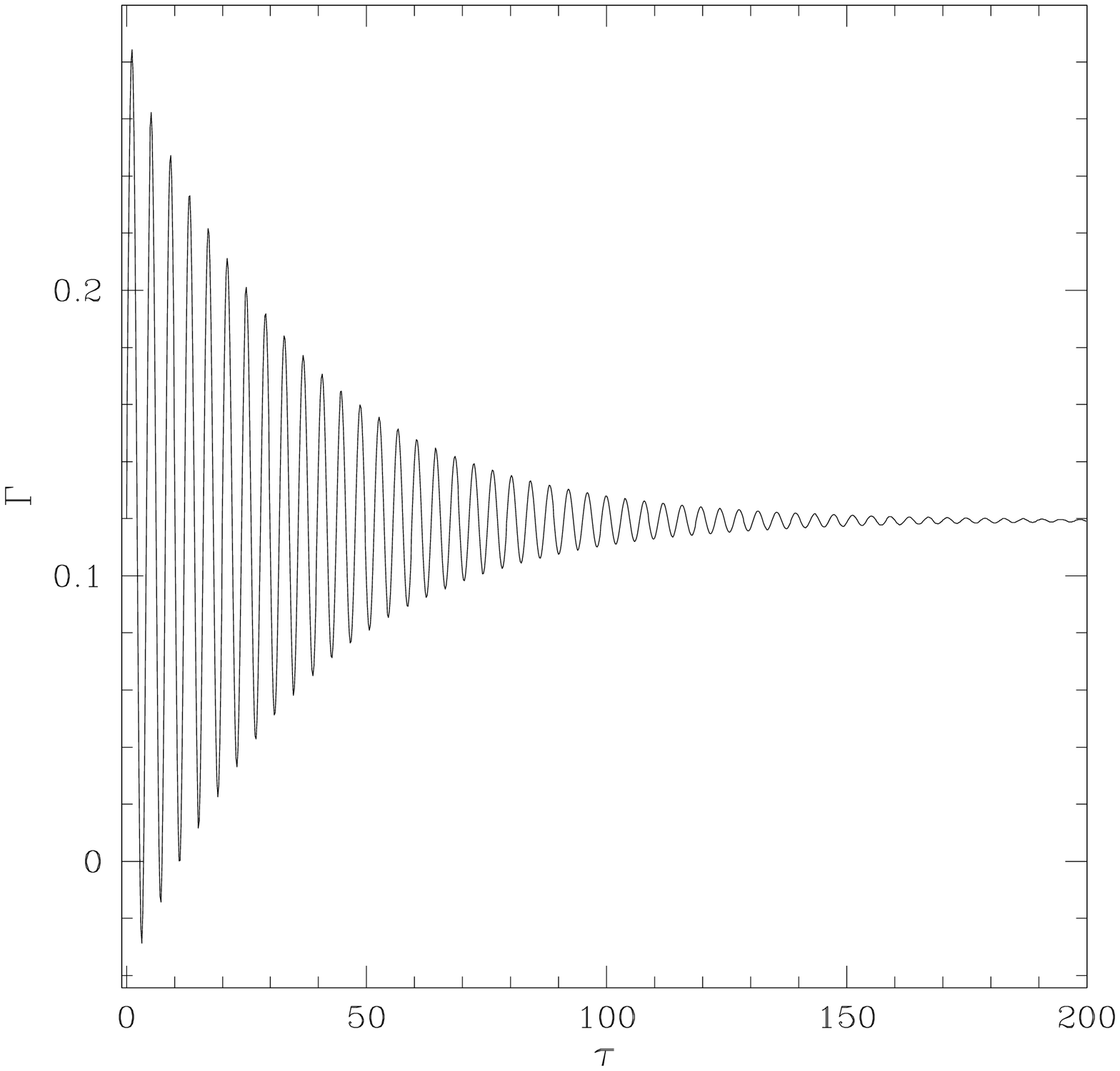}\hskip6mm\includegraphics[width=0.48\textwidth]{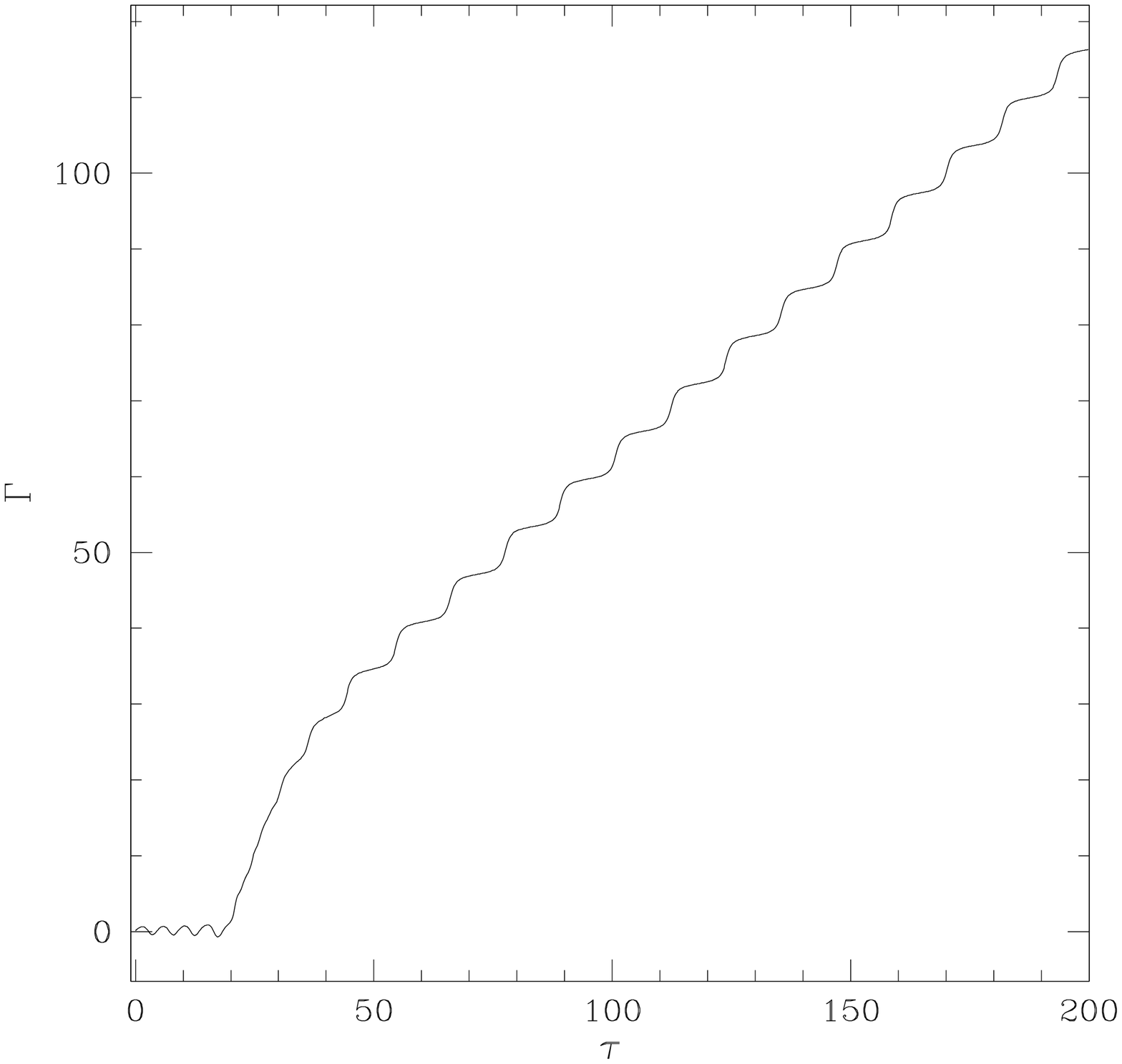}}
\FigCap{Time dependent solutions for the same coefficients as used in the left 
panels of Fig.~3, and the detuning parameter 0.65. The initial parameters are 
${A_0=0.055}$, ${A_l=0.025}$, ${\Gamma=0.12}$ (left panels) and ${A_0=0.058}$, 
${A_l=0.023}$, ${\Gamma=0.1}$ (right panels).}
\end{figure}

As we discussed in Section~4.4 there might be a very narrow range around 
${\Delta\sigma=0}$ where $a$-branch solutions are unstable. We found 
numerically an amplitude modulated solution with the period of about 60 in 
dimensionless units. The range is very narrow and we do not know how robust 
this finding is. In other cases, the asymptotic state is always a stationary 
solution. Hence we focus here only on fixed-point asymptotic solutions. 

\subsection{Regions of Attraction for the Asymptotic Solutions}
Here we present results of a more systematic survey of the time dependent 
solutions. We determine the ranges of the initial conditions (regions of 
attraction) leading to different asymptotic solutions. 

\begin{figure}[p!]
\centerline{\includegraphics[width=0.48\textwidth]{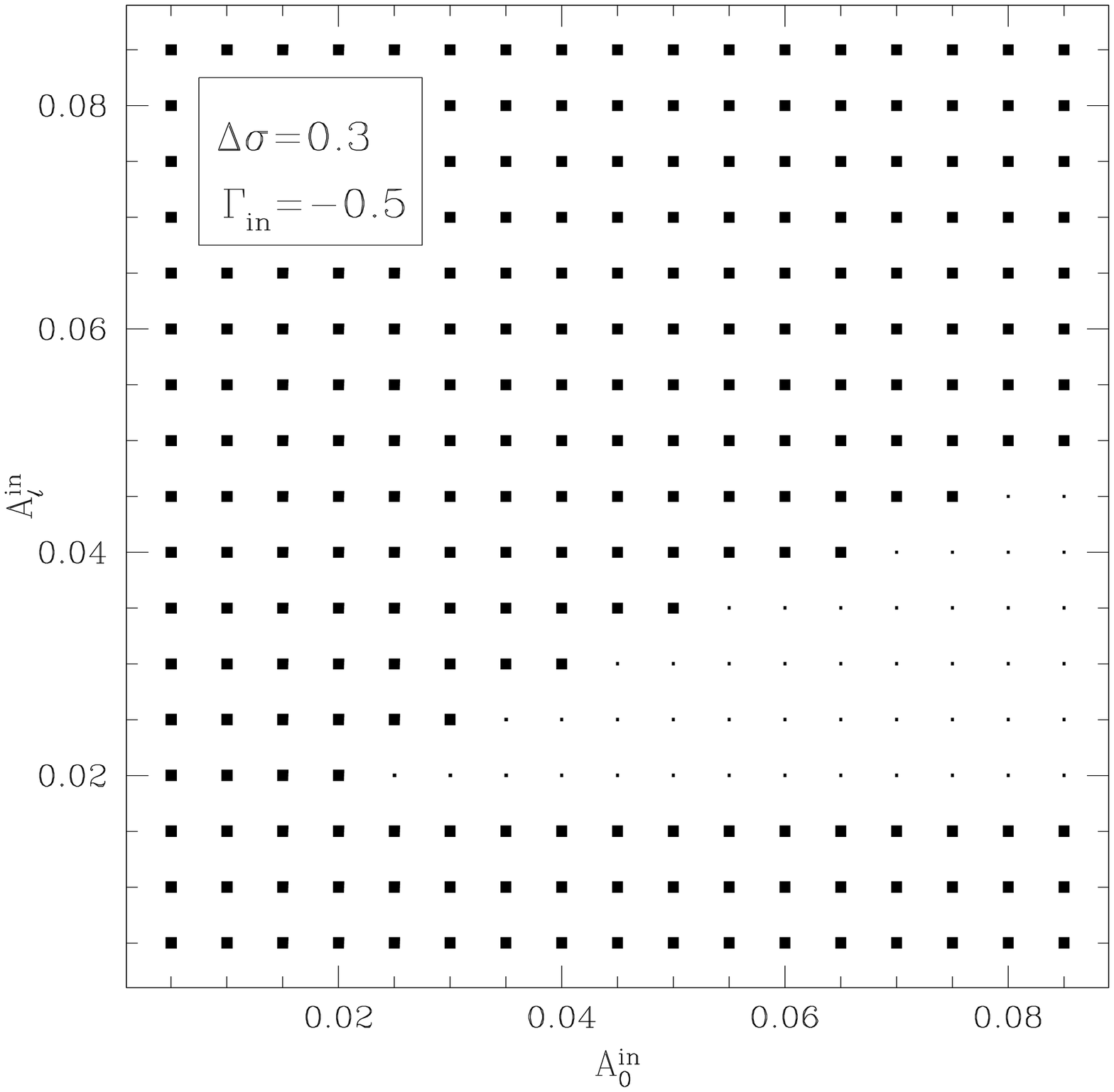}\hskip6mm\includegraphics[width=0.48\textwidth]{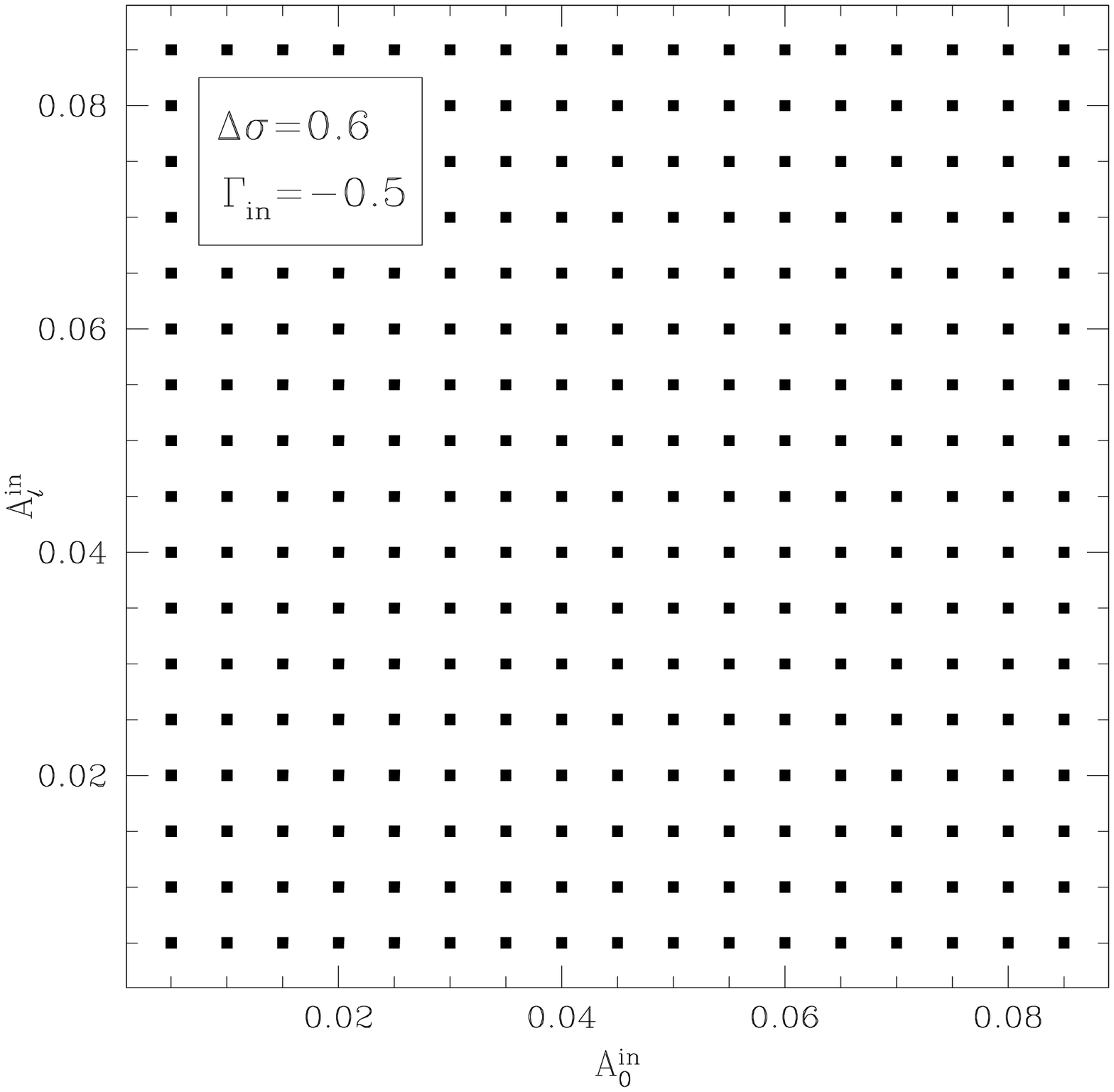}}
\centerline{\includegraphics[width=0.48\textwidth]{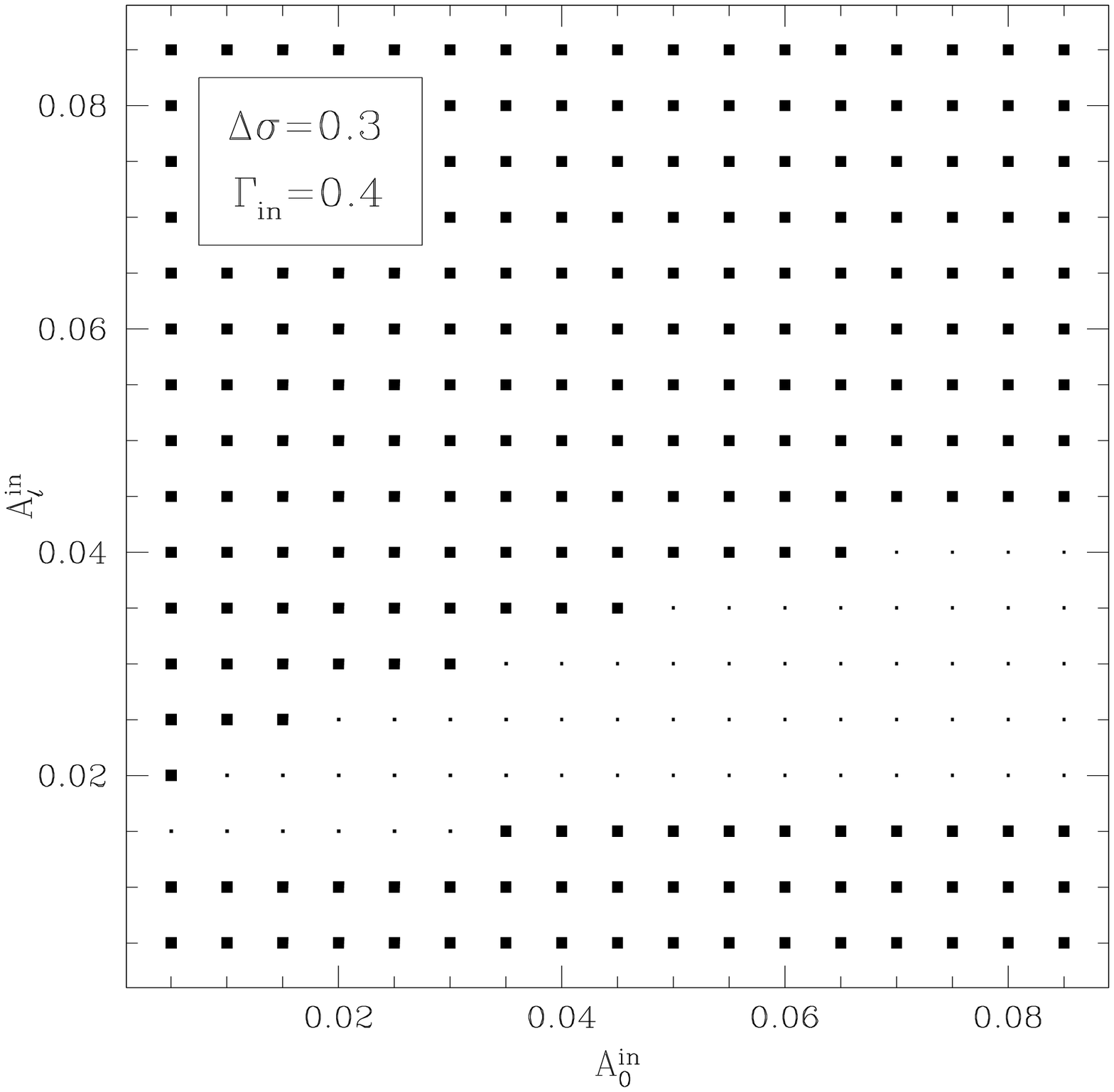}\hskip6mm\includegraphics[width=0.48\textwidth]{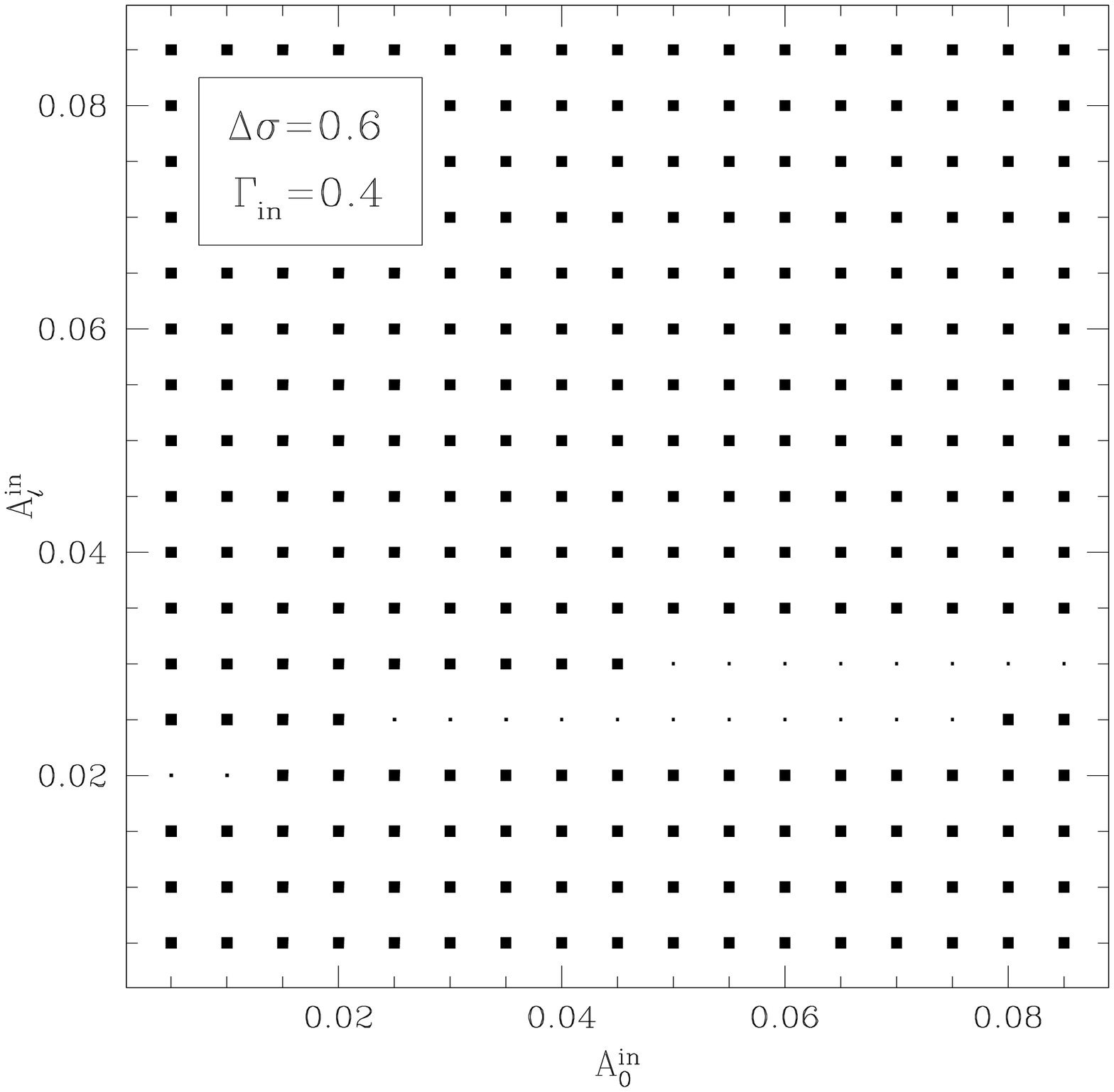}}
\centerline{\includegraphics[width=0.48\textwidth]{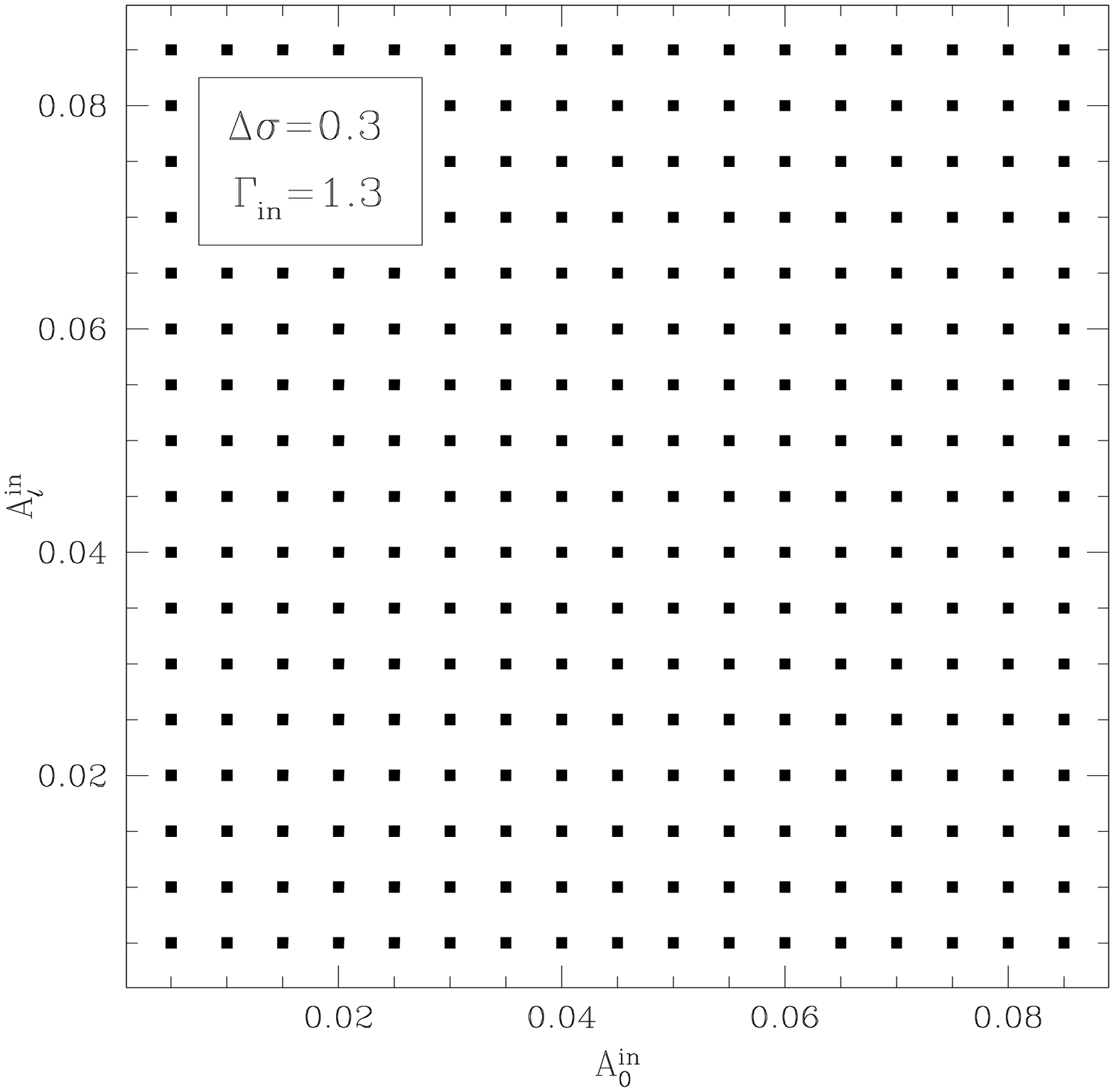}\hskip6mm\includegraphics[width=0.48\textwidth]{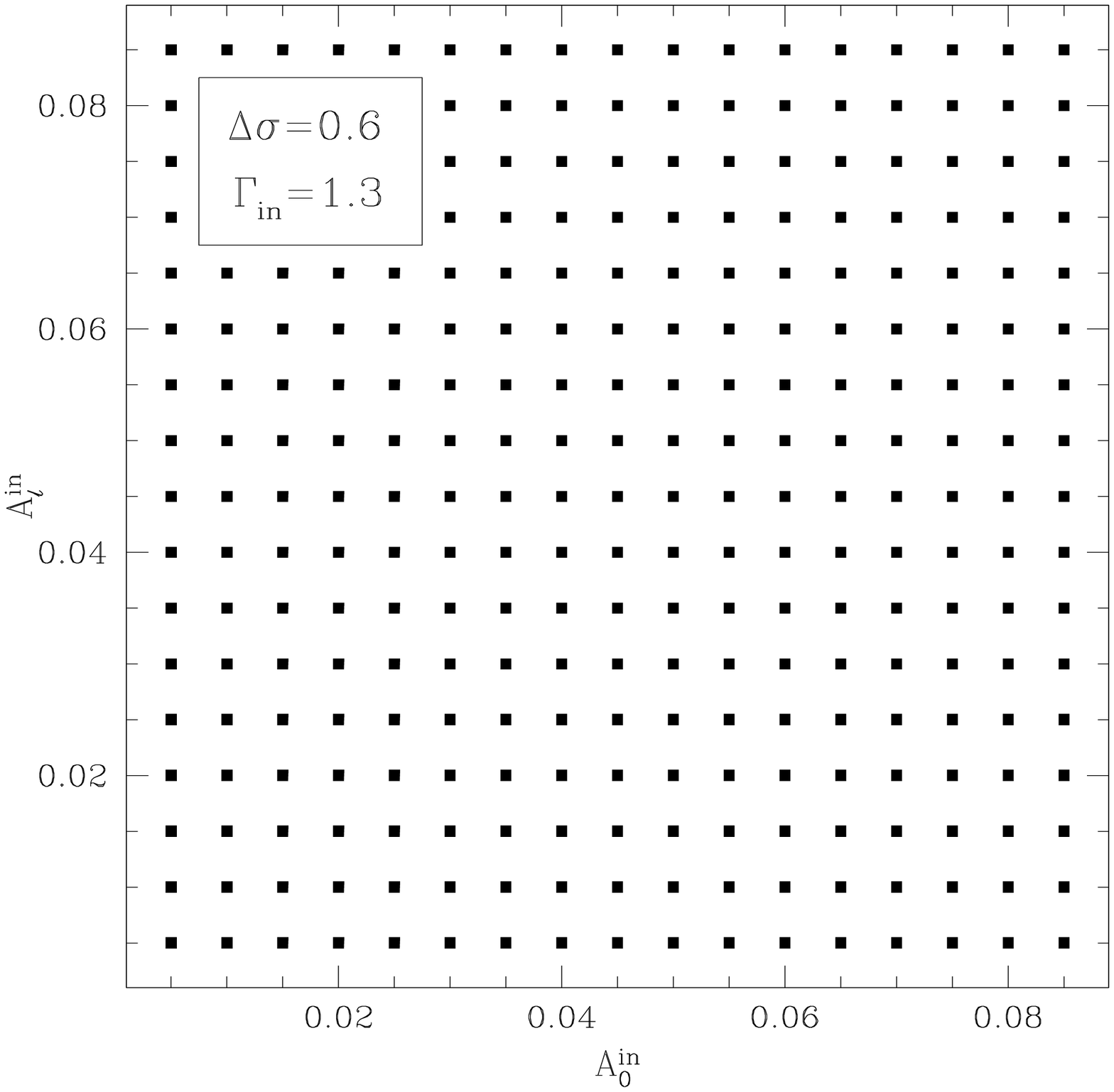}}
\FigCap{Regions of attraction for asymptotic solutions for the resonance 
between the fundamental and the ${l=1}$ modes (see Fig.~3, left panels). 
Squares denote attraction to the $a$-branch (${\Delta\sigma=0.3}$) or to the 
single-mode fixed-point (${\Delta\sigma=0.6}$) solution. Dots denote 
attraction to the $b$-branch.}
\end{figure}
\begin{figure}[p!]
\centerline{\includegraphics[width=0.48\textwidth]{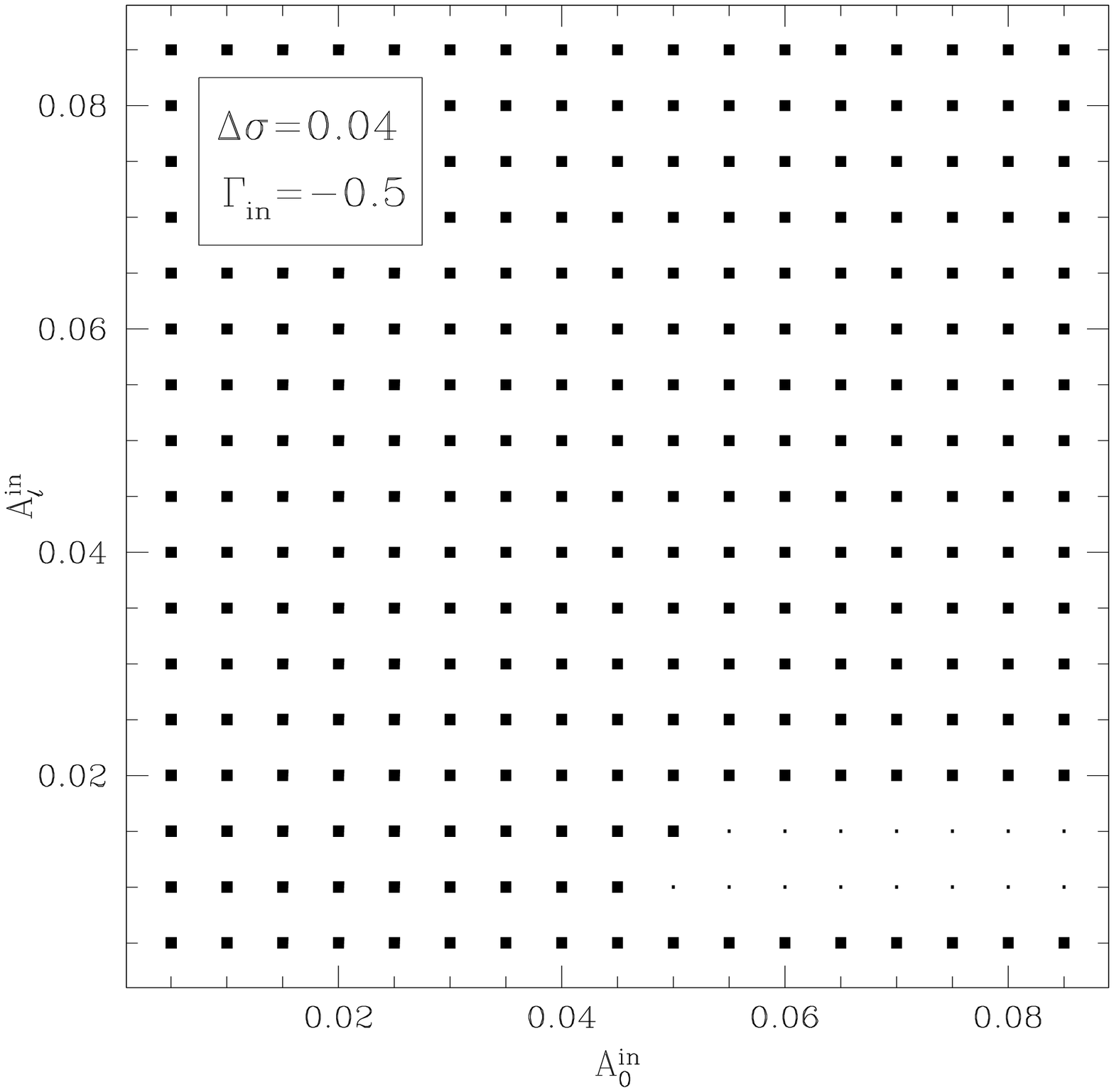}\hskip6mm\includegraphics[width=0.48\textwidth]{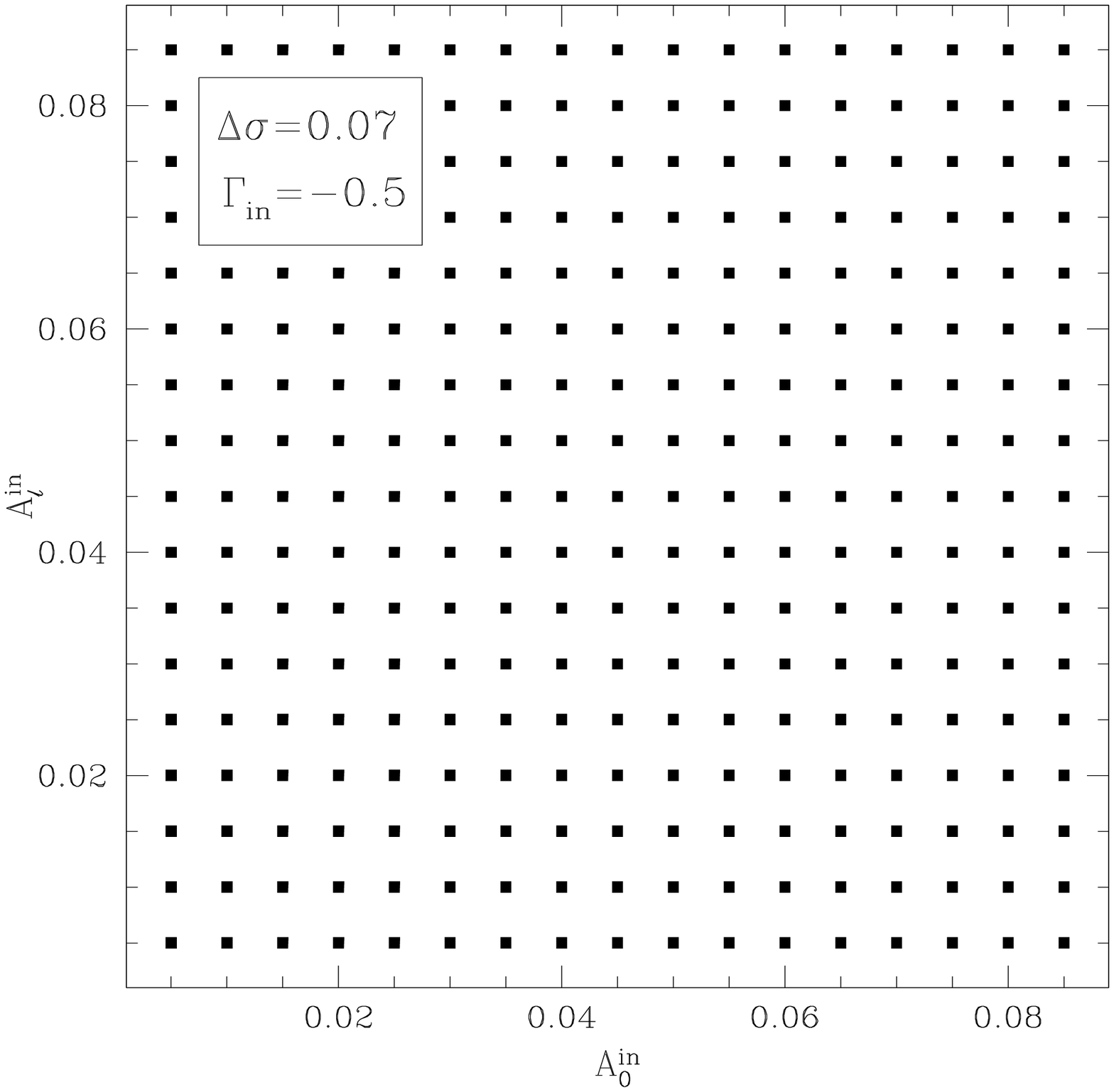}}
\centerline{\includegraphics[width=0.48\textwidth]{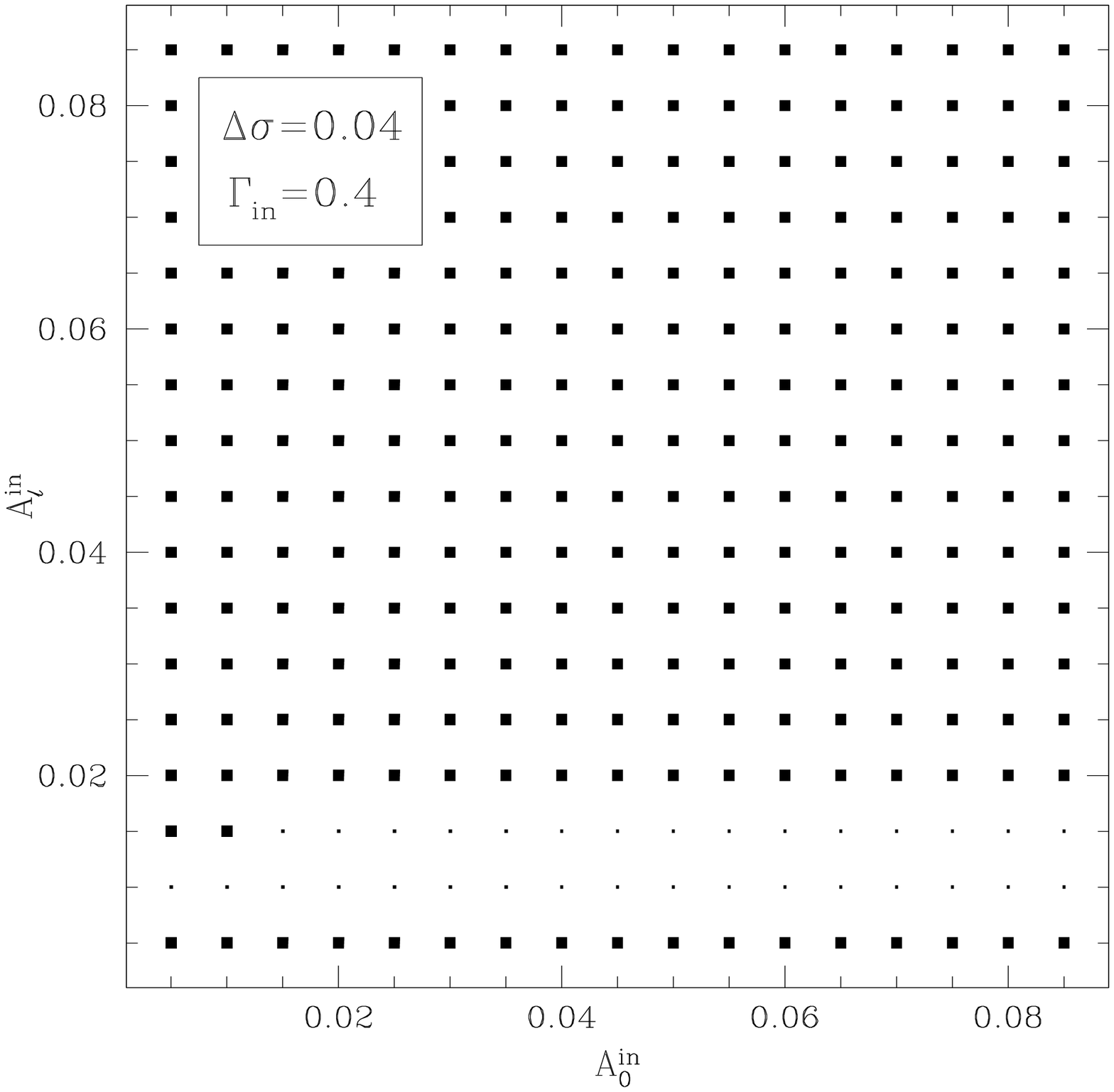}\hskip6mm\includegraphics[width=0.48\textwidth]{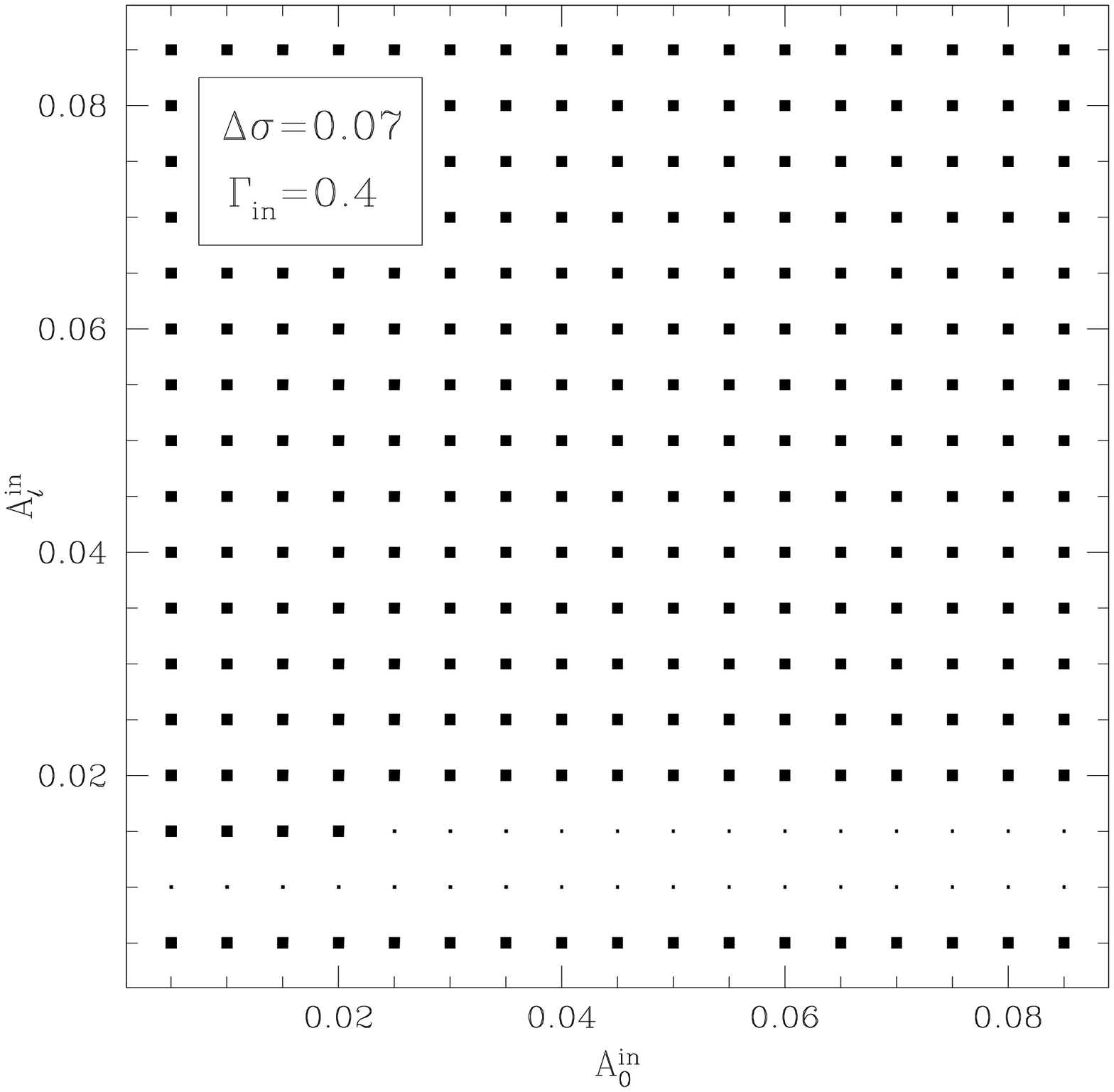}}
\centerline{\includegraphics[width=0.48\textwidth]{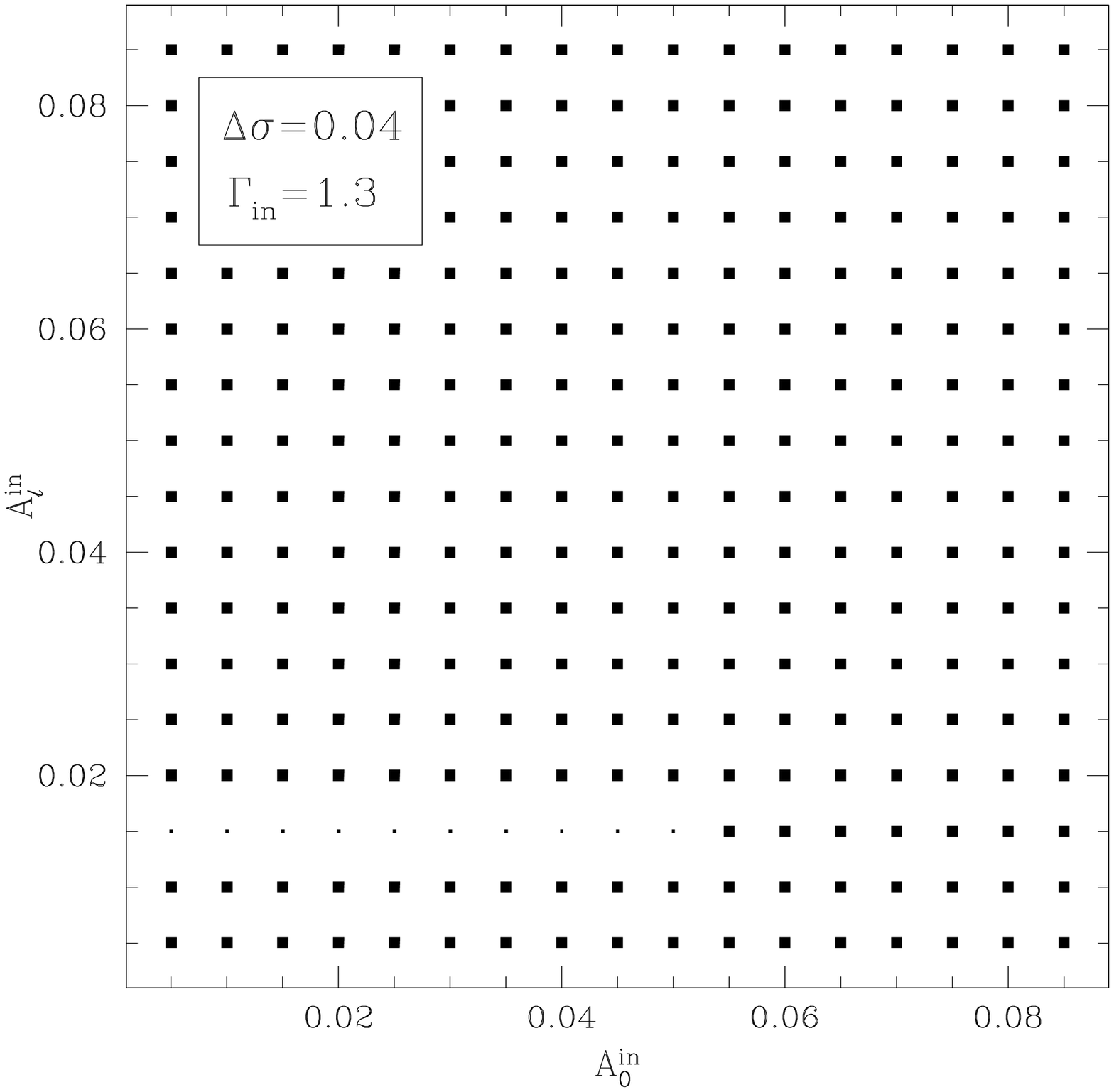}\hskip6mm\includegraphics[width=0.48\textwidth]{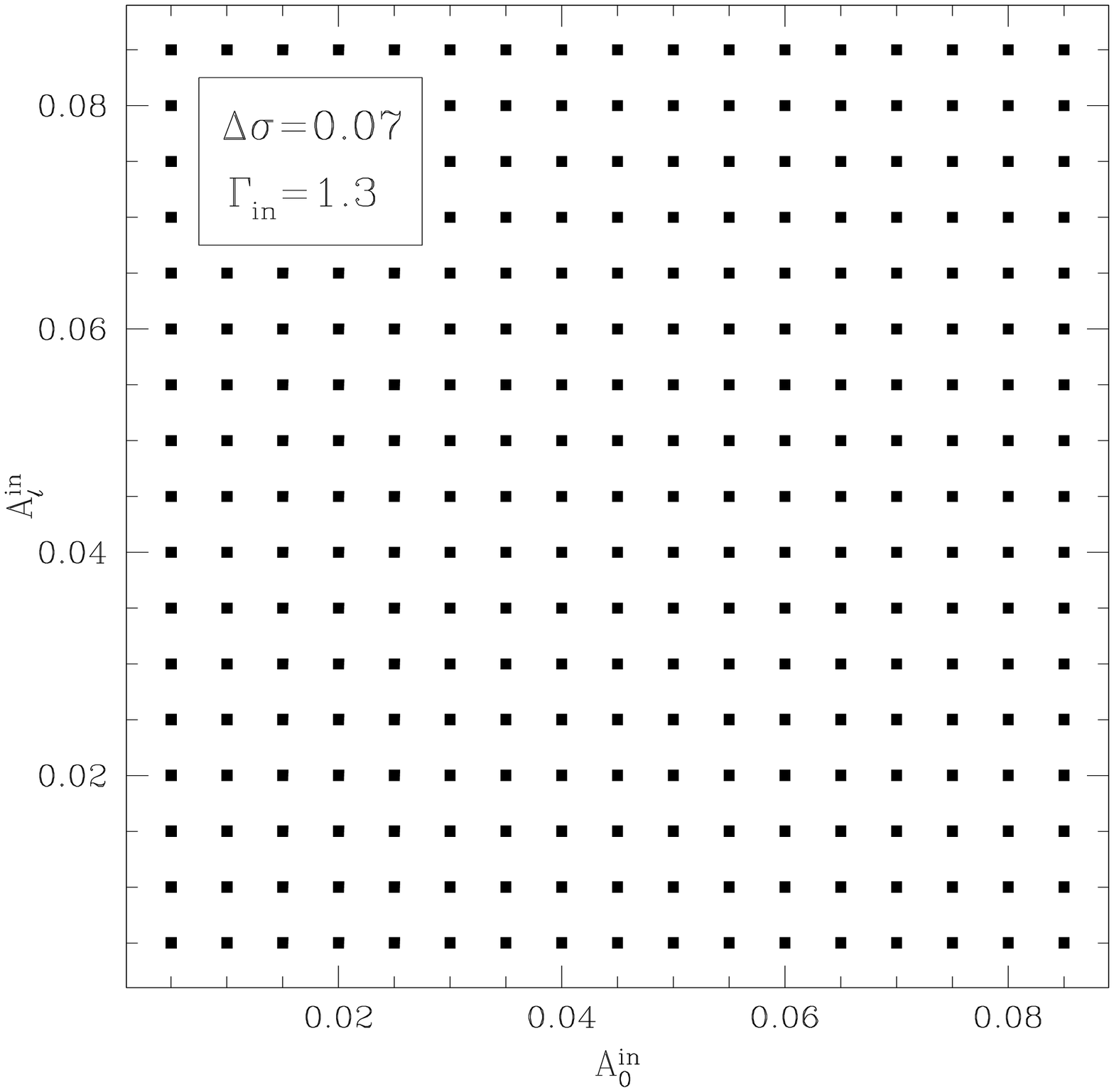}}
\FigCap{Regions of attraction for asymptotic solutions for the resonance 
between the fundamental and the ${l=2}$ modes (see Fig.~3, right panels). 
Squares and dots denote attraction to the $a$- and $b$-branch, respectively.}
\end{figure}
Fig.~8 shows the regions of attraction for the resonance between the 
fundamental radial mode and the nonradial ${l=1}$ mode. The stationary 
solutions for the adopted coefficients are shown in the left panels of Fig.~3. 
Fig.~9 shows the same solutions in the case of the resonance between the 
fundamental radial mode and the nonradial ${l=2}$ mode. The stationary 
solutions for the corresponding cases are shown in the right panels of Fig.~3. 

We made similar computations for other values of $\Delta\sigma$ and other 
pairs of modes. We found the following properties of the attraction regions. 
Except for the small region near the center of resonance in the case of the 
first overtone and $l=1$ mode, that was discussed in Sections~4.4 and 5.1, all 
the attractors are fixed-point solutions. When $b$-branch solutions are stable 
they can be attractors simultaneously with $a$-branch or single-mode 
stationary solutions. When $b$-branch solutions are unstable then the only 
asymptotic states are $a$-branch solutions if ${\Delta\sigma<\Delta\sigma_d}$ 
or single-mode solutions if ${\Delta\sigma>\Delta\sigma_d}$. 

From inspection of Figs.~8 and 9 we may conclude that the higher are the 
values of $A_l$ the smaller are the regions of attraction to the $b$-branch. 
In other words, the farther is the $b$-branch from the lower ($a$-branch or 
single mode) stationary solution, the lower is the probability of finding the 
$b$-branch pulsation. In the three dimensional space of the initial parameters 
the regions of attraction to the $b$-branch are significantly smaller than 
those to the $a$-branch or the single mode solution. Let us note also that the 
star entering the instability strip may begin its pulsation only as a single 
mode pulsator or an $a$-branch pulsator. Except for the exact resonance, a 
phase transition to a $b$-branch pulsation requires a finite energy input. 
Consequently, we will ignore $b$-branch solutions in the foregoing 
considerations. 

We see in Figs.~3--5 that if $b$-branch solutions are ignored the maximum 
values of the $A_l/A_0$ ratio is at ${\Delta\sigma=0}$. Thus, we may conclude 
that the strongest influence of the resonant nonradial mode excitation on 
pulsation is at the exact resonance. From now on we focus on this special 
case.

\Section{The Role of Nonradial Modes in Constant Amplitude Pulsation}
In this Section we calculate reduction of the radial mode amplitude, 
$A_0/A_0^0$, and the relative period change, $\Delta P/P$, caused by the 
resonant nonradial mode excitation. We also calculate the relative amplitude 
of the nonradial mode, $A_l/A_0$, whose presence may perhaps be detected by 
means of spectroscopy. 

\subsection{Analytical Expressions}
As we explained above we assume ${\Delta\sigma=0}$. Then, if 
${\cos\Gamma\ne0}$, from Eq.~(45) we immediately get 
$$(A_l^s)^2=I(A_0^s)^2,\eqno(62)$$
or
$$\frac{A_l^s}{A_0^s}=\sqrt{I}.\eqno(63)$$
If ${\cos\Gamma=0}$ then we are not in the $a$-branch for most of the cases 
considered. The exception is the case in which there is no $b$-branch. The 
example is seen in the right panels of Fig.~4. In such a situation the 
amplitude ratio is determined by Eqs.(43) and (44) where we set 
${\sin\Gamma=1}$. It happens only in a few cases in our large set of resonant 
mode pairs and we do not consider it in detail. It is sufficient to note that 
the true value of $A_l/A_0$ is always a bit smaller than $\sqrt{I}$ which we 
adopt in our calculation. 

In the case of two nonaxisymmetric modes the amplitude of the ${m=\pm1}$ modes 
is given by ${A_\pm=A_l/\sqrt2}$, which implies 
$$\frac{A_{\pm}^s}{A_0^s}=\sqrt{\frac{I}{2}}.\eqno(64)$$

From Eqs.~(43), (44) and Eq.~(62) we obtain
$$R\sin\Gamma_s=\alpha\frac{1+\gamma+2I(H_l-1+\gamma)}
{1+I(1-\gamma)}.\eqno(65)$$
When $\sin\Gamma_s$ obtained with the above expression is bigger than 1 we 
should use ${\sin\Gamma_s=1}$. This is the case of the absence of the 
$b$-branch discussed above. We will not show any numerical results for this 
case. 

After inserting solution given by Eq.~(65) into Eq.~(43) we obtain
$$(A_0^s)^2=\frac{1}{\alpha}\cdot\frac{1+I(1-
\gamma)}{1+2I(2+IH_l)},\eqno(66)$$
which implies
$$\frac{A_0^s}{A_0^0}=\sqrt{\frac{1+I(1-\gamma)}{1+2I(2+IH_l)}}.\eqno(67)$$
For the relative change of the radial mode period caused by nonradial
mode excitation we get using Eq.~(19),
$$\frac{|\Delta P_s|}{P}=\frac{|\delta\sigma_s|}{\sigma_0}=
\frac{R}{\sigma_0}|\cos\Gamma_s|(A_l^s)^2.\eqno(68)$$
The sign of $\cos\Gamma_s$ is undetermined at ${\Delta\sigma=0}$. In the rest 
of the $a$-branch the sign of $\cos\Gamma$ is opposite to the sign of 
$\Delta\sigma$. Thus, the frequency shift can be either positive or negative 
with equal probabilities. From Eq.~(65) we obtain 
$$|R\cos\Gamma_s|=\sqrt{R^2-\alpha^2\frac{\big(1+\gamma+2I(H_l-1+
\gamma)\big)^2}{\big(1+I(1-\gamma)\big)^2}}.\eqno(69)$$
For the dimensionless frequency we use 
$$\sigma_0=\frac{\omega_0}{\kappa_0}=\frac{2\pi}{P\kappa_0}.\eqno(70)$$
Finally, with the help of Eqs.~(62), (66), (69), (70) we get from Eq.~(68)
\setcounter{equation}{70}
\begin{eqnarray}
\frac{|\Delta P_s|}{P}&=&
\frac{P\kappa_0}{2\pi}\cdot\frac{I}{1+2I(1+2IH_l)}\times\nonumber\\
&&\times\sqrt{\frac{R^2}{\alpha^2}\big(1+I(1-\gamma)\big)^2
-\big(1+\gamma+2I(H_l-1+\gamma)\big)^2}.
\end{eqnarray}
In a few cases of the lack of the $b$-branch the period change is exactly 
zero. Eqs.~(67) and (71) allow us to evaluate the relative change of the 
amplitude and period, respectively, for the radial mode caused by a resonant 
excitation of nonradial mode. In the right-hand side we have quantities that 
may be calculated for specified model and mode. The nonradial mode amplitude 
can be calculated with Eq.~(63). 

\subsection{Application to Evolutionary Models of RR~Lyr Stars}
We apply here the formulae of the previous Section to our models of RR~Lyr 
stars. Like in Section~4.1 we considered the resonance between the radial and 
the nearest nonradial modes of ${l=1,2,5,6}$, for fundamental mode, and of 
${l=1,4}$, for first overtone. Using the same mode parameters we calculated 
the nonradial mode amplitudes, Eq.~(63), the reduced radial mode amplitudes, 
Eq.~(67) and the relative period changes, Eq.~(71). Results are given in 
Fig.~10. The values give the maximum effect of the resonance because -- we 
recall -- ${\Delta\sigma=0}$ was assumed. 
\begin{figure}[p!]
\centerline{\includegraphics[width=0.48\textwidth]{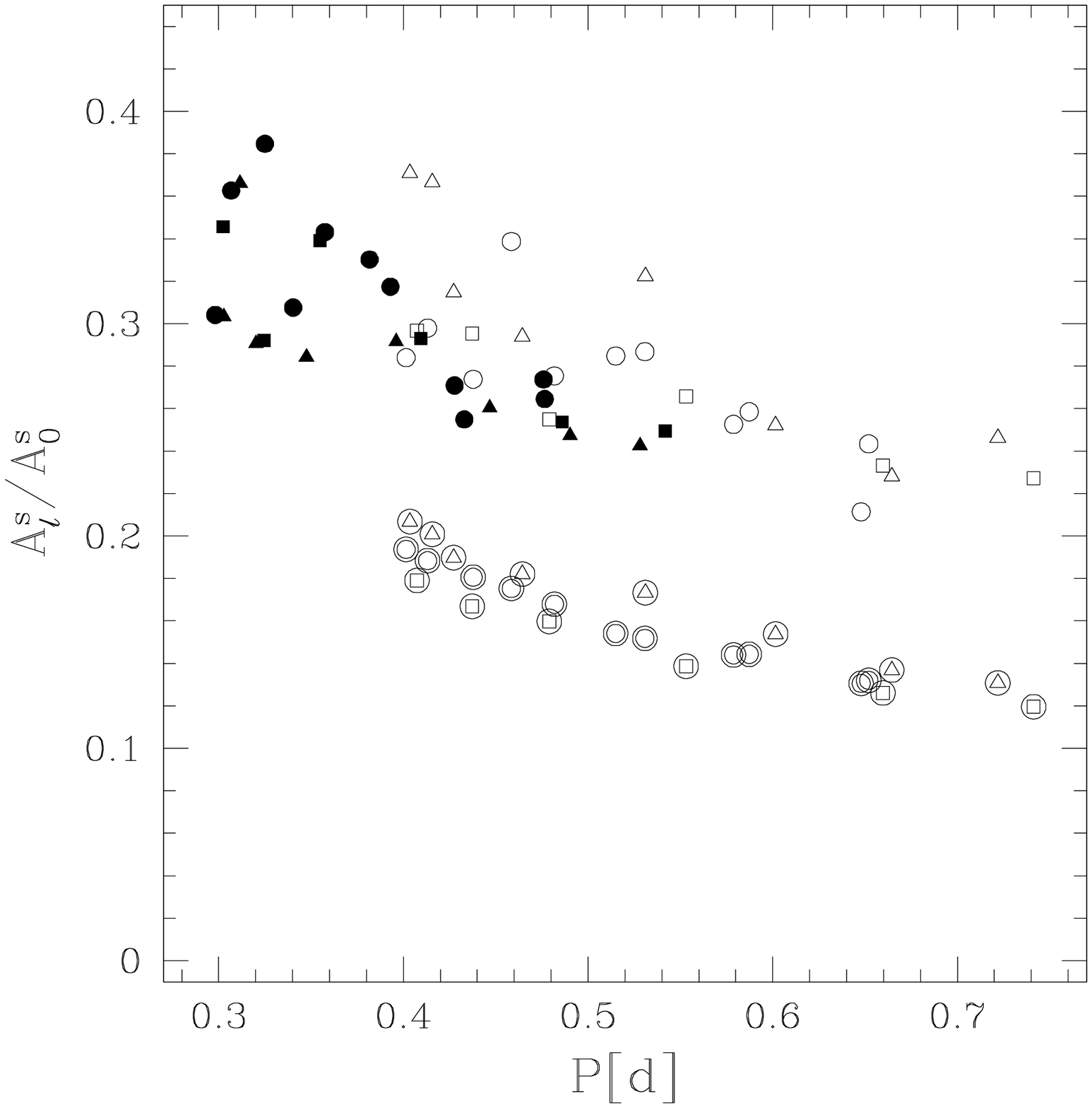}\hskip6mm\includegraphics[width=0.48\textwidth]{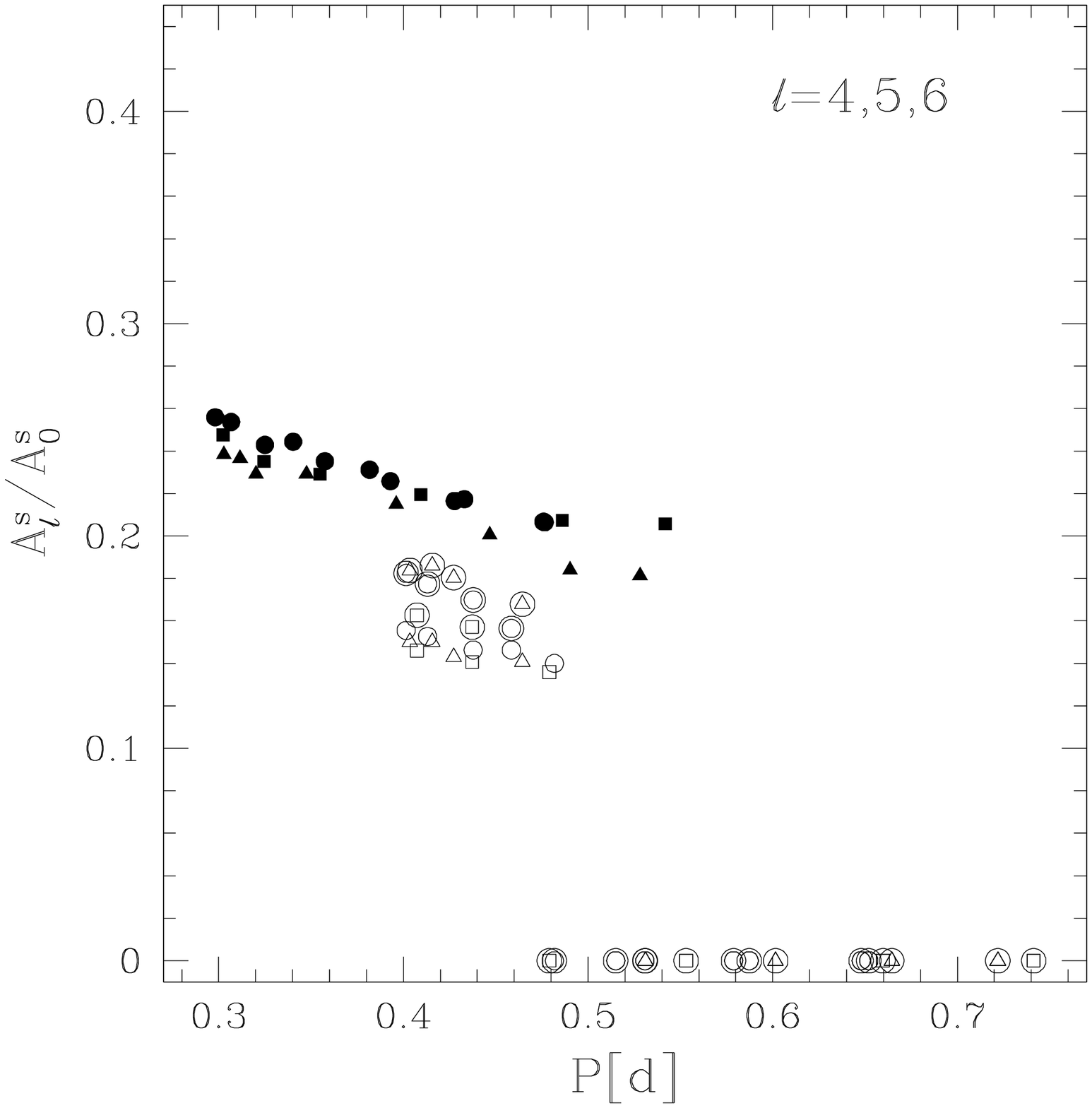}}
\centerline{\includegraphics[width=0.48\textwidth]{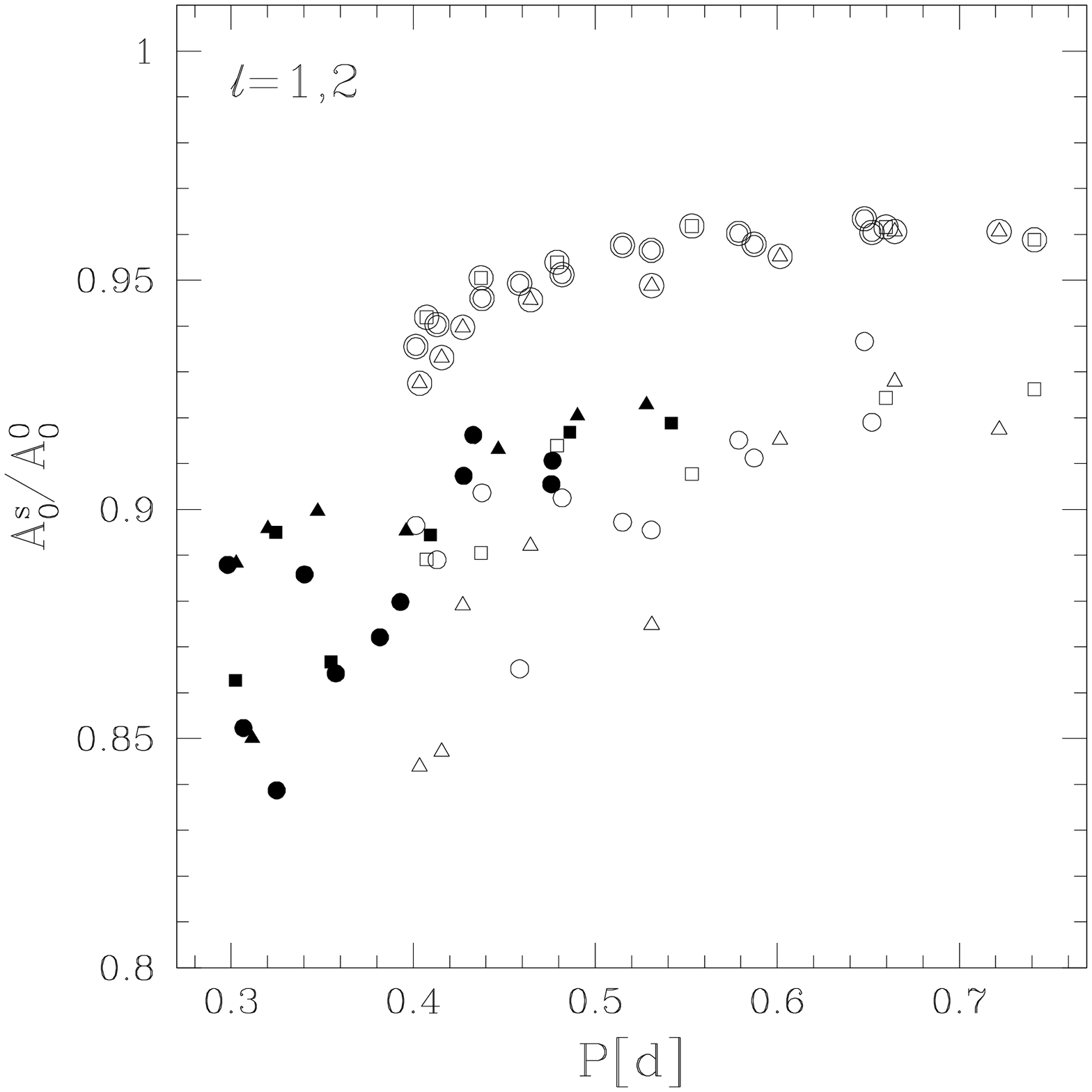}\hskip6mm\includegraphics[width=0.48\textwidth]{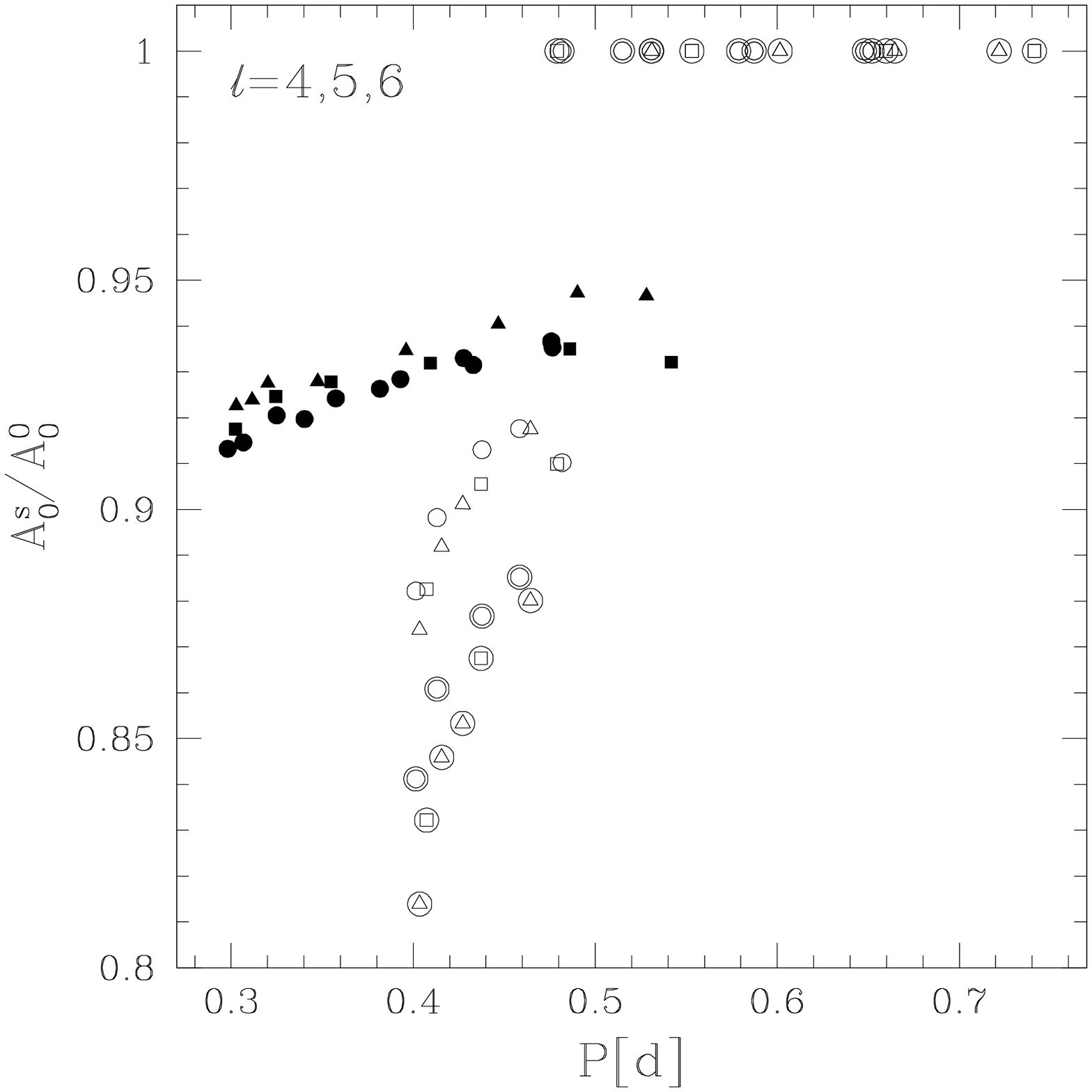}}
\centerline{\includegraphics[width=0.48\textwidth]{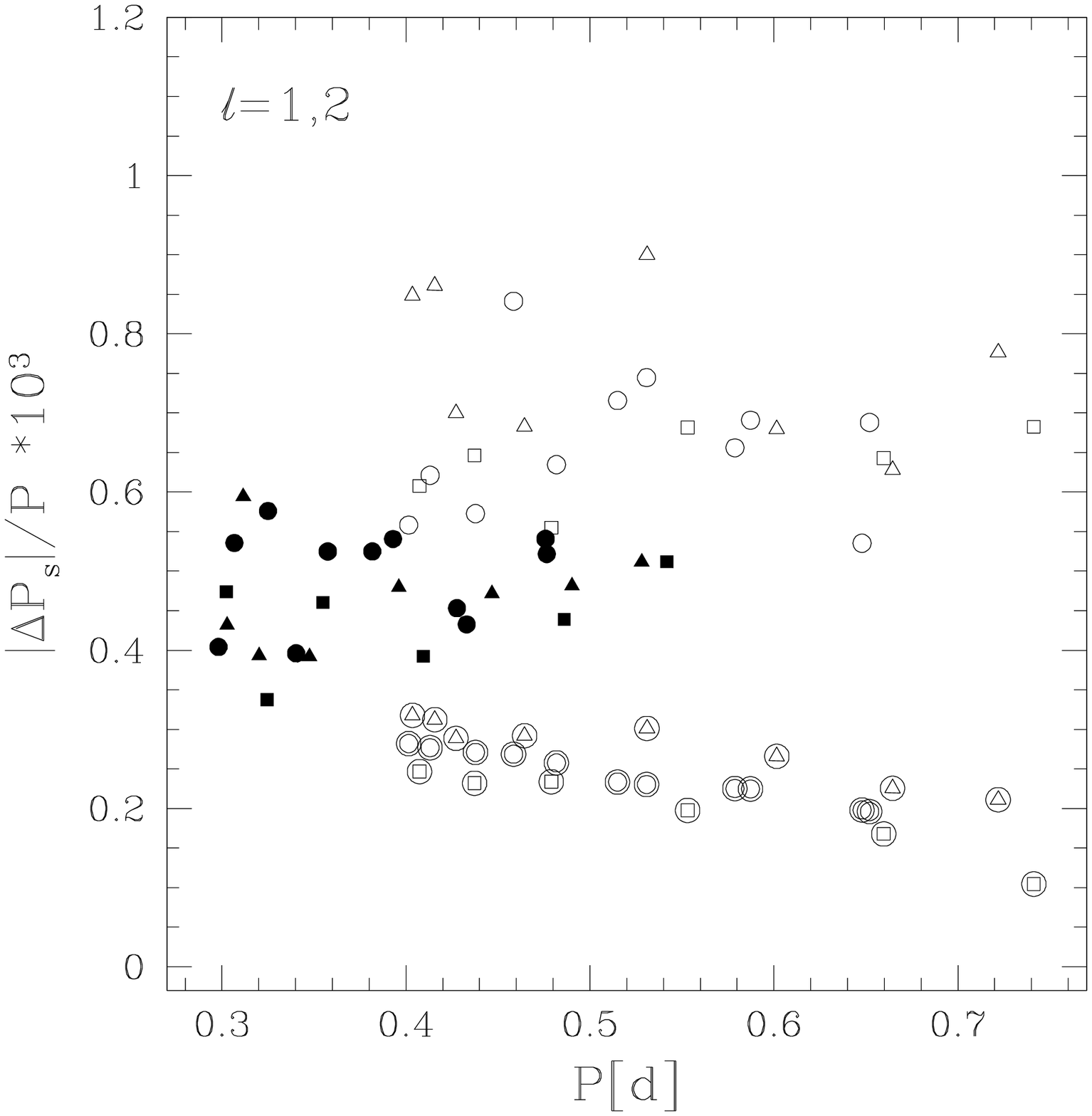}\hskip6mm\includegraphics[width=0.48\textwidth]{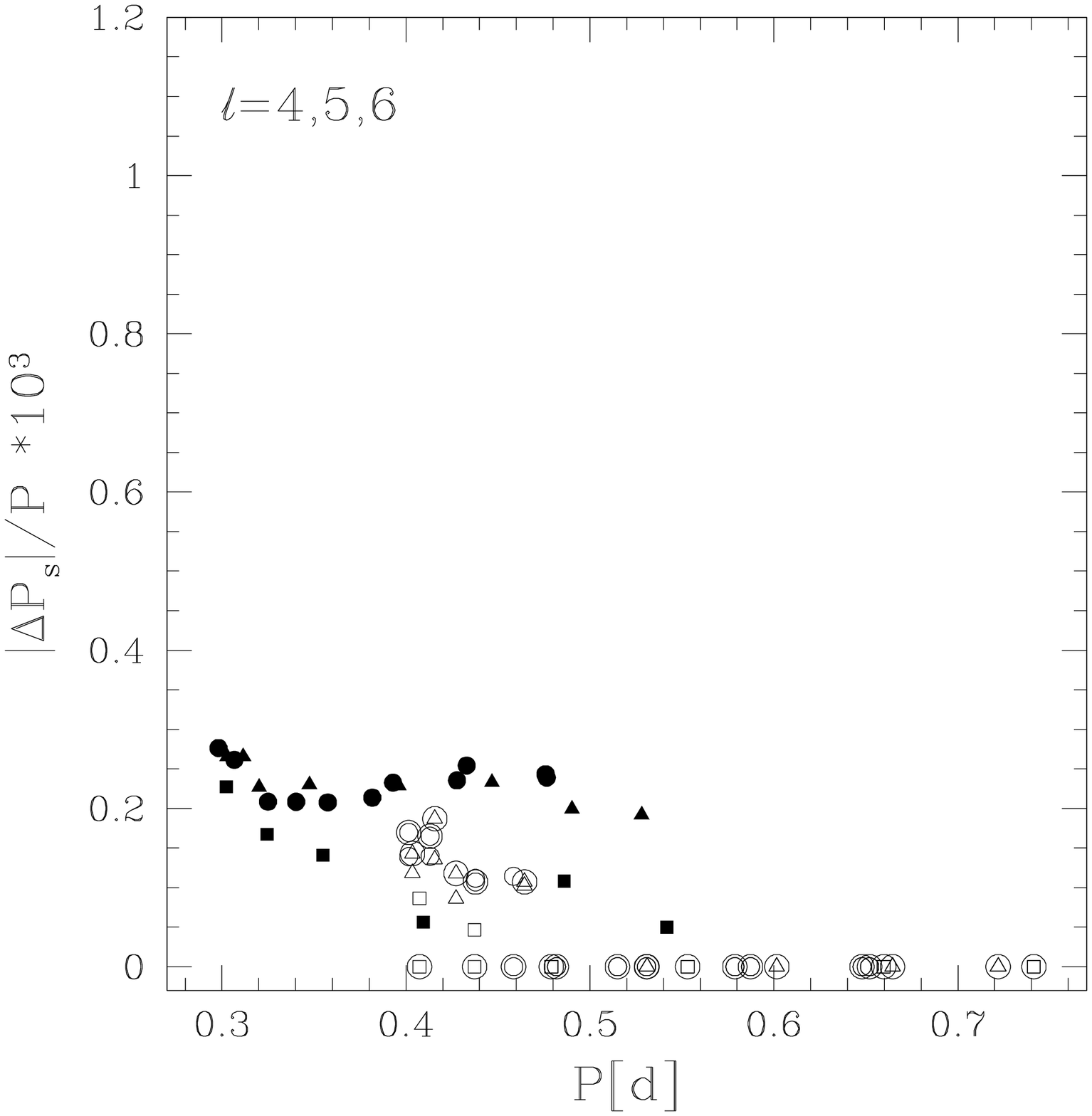}}
\FigCap{Finite amplitude consequences of the 1:1 resonance between the radial 
and the selected nonradial modes. Different symbols correspond to different 
RR~Lyr models and different modes (see Fig.~1 for more detailed explanation). 
The top panels show the relative amplitudes of the nonradial modes. The middle 
panels show the ratio of the resonant to the nonresonant radial mode 
amplitudes. The bottom panels show the relative period shift due to the 
resonance.} 
\end{figure}

We see in the top panels that the nonradial mode amplitude may be quite large. 
Still detecting the presence of nonradial modes, which would constitute an 
ultimate test for our theory, may not be easy. In the case of low degree 
nonradial modes, the determination of $l$ by means of multicolor photometry is 
feasible (see \eg Garrido 2000). We should stress, however, that the technique 
has been applied so far only to main sequence pulsators and with only a 
moderate success. For higher degree modes (${l>2}$) one may contemplate use of 
line profile variations (see \eg Aerts and Eyer 2000). Again in this case the 
technique has been used only for the main sequence pulsators. 

The amplitude reduction caused by the resonance, which is shown in the middle 
panels, is significant in some cases. Note, in particular, in the right panel 
the relatively large reduction caused by higher degree modes. The implication 
is that the pulsation amplitude must be regarded as a random quantity in 
certain ranges. It is random because the detuning parameter $\Delta\sigma$ 
which affects the amplitude is in fact a random quantity in the 
$[0,\Delta\sigma_{\rm max}]$ range. 

However, in the case of low-degree nonradial modes, the primary source of 
uncertainty of observable amplitude prediction is the orientation of the 
pulsation axis. We will now focus on $l=1$ mode and will convert amplitudes 
shown in Fig.~10 to those of light and radial velocity. 

In Appendix~B.1 we derive formulae for the radial velocity and light curve 
amplitudes. With the help of Eq.(B9) we get the ratio of radial velocity 
amplitude with the presence of the nonradial mode, $A_v$, to that without the 
nonradial mode, $A_{v,0}$. The expression is 
$$\frac{A_v}{A_{v,0}}=\sqrt{1+\epsilon_v^2+2\epsilon_v\cos(\Gamma/2)}\eqno(72)$$
where for ${l=1}$ we have 
$$\epsilon_v=1.35A_1/A_0\cos\Theta_0.\eqno(73)$$
Here $\Theta_0$ is the aspect angle, $A_1/A_0$ is the calculated amplitude 
ratio, whose maximum values are shown in the top panels of Fig.~10. 

Similarly, we get for the light amplitude ratio
$$\frac{A_{M{\rm bol}}}{A_{M{\rm bol,0}}}=\sqrt{1+\frac{\epsilon_M^2}{p_0^2}+
\frac{2\epsilon_M}{p_0^2}\!
\left[\frac{2}{f}\cos(\psi+\Gamma/2)+\cos(\Gamma/2)\right]}\eqno(74)$$
(see Eqs.~B12, B19, and B21). For ${l=1}$ 
$$\epsilon_M=1.23A_1/A_0\cos\Theta_0.\eqno(75)$$
The complex quantity $fe^{i\psi}$ is the ratio of the perturbed radiative flux 
to radial displacement at the surface. It is determined from linear 
nonadiabatic calculations. We also denoted 
$$p_0=\sqrt{1+\frac{4}{f}\cos\psi+\frac{4}{f^2}}.\eqno(76)$$

When we assume the maximum value of the $A_1/A_0$ ratio to be about 0.3 the 
maximum absolute value of $\epsilon$ is about 0.4 and 0.37 for the radial 
velocity and the light variations, respectively. The largest effect is for 
${|\cos(\Gamma/2)|=1}$. Note that for ${l=1}$ modes we get ${\Gamma\approx0}$ 
or $2\pi$. Then the {\it rhs} in Eq.~(72) is ${1\pm\epsilon_v}$. Since 
$\epsilon_v$ is a random quantity with values between ${-0.4}$ and 0.4, the 
range of the ratio $A_v/A_{v,0}$ ranges from 0.6 to 1.4. 

For the light curve we have to discuss possible values of $f$ and $\psi$. In 
our linear RR~Lyr pulsation models $f$ ranges from about 2 to over 10. The 
phase $\psi$ is correlated with $f$ so that it is close to zero for 
${f\approx2}$ and almost $\pi/2$ for ${f>10}$. For the lowest values of these 
quantities we have ${p_0=2}$ (see Eq.~76) and ${A_{M{\rm bol}}/A_{M{\rm 
bol,0}}\,=\,\sqrt{1\,+\,\epsilon_M^2/4\,+\,\epsilon_M\cos(\Gamma/2)}}$ (see 
Eq.~74). For ${|\cos(\Gamma/2)|\,=\,1}$ we obtain ${A_{M{\rm bol}}/A_{M{\rm 
bol,0}}\,=\,1\,\pm\,\epsilon_M/2}$. If ${f\,\gg\,2}$, we easily obtain 
${p_0\,=\,1}$ and $A_{M{\rm bol}}/A_{M{\rm bol,0}}=1\pm\epsilon_M$, that is 
almost as wide range as in the case of the radial velocity amplitudes. 

At higher $l$'s, the expected amplitude scatter should be much lower due to 
both the smaller $A_l/A_0$ ratio and the effect of averaging over the stellar 
disc. 

We conclude that certain scatter in the dependence of the pulsation amplitudes 
on stellar parameters should be expected. Let us note that the existence of 
such a dependence underlines empirical methods of absolute magnitude and 
metallicity determination employing light curve data (\eg Kov\'acs and Jurcsik 
1996, Jurcsik and Kov\'acs 1996). 

The period changes shown in the bottom panels of Fig.~10 are indeed rather 
small. Again they should be regarded as random quantities. The expected 
scatter is probably inconsequential for the empirical use of period data on 
RR~Lyr stars. 

\section{Amplitude and Phase Modulation Due to Nonradial Mode Excitation}
\subsection{Theoretical Prediction}
In Section~4.2 we showed that the stationary solution in the case of the 2:1+1 
resonance is an equally-spaced triplet with the central peak corresponding to 
the radial mode and the two side-peaks corresponding to the nonradial modes. 
We will show here that excitation of the triplet manifests itself, in general, 
as pulsation with amplitude and phase modulation and thus provides a plausible 
explanation of the Blazhko effect. We focus on the ${l=1}$ modes. Modes at 
${l=2}$ have significantly lower probability of excitation, as may be seen in 
Fig.~1, while those at ${l>2}$ have significantly lowered amplitudes due to 
the averaging effect, as discussed in Appendix~B. We consider first an 
excitation of the ${m=\pm1}$ modes of the same radial order $n$. 

In uniformly and slowly rotating stars the frequencies of the ${m=\pm1}$ modes 
are given by 
$$\omega_\pm=\omega_0\pm C\Omega+D\Omega^2,\eqno(77)$$
where $C$ (the Ledoux constant) and $D$ depend on the stellar structure. In 
RR~Lyr stars contribution to the Ledoux constant arises predominantly in the 
$g$-mode propagation zone and thus we can use the asymptotic value ${C=1/2}$ 
for ${l=1}$. Values of $D$ for ${l=1}$ $g$-modes are very small. As first 
noted by Buchler \etal (1995), the resonant phase-lock effect changes the 
frequencies in such a way that $D\Omega^2$ becomes zero. Eq.~(77) yields 
frequencies in the corotating system. In the inertial system we have to add 
${-m\Omega}$, which accounts for transformation of the azimuthal angle. Note 
that in our notation ${m=-1}$ corresponds to prograde modes. Thus, for the 
nonradial mode frequencies in the inertial system we have the following 
relation 
$$\omega_\pm=\omega_0\mp\lambda\eqno(78)$$
where ${\lambda=\Omega/2}$. Eq.~(78) remains valid in the case of 
depth-dependent rotation, but then $\Omega$ should be understood as the 
average rotational frequency weighted with the Brunt-V\"ais\"al\"a frequency, 
which is strongly peaked in the innermost part of the radiative interior. 

In Appendix~B.2 we show that when the amplitudes of ${l=1}$, ${m=\pm1}$ modes 
are equal and small in comparison with that of the radial mode, the radial 
velocity is given by 
$$V_r=A_{v,0}\left[1+a\cos\lambda t\right]
\cos\big(\omega t-\pi/2+b\cos\lambda t\big)\eqno(79)$$
(see Eq.~B24 and B36), where
\setcounter{equation}{79}
\begin{eqnarray}
a&=&\epsilon_v\cos(\Gamma/2),\\
b&=&\epsilon_v\sin(\Gamma/2).
\end{eqnarray}
The quantity $\epsilon_v$ is defined similarly as in Eq.~(73) but with 
$\sin\Theta_0$ instead of $\cos\Theta_0$, $A_{v,0}$ is given in Eq.~(B6). The 
maximum value of $\epsilon_v$ is about 0.4. The largest amplitude modulation, 
factor of $7/3$, is obtained for ${|\cos(\Gamma/2)|=1}$ and then there is no 
phase modulation. The opposite case of maximum phase modulation and no 
amplitude modulation corresponds to ${\cos(\Gamma/2)=0}$. It is of interest 
that for ${l=1}$ we should be close to one of these two extreme cases, since 
we have ${\sin\Gamma\approx0}$. 

In a similar manner we obtained (see Appendix~B.2 for details) the bolometric 
luminosity variations in the form 
$$\Delta M_{\rm bol}=A_{M{\rm bol},0}\left[1+c\cos\lambda t\right]
\cos\big(\omega t+\varphi_0+d\cos\lambda t\big)\eqno(82)$$
where
\setcounter{equation}{82}
\begin{eqnarray}
c=\frac{\epsilon_M}{p_0}\cos(\psi+\Gamma/2-\varphi_0),\\
d=\frac{\epsilon_M}{p_0}\sin(\psi+\Gamma/2-\varphi_0)
\end{eqnarray}
(see Eqs.~B39, B47, and B48). Quantities $p_0$ and $\varphi_0$ are given by 
Eqs.~(B13) and (B14). Explicit expression for $p_0$ is also 
given by Eq.~(76). For $\epsilon_M$ we may use {\it rhs} of Eq.~(75) replacing 
$\cos$ with $\sin$. The lowest values of $f,\psi$ are close to 2 and 0, 
respectively. From Eqs.~(B13)--(B14) we find ${p_0\approx2}$, 
${\varphi_0\approx0}$, which yields 
\begin{eqnarray}
c&\approx&\frac{\epsilon_M}{2}\cos(\Gamma/2),\\
d&\approx&\frac{\epsilon_M}{2}\sin(\Gamma/2).
\end{eqnarray}
For the highest values of ${f(\gg2)}$ we have ${\psi\approx\pi/2}$ and we find 
${p_0\approx1}$, ${\varphi_0\approx\pi/2}$, which yields 
\begin{eqnarray}
c&\approx&\epsilon_M\cos(\Gamma/2),\\
d&\approx&\epsilon_M\sin(\Gamma/2).
\end{eqnarray}
This means that we obtain the same conditions for pure amplitude and pure 
phase modulation as in the case of the radial velocity variation. Also the 
ranges of modulation are similar. However, at moderate values of $f$ and 
$\psi$ we may have significant both phase and amplitude modulations. 

The 2:1+1 resonance may lead to excitation of certain modes from consecutive 
multiplets. The two distinct cases involve the following pairs of $m$ values: 
${(1,-1)}$, ${(-1,1)}$. The resonant interaction between a $(0,0)$ pair and 
the radial mode is also possible, but this case is described by much more 
complicated AEs and should be studied separately. Thus we focus here on the 
${m=\pm1}$ pairs. Then the modulation frequency is 
$$\lambda=\left\{\begin{array}{ccc}
\frac{\Delta\omega-\Omega}{2}&\quad\mathrm{for}\quad
m\mathrm{'s}\quad&(1,-1)\\
\frac{\Delta\omega+\Omega}{2}&\quad\mathrm{for}\quad
m\mathrm{'s}\quad&(-1,1)\end{array}\right.\eqno(89)$$
where $\Delta\omega$ is the frequency difference between centers of multiplets 
of the radial order $n$ and ${n+1}$. In this case we expect unequal amplitudes 
of the side-peaks in consequence of the inequality of mode inertia. In the 
${(-1,1)}$ case the effect is systematic. It may be seen in Fig.~2 of DC that 
lower frequency modes have higher inertia typically by a factor of 1.5, 
implying amplitudes lower by about 20\% (see Eqs.~35, 34). In the ${(1,-1)}$ 
case we may have the opposite situation if ${\Omega>\Delta\omega}$. 

Excitation of nonradial modes provides a natural explanation of basic 
properties of the Blazhko-type modulation. Of course, our simple treatment of 
nonlinear effects precludes a detailed comparison with observations. The basic 
characteristics of the modulation are periods and ranges of amplitude 
variation. 

\subsection{Comparison with Observational Data}
We are now in the position to compare quantitative prediction of our model 
with observational data on the Blazhko effect. The numbers to compare are 
modulation periods and ranges of amplitude modulation. 

The modulation to pulsation period ratio, $P_B/P_0$, ranges from 20 to 1000 
(Ko\-v\'acs 1993). In our model we have 
$$\frac{P_B}{P_0}=\frac{2\omega_0}{\Delta\omega}
\frac{1}{|\Omega/\Delta\omega+k|},\eqno(90)$$
where $\Delta\omega$ is the frequency difference between centroids of 
multiplets of radial orders $n$ and ${n+1}$. We set ${k=0}$ in the case of 
modes from the same multiplet and ${k=m_n}$ in the case of modes from the 
consecutive multiplets. From Fig.~6 of DC we find the values of 
$\Delta\omega/\omega_0$ for the ${l=1}$ modes in the vicinity of the 
fundamental mode ranging from 4.5 to ${7.5\times10^{-3}}$. Let us consider 
first the case ${k\ne0}$ and ${\Omega\ll\Delta\omega}$. Then we find 
${270<P_B/P_0<450}$. Unless our models of RR~Lyr stars are grossly inadequate, 
we cannot account for large fraction of observed modulation periods. 
Interpretation of Blazhko stars with short modulation periods demands 
${\Omega\gg\Delta\omega}$. 

We are not aware of any $V_{\rm rot}\sin i$ determination for any RR~Lyr star. 
Peterson \etal (1996) provide only an upper limit of 10~km/s for 8 RRab stars. 
If we assume uniform rotation we may translate this upper limit to the minimum 
value of $P_B/P_0$. This minimum value is about 30. This is still more than 
the lowest observed modulation period. However, we should remind that $\Omega$ 
does not refer to the surface but to the deep interior and it is quite 
plausible that the interior rotates much faster than the surface. We conclude 
that our prediction of modulation periods is not inconsistent with 
observations. 

A comparison of our prediction regarding ranges of amplitude modulation with 
observational data is much more complicated. Firstly, inadequacy of our 
approximation is certainly far more important in this case; secondly, we have 
much less observational information. The only systematic survey of amplitude 
modulation in RR~Lyr stars is that from the MACHO project (Alcock \etal 2000), 
but it is restricted to RRc stars in LMC. Let us remind that the Blazhko 
phenomenon is much more common in RRab stars than in RRc. In Alcock's \etal 
(2000) list we find the cases of rather different side peak amplitudes, much 
larger than those implied by our Eq.~(64). Also there are cases of side peak 
amplitudes exceeding the central peak amplitude, contrary to our predictions. 
We believe, however, that these cases do not invalidate the very essence of 
our model, but point out to necessity of supplementing it by taking into 
account more interacting modes. Samples of Blazhko RRab stars from the MACHO 
project were analyzed by Kurtz \etal (2000). In a few examples they point out 
different degrees of phase and amplitude modulations. Our model provides a 
natural explanation of the above diversity. 

For a comparison of our simplified theory with observations, the radial 
velocity data are more suitable. The reason is that the higher order nonlinear 
effects, which we ignored, are more important in light than in radius 
variations. Unfortunately, the radial velocity data on Blazhko phenomenon are 
very limited. Recently, Chadid \etal (1999) investigated line profile 
variations in RR~Lyr itself. They find a triplet associated with the main 
frequency but it is not exactly equally spaced. The spacings are 0.027, 0.020 
c/d. We do not know whether this departure from equidistancy is significant. 
The relative side peak amplitudes are about 0.3 which is about the maximum 
value predicted by our model. 

\Section{Conclusions and Discussion}
We solved simplified amplitude equations describing resonant coupling between 
radial and nonradial modes. The equations cover the case of 1:1 resonance 
between a radial and an axisymmetric ${(m=0)}$ nonradial mode as well as the 
case of 2:1+1 resonance between a radial mode and a ${(m,-m)}$ pair. We showed 
that in the limit of slow rotation and modes of the same multiplet the two 
cases are equivalent. If the modes belong to neighboring multiplets this is 
not true, but the treatment is very similar. The crucial parameter is 
$\Delta\omega$ which is a measure of the departure from exact resonance. We 
showed that if radial pulsations are unstable, which requires $\Delta\omega$ 
to be less than certain critical value, then there are typically two branches 
of stationary solutions which merge at ${\Delta\omega=0}$. The branch 
corresponding to lower relative amplitude of the nonradial mode is almost 
always stable. The other branch may also be stable in a certain range around 
${\Delta\omega=0}$, but the region of attraction to solutions in the latter 
branch is relatively small. For this reason we use the solutions corresponding 
to ${\Delta\omega=0}$ as representing the maximum departure from pure radial 
pulsation. 

Excitation of an axisymmetric nonradial mode leads to pulsation with constant 
amplitude. The presence of a nonradial component affects both period and 
amplitude. The most significant effect is an amplitude change in the case of 
excitation of a dipole mode. The size of the expected amplitude change depends 
on $\Delta\omega$ and the aspect angle of the pulsation axis. Both must be 
regarded as random quantities. The former, in principle, is determined by 
stellar parameters, but it is so sensitive that we may specify only its upper 
limit which is ${(\omega_{n,l}-\omega_{n+1,l})/2}$. We found that for the 
specified model and pulsation mode the uncertainty of the amplitude may reach 
up to 40\%. The consequence is that pulsation amplitude and other 
characteristics of light curve are not uniquely determined by mean stellar 
parameters. This is not a good news for diagnostic applications of the light 
curve data. 

Amplitude modulated pulsation is expected in the case of excitation of 
nonaxisymmetric pair of nonradial modes of low degree $l$. Here again the case 
of ${l=1}$ is most likely and most interesting. In this case we find that the 
maximum range of amplitude modulation is about 40\%. The modulation period is 
determined by the characteristic of the deepest layers of the radiative 
stellar interior. Specifically, it is given by ${4\pi/|\Omega_{\rm 
rot}+k(\omega_{n,l}-\omega_{n+1,l})|}$, where ${k=0,\pm1}$, $\Omega_{\rm rot}$ 
is the mean rotation rate in those layers, and ${\omega_{n,l}-\omega_{n+1,l}}$ 
is the distance between the centers of the consecutive multiplets, $n$ at 
fixed $l$. Hence, if this is the modulation seen as the Blazhko effect then 
the Blazhko period is a probe of deep interior properties of Horizontal Branch 
stars. The prospect of sounding interiors of HB stars is quite interesting. 

The reader must be reminded of simplifications used in this paper. We rely on 
the formalism which assumes that the amplitudes are small. This is not true. 
Especially our prediction of the light amplitude must be regarded unreliable. 
Higher order nonlinear effects should be included. Our treatment of mode 
coupling was simplified by adopting a simple scaling of the coefficients and 
by assuming that they are either pure real or pure imaginary. Taking into 
account their complex character may imply an instability of constant amplitude 
solution and imply periodic limit cycle (Moskalik 1986). In this case, the 
modulation period would be determined by the properties of the outermost 
layers. Validity of our formalism is restricted to the cases of 1:1 and 2:1+1 
resonances. Furthermore, couplings which are not covered by our formalism may 
also play a role. These include coupling between the nonradial modes within 
the multiplet as well as the coupling between the radial mode and two 
axisymmetric nonradial modes of consecutive multiplets. We plan to consider 
these cases in our subsequent paper. 

We have realized that there is a considerable paucity of observational data on 
the Blazhko effect. The crucial signature of the type of mode coupling 
considered in this paper is the occurrence of the equidistant triplet in  
pulsation spectra. Such spectra, based on modern photometry of RR~Lyr stars, 
are available only for RRc stars and not for RRab stars, for which the Blazhko 
effect is claimed to be much more common. From the point of view of testing 
our model of the Blazhko effect, spectroscopy is more suitable for detecting 
nonradial modes than photometry. Further, if such modes are detected we may 
extract the information about mode order and the aspect angle of the pulsation 
axis. With this information we may learn about differential rotation in the 
interiors of Horizontal Branch stars. Spectroscopic studies of Blazhko effect 
began only very recently (Chadid \etal 1999). 

\Acknow{This work was supported in part by the KBN grant\\ 2-P03D-14.}

\appendix
\section*{Appendix}
\setcounter{equation}{0}
\renewcommand{\theequation}{{\em\thesection}\arabic{equation}}
\section{Amplitude Equations for Two- and Three-Mode Resonance}
In this Appendix we outline the derivation of amplitude equations for the 1:1 
resonance (the radial and the ${m=0}$ nonradial mode) and the 2:1+1 resonance 
(the radial and the pair ${\pm m}$ of nonradial modes). We briefly sketch how 
the basic form of amplitude equations and the relations between coupling 
coefficients can be obtained. Our main aim is to capture the difference in the 
coupling coefficients for both cases by considering typical third order terms. 

The oscillatory mode is described by the displacement given in Eq.~(4). The 
nonlinear equations for the mode amplitudes contain operators acting on 
eigenfunctions and integrated over the volume of a star. For simplicity we 
will be concerned only with the radial component of the displacement. Taking 
into account angular components introduces additional terms in integrals but 
does not change the form of the equations and the relations between 
coefficients. Thus we can write 
$$\xi_{r,j}(\boldsymbol{r},t)=\frac{1}{2}a_j(t)w_j(\boldsymbol{r},t)+cc,\eqno(A1)$$
where
$$w_j(\boldsymbol{r},t)=h_j(r)Y_l^m(\theta,\phi){\rm e}^{{\rm i}\omega_jt}.\eqno(A2)$$
Degrees $l,m$ correspond to mode $j$. Spherical harmonics are expressed by the 
normalized associated Legendre functions 
$$Y_l^m(\theta,\phi)=(-1)^m\frac{1}{\sqrt{2\pi}}\tilde{P}_{l,m}(\cos\theta)
{\rm e}^{{\rm i}m\phi},\eqno(A3)$$
$$\tilde{P}_{l,m}(x)=\sqrt{\frac{(2l+1)(l-m)!}{2(l+m)!}}P_{l,m}(x),\eqno(A4)$$
and
$$P_{l,m}(x)=\frac{1}{2^l l!}(1-x^2)^{m/2}\frac{{\rm d}^{l+m}}{{\rm d} 
x^{l+m}}(x^2-1)^l\eqno(A5)$$
is the associated Legendre function. Normalizations are
$$\int\!|Y_{l,m}|^2{\rm d}\Omega=1,\eqno(A6)$$
$$\int\limits_{-1}^1\!\tilde{P}_{l,m}^2(x)\,{\rm d} x=1.\eqno(A7)$$
We will assume the dominant nonlinear terms to be of the third order. These 
terms in equation for the time derivative of the amplitude $a_j$ are 
determined by the expression 
$$T_j=\frac{1}{I_j}\langle w_j|N|\xi_1,\xi_2,\xi_3\rangle,\eqno(A8)$$
where $I_j$ is the mode inertia, $N$ is some operator that is linear with 
respect to each argument and symmetrical with respect to interchanges of 
arguments and 1, 2, 3 correspond to all modes that gives nonvanishing resonant 
terms. The bracket denotes integration over the volume of the star. The 
operator in Eq.~(A8) can be presented in the form 
\setcounter{equation}{8}
\begin{eqnarray}
\lefteqn{\langle w_j|N|\xi_1,\xi_2,\xi_3\rangle=}\nonumber\\
&&=a_1a_2a_3\langle h_j|N_r|h_1,h_2,h_3\rangle
\langle Y_j|N_{\theta,\varphi}|Y_1,Y_2,Y_3\rangle
{\rm e}^{{\rm i}(\omega_1+\omega_2+\omega_3-\omega_j)t}.
\end{eqnarray}
The angular part of the operator may be simplified to the factor $\int 
Y_j^*Y_1Y_2Y_3{\rm d}\Omega$. 

We keep only resonant terms, \ie terms slowly varying or constant in time. 
This means one of the frequencies $\omega_1,\omega_2,\omega_3$ has to be taken 
with minus sign, which implies one of the eigenfunctions $\xi_1,\xi_2,\xi_3$ 
should be taken complex conjugated. Additionally, integration of the product 
of spherical harmonics includes integration of factor $\exp[{\rm 
i}\varphi(m_1+m_2+m_3-m_j)]$ where one of the numbers $m_1,m_2,m_3$ is taken 
with negative sign (complex conjugation of spherical harmonic) and for 
nonvanishing terms the sum of $m$'s in the exponent has to be zero. 

\subsection{The 1:1 Resonance}
The nonvanishing terms are
\begin{eqnarray}
\langle w_0|N|\xi_0,\xi_0,\xi_0^*\rangle&+&
\langle w_0|N|\xi_0,\xi_0^*,\xi_0\rangle+
\langle w_0|N|\xi_0^*,\xi_0,\xi_0\rangle=\nonumber\\
&=&3\langle h_0|N_r|h_0,h_0,h_0^*\rangle\!
\int\!(Y_0^*Y_0)^2{\rm d}\Omega\, |a_0|^2a_0,
\end{eqnarray}
and similarly
\begin{eqnarray}
\langle w_0|N|\xi_0,\xi_l,\xi_l^*\rangle&+&{\rm permutations}=\nonumber\\
&=&6\langle h_0|N_r|h_0,h_l,h_l^*\rangle
\!\int\! |Y_0|^2|Y_l|^2{\rm d}\Omega\, |a_l|^2a_0,\\
\langle w_0|N|\xi_0^*,\xi_l,\xi_l\rangle&+&{\rm permutations}=\nonumber\\
&=&3\langle h_0|N_r|h_0^*,h_l,h_l\rangle
\!\int\! (Y_0^*)^2Y_l^2{\rm d}\Omega\,{\rm e}^{2{\rm i}t(\omega_l-\omega_0)}a_l^2a_0^*.
\quad
\end{eqnarray}
All $m$'s are equal to zero. For even $l$ terms involving $\int Y_l^3$ do not 
vanish, but we neglect them as was explained in Section~3.1. Symmetrically 
\begin{eqnarray}
\langle w_l|N|\xi_l,\xi_l,\xi_l^*\rangle&+&{\rm permutations}=\nonumber\\
&=&3\langle h_l|N_r|h_l,h_l,h_l^*\rangle
\!\int\! (Y_lY_l^*)^2{\rm d}\Omega\, |a_l|^2a_l,\\
\langle w_l|N|\xi_l,\xi_0,\xi_0^*\rangle&+&{\rm permutations}=\nonumber\\
&=&6\langle h_l|N_r|h_l,h_0,h_0^*\rangle
\!\int\! |Y_0|^2|Y_l|^2{\rm d}\Omega\, |a_0|^2a_l,\\
\langle w_l|N|\xi_l^*,\xi_0,\xi_0\rangle&+&{\rm permutations}=\nonumber\\
&=&3\langle h_l|N_r|h_l^*,h_0,h_0\rangle
\!\int\! Y_0^2(Y_l^*)^2{\rm d}\Omega\,{\rm e}^{-2{\rm i}t(\omega_l-\omega_0)}a_0^2a_l^*.
\qquad
\end{eqnarray}
Radial integration may be done only over the envelope. The justification of it 
is that the radial mode eigenfunction is nearly zero in the interior and most 
terms involve it. The term involving only the nonradial mode eigenfunction is 
the saturation term and we are interested only in its real part, as was 
explained in Section~3.1. The real parts of saturation terms are determined by 
nonadiabacity of oscillations which is very weak in the deep interior and it 
is sufficient to restrict integration to nonadiabatic envelope. But in the 
envelope all radial eigenfunctions are nearly the same. It means the brackets 
on the right-hand sides of Eqs.~(A10)--(A15) are the same. 

Let us introduce the quantities
\begin{eqnarray}
J_{00}&=&3\langle h_0|N_r|h_0,h_0,h_0\rangle\!\int\!Y_0^4{\rm d} \Omega=
\frac{3}{4\pi}\langle h_0|N_r|h_0,h_0,h_0\rangle,\\
J_{0l}&=&3\langle h_0|N_r|h_0,h_l,h_l\rangle\!\int\!Y_0^2Y_l^2{\rm d} \Omega=
\frac{3}{4\pi}\langle h_0|N_r|h_0,h_0,h_0\rangle=J_{00},\\
J_{ll}&=&3\langle h_l|N_r|h_l,h_l,h_l\rangle\!\int\!Y_l^4{\rm d} \Omega=
\nonumber\\
&=&\frac{3}{2\pi}
\langle h_0|N_r|h_0,h_0,h_0\rangle\!\int_{-1}^1\!\tilde{P}_l^4(x){\rm d} x=2J_{00}H_l,
\end{eqnarray}
where we used normalization of spherical harmonics, the fact that they are 
real, ${h_0(r)=h_l(r)}$ in the envelope, and the easy-to-show property 
$$\int\!Y_1Y_2Y_3Y_4{\rm d} \Omega=\frac{1}{2\pi}\int_{-1}^1\!\tilde{P}_1
\tilde{P}_2\tilde{P}_3\tilde{P}_4{\rm d} x.$$
The quantity $H_l$ is equal $\int\limits_{-1}^1\!\tilde{P}_l^4(x){\rm d} x$. 
With Eqs.~(A16)--(A18) one can write more explicit expressions for the 
nonlinear terms~(A8) 
\begin{eqnarray}
I_0T_0&=&J_{00}|a_0|^2a_0+2J_{00}|a_l|^2a_0+
J_{00}\,a_l^2a_0^*\,{\rm e}^{2{\rm i}(\omega_l-\omega_0)t},\\
I_lT_l&=&2J_{00}|a_0|^2a_l+2J_{00}H_l|a_l|^2a_l+
J_{00}\,a_0^2\,a_l^*{\rm e}^{-2{\rm i}(\omega_l-\omega_0)t}.
\end{eqnarray}
With above expressions the amplitude equations are
\begin{eqnarray}
\frac{{\rm d} a_0}{{\rm d} t}&=&\kappa_0a_0+T_0,\\
\frac{{\rm d} a_l}{{\rm d} t}&=&\kappa_la_l+T_l.
\end{eqnarray}
Relations given by Eq.~(13) and the scaling-with-inertia properties (Eqs.~14 
and 15) are obtained in a straightforward manner from Eqs.~(A19) and (A20). 


\subsection{The 2:1+1 Resonance}
In this case we have nonvanishing resonant terms
\begin{eqnarray*}
\langle w_0|N|\xi_0,\xi_0,\xi_0^*\rangle&:&\quad\textrm{3 permutations},\\
\langle w_0|N|\xi_0,\xi_+,\xi_+^*\rangle&:&\quad\textrm{6 permutations},\\
\langle w_0|N|\xi_0,\xi_-,\xi_-^*\rangle&:&\quad\textrm{6 permutations},\\
\langle w_0|N|\xi_0^*,\xi_+,\xi_-\rangle&:&\quad\textrm{6 permutations},
\end{eqnarray*}
\begin{eqnarray*}
\langle w_+|N|\xi_+,\xi_+,\xi_+^*\rangle&:&\quad\textrm{3 permutations},\\
\langle w_+|N|\xi_+,\xi_0,\xi_0^*\rangle&:&\quad\textrm{6 permutations},\\
\langle w_+|N|\xi_+,\xi_-,\xi_-^*\rangle&:&\quad\textrm{6 permutations},\\
\langle w_+|N|\xi_-^*,\xi_0,\xi_0\rangle&:&\quad\textrm{3 permutations},
\end{eqnarray*}
and symmetrically for the mode $`-'$. All properties of the radial 
eigenfunctions are the same as in the previous case. The only $m$-dependent 
quantities are eigenfrequencies and spherical harmonics. Then quantities given 
by Eqs.~(A16)--(A18) are defined analogically, but the product $Y_l^mY_l^{-m}$ 
should be used instead of $Y_l^2$. Quantity $J_{00}$ is the same as in the 
${m=0}$ case, as well as $J_{0l}^m$ which is equal $J_{00}$, too. Finally, 
${J_{ll}^m=2J_{00}H_l^m}$, where ${H_l^m=\int\limits_{-1}^1\tilde{P}_{l,m}^4
(x){\rm d}x}$. Then the nonlinear terms~(A8) are 
\begin{eqnarray}
I_0T_0&=&J_{00}|a_0|^2a_0+2J_{00}\left(|a_+|^2+|a_-|^2\right)a_0+\nonumber\\
&&+2J_{00}\,a_-a_+a_0^*\,e^{i(\omega_++\omega_--2\omega_0)t},\\
I_\pm T_\pm&=&2J_{00}|a_0|^2a_\pm+
2J_{00}H_l^m\left(|a_\pm|^2+2|a_\mp|^2\right)a_\pm+\nonumber\\
&&+J_{00}\,a_0^2\,a_\mp^*e^{-i(\omega_++\omega_--2\omega_0)t}.
\end{eqnarray}
When we assume equality of nonradial modes inertia, which implies Eq.~(24), we 
obtain, similarly as in the previous case, Eqs.~(25)--(30). 


\section{Observed Radial Velocity and Light Curves}
We use here a simple analytical expressions (Dziembowski 1977) which should be 
sufficient for crude estimates presented in this paper. The expression for 
observable radial velocity due to an individual oscillation mode is given by 
$$V_{\rm rad}=\omega_{l,m}R_{ph}A_{l,m}\tilde{P}_{l,m}(\cos\Theta_0)
\cos(\omega_{l,m}t-\pi/2-m\Phi_0)(u_l+\alpha_Hv_l)\eqno(B1)$$
where $l,m$ are quantum numbers of the mode; $\omega_{l,m}$ is its frequency; 
$R_{\rm ph}$ is the photospheric stellar radius; $A_{l,m}$ and 
$\tilde{P}_{l,m}$ are  mode amplitude and the normalized associated Legendre 
function, respectively; $\Theta_0,\Phi_0$ determine the direction to the 
observer in the system of coordinates connected with the star; $u_l,v_l$ are 
certain limb darkening dependent quantities which take into account averaging 
over the stellar disc; and $\alpha_H=GM/(\omega^2R_{\rm ph}^3)$. For RR~Lyr 
stars this quantity is of the order of 0.1. Similarly we have the expression 
for the bolometric magnitude change 
\setcounter{equation}{1}
\begin{eqnarray}
\Delta M_{\rm bol}&=&-1.086A_{l,m}\tilde{P}_{l,m}(\cos\Theta_0)
\big[f_l\cos(\omega_{l,m}t+\psi_l-m\Phi_0)b_l+
\nonumber\\
&&+\cos(\omega_{l,m}t-m\Phi_0)(2b_l-c_l)\big]
\end{eqnarray}
where $b_l,c_l$ are again certain limb darkening dependent integrals over the 
stellar disc. Quantities $f_l,\psi_l$ are obtained with solving nonadiabatic 
oscillation equations. They allow us to translate the radius changes into the 
luminosity changes. The first term in the square bracket in Eq.~(B2) 
describes emerging flux oscillations while the second term describes the 
radius changes connected with oscillatory mode. The values of 
$b_l,c_l,u_l,v_l$ are given by Dziembowski (1977b) for Eddington's limb 
darkening law. For ${l>2}$ they are small due to the averaging effect. The 
highest values are obviously for ${l=1}$ and for radial modes and we focus on 
these modes. Then we have 
$$\tilde{P}_{0,0}=\sqrt{1/2},\qquad
\tilde{P}_{1,0}(\cos\Theta_0)=\sqrt{3/2}\cos\Theta_0,$$
$$\tilde{P}_{1,1}(\cos\Theta_0)=\sqrt{3/4}\sin\Theta_0.$$
The averaging coefficients are ${u_0=0.708}$, ${u_1=0.55}$, ${v_0=0}$, 
${v_1=0.45}$, ${b_0=1}$, ${b_1=0.708}$, ${c_0=0}$, and ${c_1=1.416}$. Since 
$\alpha_H$ is small we will neglect terms $\alpha_Hv_l$.


\subsection{The 1:1 Resonance}
Let us consider a superposition of a radial and a ${m=0}$ nonradial resonant 
modes satisfying the double-mode fixed-point solution of AEs. Then their real 
frequencies are exactly equal due to the phase-lock phenomenon. The relative 
phase $\Gamma$ is given by Eq.~(12) and we see that for phase-locked solution 
we have 
$$\Gamma=2(\phi_{l,0}-\phi_{0,0})$$
where $\phi_{j,0}$ are initial phases in the expression
$$\phi_j=\frac{{\rm d}\phi_j}{{\rm d} t}t+\phi_{j,0}.$$
For individual mode the time dependence is given by 
$\cos(\omega_j^lt+\phi_j)=\cos(\omega_jt+\phi_{j,0})$, where $\omega_j^l$ 
and $\omega_j$ are the linear and nonlinear frequencies, respectively. We may 
choose the initial phase $\phi_{0,0}$ to be zero. Then we have 
${\phi_{j,0}=\Gamma/2}$ and with the use of Eq.~(B1) for ${(l,m)=(0,0)}$ and 
$(l,0)$, we get 
\setcounter{equation}{2}
\begin{eqnarray}
V_{\rm rad}&=&A_0\omega_0R_{ph}u_0\tilde{P}_{0,0}\cos(\omega_0t-\pi/2)+\nonumber\\
&&+A_l\omega_lR_{ph}u_l\tilde{P}_{l,0}(\cos\Theta_0)\cos(\omega_lt-\pi/2+\Gamma/2)=
\nonumber\\
&=&A_0\omega R_{ph}\tilde{P}_{0,0}u_0
\left[\cos(\omega t-\pi/2)+\epsilon_v\cos(\omega t-\pi/2+\Gamma/2)\right],
\end{eqnarray}
where we adopted ${\omega_0=\omega_l\equiv\omega}$ and introduced
$$\epsilon_v=\frac{\tilde{P}_{l,0}(\cos\Theta_0)}{\tilde{P}_{0,0}}\frac{u_l}{u_0}
\frac{A_l}{A_0}.\eqno(B4)$$
Eq.~(B3) may be written in the form
\setcounter{equation}{4}
\begin{eqnarray}
V_{\rm rad}&=&A_{v,0}\left[\cos(\omega t-\pi/2)\left(1+
\epsilon_v\cos(\Gamma/2)\right)
-\epsilon_v\sin(\Gamma/2)\sin(\omega t-\pi/2)\right]=\nonumber\\
&=&A_{v,0}p\cos(\omega t-\pi/2+\varphi)
\end{eqnarray}
where 
$$A_{v,0}=A_0\omega R_{ph}u_0\tilde{P}_{0,0}\eqno(B6)$$
is the radial velocity amplitude for the radial mode,
$$p\cos\varphi=1+\epsilon_v\cos(\Gamma/2)\eqno(B7)$$
and
$$p\sin\varphi=\epsilon_v\sin(\Gamma/2).\eqno(B8)$$
From Eqs.~(B5), (B7), and (B8) we see that the
radial velocity amplitude is given by
$$A_v=A_{v,0}\sqrt{1+\epsilon_v^2+2\epsilon_v\cos(\Gamma/2)},\eqno(B9)$$
which is equivalent to Eq.~(72).

With the same choice of phases $\phi_{j,0}$ we obtain for the bolometric 
magnitude variations 
\setcounter{equation}{9}
\begin{eqnarray}
\Delta M_{\rm bol}&=&CA_0\tilde{P}_{0,0}\left[f_0\cos(\omega t+\psi_0)b_0+\cos\omega t(2b_0-c_0)\right]
+\nonumber\\
&&+CA_1\tilde{P}_{1,0}(\cos\Theta_0)f_1b_1\cos(\omega t+\psi_1+\Gamma/2)=\nonumber\\
&=&CA_0\tilde{P}_{0,0}f\!\left[\frac{2}{f}\cos\omega t+\cos(\omega t+\psi)+
\epsilon_M\cos(\omega t+\psi+\Gamma/2)\right],\qquad
\end{eqnarray}
where, after Dziembowski (1997b), we assume ${f_l=f_0\equiv f}$ and 
${\psi_l=\psi_0\equiv\psi}$, for our low degree modes. The quantity 
$\epsilon_M$ is defined in the same way as $\epsilon_v$ but with $b_l$ instead 
of $u_l$. The radius changes vanish for ${l=1}$ modes. The contribution to 
light variation from the radial mode is given by 
$$\Delta M_{{\rm bol},0}=A_{M{\rm bol},0}\cos(\omega t+\varphi_0)\eqno(B11)$$
where
$$A_{M{\rm bol},0}=CA_0\tilde{P}_{0,0}fp_0,\eqno(B12)$$
\setcounter{equation}{12}
\begin{eqnarray}
p_0\cos\varphi_0&=&\frac{2}{f}+\cos\psi,\\
p_0\sin\varphi_0&=&\sin\psi,
\end{eqnarray}
and
$$p_0^2=1+\frac{4}{f^2}+\frac{4}{f}\cos\psi.\eqno(B15)$$
After adding the contribution from the nonradial mode we obtain
\setcounter{equation}{15}
\begin{eqnarray}
\Delta M_{\rm bol}&=&CA_0\tilde{P}_{0,0}f\big\{
\cos\omega t\left[p_0\cos\varphi_0+\epsilon_M\cos(\psi+\Gamma/2)\right]-
\nonumber\\
&&-\sin\omega t\left[p_0\sin\varphi_0+\epsilon_M\sin(\psi+\Gamma/2)\right]\big\}
=\nonumber\\
&=&CA_0\tilde{P}_{0,0}fp_1\cos(\omega t+\varphi_1)=A_{M{\rm bol}}\cos(\omega t+\varphi_1),
\end{eqnarray}
where
\begin{eqnarray}
p_1\cos\varphi_1&=&p_0\cos\varphi_0+\epsilon_M\cos(\psi+\Gamma/2),\\
p_1\sin\varphi_1&=&p_0\sin\varphi_0+\epsilon_M\sin(\psi+\Gamma/2),
\end{eqnarray}
and 
$$A_{M{\rm bol}}=CA_0\tilde{P}_{0,0}fp_1.\eqno(B19)$$
Eqs.~(B13), (B14), (B17), and (B18) yield
$$p_1^2=p_0^2+\epsilon_M^2+2\epsilon_M
\left[\frac{2}{f}\cos(\psi+\Gamma/2)+\cos(\Gamma/2)\right].\eqno(B20)$$
Using Eqs.~(B12), (B19), and (B20) we see that the presence of nonradial mode 
changes the observed amplitude by the factor 
$$\frac{p_1}{p_0}=\sqrt{1+\frac{\epsilon_M^2}{p_0^2}+\frac{2\epsilon_M}{p_0^2}
\left[\frac{2}{f}\cos(\psi+\Gamma/2)+\cos(\Gamma/2)\right]}.\eqno(B21)$$
This equation is equivalent to Eq.~(74).


\subsection{The 2:1+1 Resonance}
We consider here a superposition of a radial mode and a pair of nonradial 
modes. The phase-lock phenomenon in this case causes the three frequencies to 
be equidistant, with the radial mode frequency in the center. The relative 
phase $\Gamma$ is given by Eq.~(23) and, similarly as in Appendix~B.1, for the 
constant amplitude solution, it is given by 
$$\Gamma=\phi_{m,0}+\phi_{-m,0}-2\phi_{0,0}.$$

Again we choose ${\phi_{0,0}=0}$. We denote amplitudes and resonant 
frequencies of the radial, the $(l,m)$, and ${(l,-m)}$ modes as $A_0$, $A_-$, 
$A_+$, and $\omega$, ${\omega-\lambda}$, ${\omega+\lambda}$, respectively. Now 
we have 
\setcounter{equation}{21}
\begin{eqnarray}
V_{rad}^m&=&A_0\omega R_{ph}u_0\tilde{P}_{0,0}\cos(\omega t-\pi/2)+\nonumber\\
&&+A_-(\omega-\lambda)R_{ph}u_l\tilde{P}_{l,m}(\cos\Theta_0)
\cos(\omega t-\lambda t-\pi/2+\phi_{m,0})+\nonumber\\
&&+A_+(\omega+\lambda)R_{ph}u_l\tilde{P}_{l,m}(\cos\Theta_0)
\cos(\omega t+\lambda t-\pi/2+\phi_{-m,0})=\nonumber\\
&=&A_{v,0}
[\cos(\omega t-\pi/2)+\epsilon_{v,-}\cos(\omega t-\lambda t-\pi/2+\phi_{-,0})
+\nonumber\\
&&+\epsilon_{v,+}\cos(\omega t+\lambda t-\pi/2+\phi_{+,0})]
\end{eqnarray}
where
$$\epsilon_{v,\pm}=\frac{A_\pm}{A_0}\frac{\omega\pm\lambda}{\omega}\frac{u_l}{u_0}
\frac{\tilde{P}_{l,m}(\cos\Theta_0)}{\tilde{P}_{0,0}}\approx
\frac{u_l}{u_0}\frac{\tilde{P}_{l,m}(\cos\Theta_0)}{\tilde{P}_{0,0}}\frac{A_\pm}{A_0}.\eqno(B23)$$
We made use here of ${\lambda\ll\omega}$. The amplitude $A_{v,0}$ is given in 
Eq.~(B6). Keeping only linear terms in $\epsilon_{v,\pm}$ and making use of 
certain well-known trigonometric relations we get 
$$V_{\rm rad}^m=A_{v,0}[1+a\cos(\lambda t+\beta)]
\cos\big(\omega t-\pi/2+b\cos(\lambda t+\gamma)\big)\eqno(B24)$$
where the modulation coefficients, $a,b$, and the phase shifts, 
$\beta,\gamma$, are given by 
\setcounter{equation}{24}
\begin{eqnarray}
a\cos\beta&=&\epsilon_{v,-}\cos\phi_{-,0}+\epsilon_{v,+}\cos\phi_{+,0},\\
a\sin\beta&=&-\epsilon_{v,-}\sin\phi_{-,0}+\epsilon_{v,+}\sin\phi_{+,0},\\
b\sin\gamma&=&\epsilon_{v,-}\cos\phi_{-,0}-\epsilon_{v,+}\cos\phi_{+,0},\\
b\cos\gamma&=&\epsilon_{v,-}\sin\phi_{-,0}+\epsilon_{v,+}\sin\phi_{+,0}.
\end{eqnarray}
For ${l=1}$ and ${A_-=A_+\equiv A_1/\sqrt{2}}$ we obtain
$$\epsilon_{v,\pm}=\frac{\sqrt{3}}{2}\sin\Theta_0\frac{u_1}{u_0}\frac{A_1}{A_0}
\equiv\frac{\epsilon_{v,m}}{2}.\eqno(B29)$$
Inserting it to Eqs.~(B25)--(B28) leads to
\setcounter{equation}{29}
\begin{eqnarray}
a\cos\beta&=&\frac{\epsilon_{v,m}}{2}(\cos\phi_{-,0}+\cos\phi_{+,0})=
\epsilon_{v,m}\cos(\Gamma/2)\cos(\Delta/2),\\
a\sin\beta&=&\epsilon_{v,m}\cos(\Gamma/2)\sin(\Delta/2),\\
b\sin\gamma&=&\epsilon_{v,m}\sin(\Gamma/2)\sin(\Delta/2),\\
b\cos\gamma&=&\epsilon_{v,m}\sin(\Gamma/2)\cos(\Delta/2),
\end{eqnarray}
where ${\Gamma=\phi_{+,0}+\phi_{-,0}}$ and ${\Delta=\phi_{+,0}-\phi_{-,0}}$.
Finally we get
\begin{eqnarray}
a&=&\epsilon_{v,m}\cos(\Gamma/2),\\
b&=&\epsilon_{v,m}\sin(\Gamma/2),\\
\beta&=&\gamma=\Delta/2.
\end{eqnarray}
Since the modulation frequency, $\lambda$, is much lower than the oscillation 
frequency, $\omega$, the choice of initial modulation phase is unimportant, 
and we may put $\Delta=0$ in Eq.~(B36) and thus get Eq.~(79). 

In a very similar manner we get the following expression for the bolometric 
light variations 
\begin{eqnarray}
\Delta M_{\rm bol}^m=A_{M{\rm bol},0}\big\{\cos\omega t\big[\cos\varphi_0&\!+&\nonumber\\
+\frac{\cos\lambda t}{p_0}&\!\!\big(&\!\!
\cos\psi(\epsilon_{M,-}\cos\phi_{-,0}+\epsilon_{M,+}\cos\phi_{+,0})-\nonumber\\
&\!-&\!\sin\psi(\epsilon_{M,-}\sin\phi_{-,0}+\epsilon_{M,+}\sin\phi_{+,0})\big)+
\nonumber\\
+\frac{\sin\lambda t}{p_0}&\!\!\big(&\!\!
\cos\psi(\epsilon_{M,-}\sin\phi_{-,0}-\epsilon_{M,+}\sin\phi_{+,0})+\nonumber\\
&\!+&\!\sin\psi(\epsilon_{M,-}\cos\phi_{-,0}-\epsilon_{M,+}\cos\phi_{+,0})\big)
\big]-\nonumber\\
-\sin\omega t\big[\sin\varphi_0&\!+&\nonumber\\
+\frac{\cos\lambda t}{p_0}&\!\big(&\!
\sin\psi(\epsilon_{M,-}\cos\phi_{-,0}+\epsilon_{M,+}\cos\phi_{+,0})+\nonumber\\
&\!+&\!\cos\psi(\epsilon_{M,-}\sin\phi_{-,0}+\epsilon_{M,+}\sin\phi_{+,0})\big)+
\nonumber\\
+\frac{\sin\lambda t}{p_0}&\!\!\big(&\!\!
\sin\psi(\epsilon_{M,-}\sin\phi_{-,0}-\epsilon_{M,+}\sin\phi_{+,0})-\nonumber\\
&\!-&\!\cos\psi(\epsilon_{M,-}\cos\phi_{-,0}-\epsilon_{M,+}\cos\phi_{+,0})\big)
\big]\big\}\nonumber\\
&&
\end{eqnarray}
where $A_{M{\rm bol},0}$ is given by Eq.~(B12), $p_0,\varphi_0$ are given by 
Eqs.~(B13), (B14), and 
$$\epsilon_{M,\pm}=\sqrt{\frac{3}{2}}\sin\Theta_0\frac{b_1}{b_0}\frac{A_\pm}{A_0}.\eqno(B38)$$
For ${\epsilon_{M,\pm}\ll1}$ we get approximately 
$$\Delta M_{\rm bol}^m=A_{M{\rm bol},0}[1+c\cos(\lambda t+\beta')]
\cos\big(\omega t+\varphi_0+d\cos(\lambda t+\gamma')\big)\eqno(B39)$$
where
\setcounter{equation}{39}
\begin{eqnarray}
p_0(c\cos\varphi_0\cos\beta'&-&d\sin\varphi_0\cos\gamma')=\nonumber\\
&=&\epsilon_{M,-}\cos(\psi+\phi_{-,0})+\epsilon_{M,+}\cos(\psi+\phi_{+,0}),\\
p_0(-c\cos\varphi_0\sin\beta'&+&d\sin\varphi_0\sin\gamma')=\nonumber\\
&=&\epsilon_{M,-}\sin(\psi+\phi_{-,0})-\epsilon_{M,+}\sin(\psi+\phi_{+,0}),\\
p_0(c\sin\varphi_0\cos\beta'&+&d\cos\varphi_0\cos\gamma')=\nonumber\\
&=&\epsilon_{M,-}\sin(\psi+\phi_{-,0})+\epsilon_{M,+}\sin(\psi+\phi_{+,0}),\\
p_0(c\sin\varphi_0\sin\beta'&+&d\cos\varphi_0\sin\gamma')=\nonumber\\
&=&\epsilon_{M,-}\cos(\psi+\phi_{-,0})-\epsilon_{M,+}\cos(\psi+\phi_{+,0}).
\end{eqnarray}
For ${A_+=A_-\equiv A_1/\sqrt{2}}$ we have again
$$\beta'=\gamma'=\Delta/2\eqno(B44)$$
and then
\setcounter{equation}{44}
\begin{eqnarray}
c\cos\varphi_0-d\sin\varphi_0&=&\frac{\epsilon_{M,m}}{p_0}\cos(\psi+\Gamma/2),\\
c\sin\varphi_0+d\cos\varphi_0&=&\frac{\epsilon_{M,m}}{p_0}\sin(\psi+\Gamma/2)
\end{eqnarray}
where $\epsilon_{M,m}\equiv2\epsilon_{M,+}=2\epsilon_{M,-}$. The above equations 
lead to
\begin{eqnarray}
c&=&\frac{\epsilon_{M,m}}{p_0}\cos(\psi+\Gamma/2-\varphi_0),\\
d&=&\frac{\epsilon_{M,m}}{p_0}\sin(\psi+\Gamma/2-\varphi_0).
\end{eqnarray}
Again choosing $\Delta=0$ in Eq.~(B44) we obtain Eq.~(B39) to be equivalent to 
Eq.~(82) 

\begin{references}
\refitem{Aerts, C., and Eyer, L.}{2000}{~}{~}{``Delta Scuti and Related 
Stars'', {\it ASP Conf. Ser.}, 210, Eds. M. Breger and M. Montgomery, p.~113}
\refitem{Alcock, C., \etal}{2000}{\ApJ}{542}{257}
\refitem{Blazhko, S.}{1907}{Astron. Nachr.}{175}{325}
\refitem{Buchler, J.R., and Goupil, M.J.}{1984}{\ApJ}{279}{394}
\refitem{Buchler, J.R., Goupil, M.J., and Hansen, C.J.}{1997}{\AA}{321}{159}
\refitem{Buchler, J.R., Goupil, M.J., and Serre, T.}{1995}{\AA}{296}{405}
\refitem{Chadid, M., Kolenberg, K., Aerts, C., and Gillet, D.}{1999}{\AA}{352}{201}
\refitem{Dziembowski, W.A.}{1977a}{\Acta}{27}{95}
\refitem{Dziembowski, W.A.}{1977b}{\Acta}{27}{203}
\refitem{Dziembowski, W.A.}{1982}{\Acta}{32}{147}
\refitem{Dziembowski, W.A., and Cassisi, S.}{1999}{\Acta}{49}{371}
\refitem{Dziembowski, W.A., and Kr\'olikowska, M.}{1985}{\Acta}{35}{5}
\refitem{Garrido, R.}{2000}{~}{~}{``Delta Scuti and Related Stars'', {\it ASP 
Conf. Ser.}, 210, Eds. M. Breger and M. Montgomery, p.~67}
\refitem{Jurcsik, J., and Kov\'acs, G.}{1996}{\AA}{312}{111}
\refitem{Kov\'{a}cs, G.}{1993}{~}{~}{``Stochastic Processes in Astrophysics'', 
{\it Annals of New York Academy of Sciences}, 706, Eds. J.R. Buchler and H.E. 
Kandrup, p.~70}
\refitem{Kov\'{a}cs, G.}{2000}{~}{~}{``Nonlinear Studies of Stellar 
Pulsation'', Eds. M. Takeuti and D.D. Sasselov, {\it Astrophysics and Space 
Science Library Series}, Kluwer (in press)}
\refitem{Kov\'{a}cs, G., and Buchler, J.R.}{1988}{\ApJ}{324}{1026}
\refitem{Kov\'acs, G., and Jurcsik, J.}{1996}{\ApJL}{466}{L17}
\refitem{Kurtz, D.W., \etal}{2000}{~}{~}{``The Impact of Large-Scale 
Surveys on Pulsating Star Research'', {\it ASP Conf. Ser.}, 203, Eds. L. 
Szabados and D. Kurtz, p.~291}
\refitem{Moskalik, P.}{1986}{\Acta}{36}{333}
\refitem{Osaki, Y.}{1977}{PASJ}{29}{234}
\refitem{Peterson, R.C., Carney, B.W., and Latham D.W.}{1996}{\ApJL}{465}{L47}
\refitem{Shibahashi, H.}{1995}{~}{~}{in: "GONG'94: Helio- and 
Asteroseismology", {\it ASP Conf. Ser}, Vol.~76, Eds. R.K. Ulrich, E.J. 
Rhodes, Jr., and W. D{\'a}ppen, p.~618}
\refitem{Shibahashi, H.}{2000}{~}{~}{``The Impact of Large-Scale Surveys on 
Pulsating Star Research'', {\it ASP Conf. Ser.}, 203, Eds. L. Szabados and 
D. Kurtz, p.~299}
\refitem{Szeidl, B.}{1988}{~}{~}{``Multimode Stellar Pulsations'', Eds. 
G. Kov\'{a}cs, L. Szabados, B. Szeidl, Konkoly Observatory, Kultura, 
Budapest, p.~45}
\refitem{Unno, W., Osaki, Y., Saio, H., and Shibahashi, H.}{1989}{~}{~}
{``Nonradial Oscillations of Stars'', University of Tokyo Press}
\refitem{Van~Hoolst, T.}{1992}{~}{~}{``Nonlinear, Nonradial Oscillations of 
Stars. Resonances Between Two Modes with Nearly Equal Frequencies'',
PhD thesis, Katholieke Universiteit, Leuven}
\refitem{Van~Hoolst, T.}{1994}{\AA}{292}{471}
\refitem{Van~Hoolst, T., Dziembowski, W.A., and Kawaler, S.D.}{1998}{MNRAS}{297}{536}
\refitem{Van~Hoolst, T., and Waelkens, C.}{1995}{\AA}{295}{361}
\refitem{Wersinger, J.M., Finn, J.M., and Ott, E.}{1980}{Physics of Fluids}{23}{1142}
\end{references}
\end{document}